\documentclass[namedreferences,hyperref,optionalrh,solaromanenum]{spr-sola}

\usepackage{graphicx}              % For eps figures, newer & more powerfull
\usepackage{color}                 % For color text: \color command
\usepackage{breakurl}              % For breaking URLs easily trough lines in DVI mode
\usepackage{mathtools}
\usepackage{multirow}
\usepackage{amsmath}
\usepackage{url}
\renewcommand{\vec}[1]{\mathbfit{#1}}
\chardef\us=`\_

\begin{document}

%\linenumbers

\begin{frontmatter}

\title{The McNish and Lincoln Solar Activity Predictions: The Method and its Performance}

\author[addressref={inst1},corref,email={frederic.clette@oma.be}]
{\inits{F.}\fnm{Fr{\'e}d{\'e}ric}~\lnm{Clette}\orcid{0000-0002-3343-5153} }

\author[addressref={inst2},email={shantanu.jain@skoltech.ru}]
{\inits{S.}\fnm{Shantanu}~\lnm{Jain}\orcid{0000-0003-3883-8960} }

\author[addressref={inst2},email={t.podladchokova@skol.tech}]
{\inits{T.}\fnm{Tatiana}~\lnm{Podladchikova}\orcid{0000-0002-9189-1579} }

\address[id=inst1]{World Data Center SILSO, Royal Observatory of Belgium, 1180 Brussels, Belgium}
\address[id=inst2]{Skolkovo Institute of Science and Technology, Skolkovo Innovation Center, Building 3, Moscow 143026, Russia}

\runningauthor{F.\,Clette, S.\,Jain and T.\,Podladchikova}
\runningtitle{The McNish and Lincoln Prediction Method}

\begin{abstract}
The McNish and Lincoln (ML) method, introduced in 1949,  was one of the first attempts to produce mid-term forecasts of solar activity, up to 12 months ahead. However, it has been poorly described and evaluated in the past literature, in particular its actual operational implementation by NOAA. 	
Here, we reconstruct the exact formulation of the method, as it was applied since the early 1970s, and we provide a full mathematical derivation of the prediction errors. For bench-marking the method, we also produce monthly predictions over the past 190 years, from 1833 (Cycle 8) to 2023 (Cycle 25), and develop statistics of the differences between the predictions and the observed 13-month smoothed sunspot number (SSN) time series, according to the phase in the solar cycle.
Our analysis shows that the ML method is heavily constrained because it is primarily based on the mean of all past cycles, which imposes a fixed amplitude and length and suffers from a temporal smearing that grows towards the end of the solar cycle.  We find that predictions are completely unreliable in the first 12 months of the cycle, and over the last two years preceding the ending minimum (around 130 months), and beyond this minimum. By contrast, in the course of the cycle (months 18 to 65), ML predictions prove to be reliable over a time range of up to 50 months (4.2 years), thus much longer than the 12-month conventional range used so far. However, we find that predictions then suffer from systematic under-(over-)estimates for cycles that have a higher (lower) amplitude than the base mean cycle. 
Overall, we conclude that although the ML method provides valid prediction errors, it suffers from strong limitations, with very little room for improvement, as it indifferently merges all past cycles into a single fixed statistics.
\end{abstract}

% \keywords{Sun -- Solar activity -- Solar indices -- Solar irradiance (radio) -- Solar cycle}

\end{frontmatter}

%________________________________________________________________________

\section{Introduction}
Our planet is permanently exposed to the influence of the Sun, which is modulated by the 11-year cycle \citep{GrayEtal2010, SolankiEtal2013AnnRev, ChatzistergosEtal2023JASTP}. Over recent decades, the development of new technologies (manned and unmanned space missions, telecommunications, space-based navigation systems) and the ever-growing dependency of modern societies on electric energy have strongly increased the vulnerability of human activities to variations of the solar magnetic activity. Many of the risks are not only associated to the short-term predictions of individual solar eruptive events, but on the mid- and long-term evolution of the occurrence rate and of the power of such solar eruptions, and also on cumulative effects. This includes for instance, the low-Earth orbit satellite lifetime (atmospheric drag), the maintenance of major ground infrastructures (e.g. corrosion of pipelines) or health effects on aviation staff (cumulated radiation doses).

Already by the mid-$20^{th}$ century, the necessity of mid- and long-term forecasts of the evolution of the solar cycles prompted solar physicists to develop the first operational prediction methods. Given the limited computing means and the absence of any physical model of the sub-surface dynamo process producing the magnetic fields that emerge at the solar surface, the first strategy consisted in using the past history of the solar cycle. Those early methods are now included in the ``climatology'' category of predictions \citep{Pesnell2020,Petrovay2020LRSP}, as they rely on the statistics of past solar cycles to derive the most probable future evolution, given the most recent observed progress of the current solar cycle.

The first operational methods were those of McNish and Lincoln \citep{McNishLincoln1949} (hereafter ML) and Waldmeier \citep{Waldmeier1968}, the latter having actually its roots in a set of standard solar-cycle curves created in the 1930s \citep{Waldmeier1935, Waldmeier1937PhDT}. As, at that time, the only multi-century record of solar cycle is provided by the international sunspot number \citep{Wolf1856, CletteLefevre2016} (hereafter SN), both methods are entirely based on this reference time series. Likewise, the forecasts are also expressed in terms of the SN, as the latter is also used as the standard measure of solar activity for most long-term research applications. 

The SN series is continuously maintained and extended by the World Data Center (WDC) ``Sunspot Index and Long-term Solar Observations'' (SILSO; \url{https://www.sidc.be/SILSO/home}), and following an end-to-end re-calibration in 2015, it was recently upgraded to Version 2. As the details of the day-to-day variations of solar activity have a strong random and local character unrelated to the underlying global dynamo process \citep{DudokEtAl2016}, and as those forecasts have a long-term scope, the SN is mostly used in the form of 13-month smoothed monthly mean sunspot numbers (noted hereafter SSN), from which the times of the minima and maxima of the cycles are also derived. 

Since 1998, the time series of the group sunspot number \citep{HoytSchatten1998}, which extends further back in time to the early $17^{th}$ century, has become available next to the SN. However, given the cruder observations and large uncertainties in the early data, adding more solar cycles does not bring any gain in predictive accuracy. Moreover, currently, only the SN is continuously kept up to date until the present, which is necessary for operational predictions. Therefore, even today, most prediction methods, which now call on a broader array of more advanced approaches \citep{Petrovay2020LRSP,Pesnell2020}, still rely on the homogeneous sunspot data after the mid-$19^{th}$ century, as provided by the SN.

In this article, we will focus on the ML method, which was developed at the National Bureau of Standards in 1949, in support to the NOAA (National Oceanic and Atmospheric Administration) in the USA. The ML earned its importance firstly because it led to all mid-term solar activity predictions published during several decades in the monthly issues of the Solar Geophysical Data bulletin, an operational compilation of solar-terrestrial data that was published on a monthly basis from 1955 to 2009 (\url{ftp://ftp.ngdc.noaa.gov/STP/SOLAR_DATA/SGD_PDFversion}). For a few more years, until 2016, those predictions were continued as an on-line product by the NGDC at NOAA (National Geophysical Data Center, now part of the NCIE, National Centers for Environmental Information), and then were taken over by the WDC SILSO, where they are still produced on a monthly basis, next to two other prediction methods, together with optimized versions of those predictions based on an adaptive Kalman filter \citep{PodladchikovaRVDL2012}. They are distributed via the SILSO Web site (\url{https://www.sidc.be/SILSO/forecasts}). 

Given this longevity, many operational applications and sciences analyses have been based on the ML method over a very long period. When revisiting past published results, a proper understanding of the ML method is thus often required. Moreover, nowadays, in hindsight, this rather basic ``climatology'' method forms an important benchmark for a large diversity of more advanced prediction methods. Indeed, any new method must demonstrate in a quantitative way to what extent it provides an effective improvement in precision and reliability against baseline predictions from base classical methods like ML. Therefore, a proper quantification of the performance of the ML method is necessary for such bench-marking.

Unfortunately, very little was published about the ML method besides the initial paper \citep{McNishLincoln1949}, which itself only describes the base principle and a simple sample application that is different from the later operational implementation. A few more recent articles explored a bit the limitations of the ML methods \citep{StewardOstrow1970, HollandVaughan1984, Hildner1990, Niehuss1996, Fessant1996, Lantos2006}, but they only studied the improvements obtained by bringing partial modifications to the original method, for the predictions of only one or two sample cycles. 

In order to fill those past gaps, the goal of this article is first to fully document the primary operational ML method as implemented by the NGDC/NOAA, based directly on the original FORTRAN programs that were communicated by NOAA to us at WDC SILSO in 2016. In Section\,\ref{Sec_History}, we first  retrace the history of the ML method implementations. In Section\,\ref{Sec_MLmethod}, we then interpret the actual formulae in relation with the original formulation by \citet{McNishLincoln1949}. In Section\,\ref{Sec_PredError}, we provide a full mathematical derivation of the error formula embedded in the programmed method. In Section\,\ref{Sec_KeyProp}, we extract the key properties of the primary components of the ML predictions. After generating a large set of predictions for all usable past solar cycles (number 8 to 25, with the conventional numbering where Cycle 1 is the one starting in 1755), in Section\,\ref{Sec_Statistics}, we then derive statistics of the differences between the predictions and actual SSN values, in order to quantify the uncertainties of the ML predictions and compare them to the above mathematical error. Finally, in Sections\,\ref{Sec_Discussion} and \ref{Sec_Conclusions}, we draw interpretations and several key conclusions about the temporal ranges over which the ML method is applicable and about irreducible limitations imposed by the base underlying principles of this basic method.
 
\section{A Brief History: The Original Version of the ML Method and its Later Adaptations} \label{Sec_History}

\subsection{The Original Foundations} \label{Sec_Fundations}
In their original 1949 article, McNish and Lincoln introduce their two base assumptions:
\begin{enumerate}
	\item Given the cyclic nature of solar activity, the base prediction for any time in a future cycle is simply the mean value of all observed past cycles at the same relative time, counted from the start of each cycle. For their calculation, all cycles are thus aligned on their starting minimum, with a common time axis, counted relative to this starting date.
	\item  This first mean estimate can be improved by adding a correction proportional to the departure of the most recent observed SSN values of the current cycle from the corresponding mean value, multiplied by a proportionality factor determined by the method of least squares.
\end{enumerate}
In order to achieve a predictive capability, they derive this correction factor based on a least-square regression over all past pairs of years separated by the same interval in all previous cycles. We refer the reader to the original text for the details of this first implementation \citep{McNishLincoln1949}. In the next Section, we will provide the complete formulation, but historically, the relevant elements emerging from this original reference are the following. 

In their initial formulation, McNish and Lincoln only worked with yearly mean SNs and derived a prediction only for the next year after the current year. In the final part of their article, where they apply the method to Cycle 18, they also briefly experiment predictions based on shorter 3-month running means and obtained at 8-month intervals, in order to reduce the operational delay between the time of the last available SSN and the time of prediction. 

One of their main findings is that the best predictions are already obtained by using only one past SN value in the current cycle, namely the 12-month mean from the previous year (year$-1$). The inclusion of year$-2$ or year$-3$ does not bring any significant gain in accuracy. This choice of basing predictions only on the most recent observed SN will be kept in all subsequent implementations of the ML method. Based on a Chi-square test, McNish and Lincoln also conclude that historical SN values before Cycle 8 are less accurate and could degrade the quality of the predictions. They thus only use data from Cycle 8 up to the last cycle before the cycle in progress at the time of prediction. This choice of ignoring all cycles before Cycle 8 also persisted in most subsequent implementations. We finally point out that no calculation of the prediction error was made in this initial article.

\subsection{Multiple Adaptations} \label{Sec_Adaptations}
As the time delay of one year between successive predictions and a temporal resolution of one year was too coarse for actual applications, in 1970, \citet{StewardOstrow1970} transformed the original ML formulation to produce predictions at monthly intervals using monthly values the SSN. From then on, this became the base method used by the Space Environment Services Center (SESC) and later on by the NOAA and the Paris-Meudon Observatory. It survived up to us nowadays through the heritage FORTRAN programs used by NGDC/NOAA until 2016. 

Though quite different from the original yearly scheme, this upgrade keeps the two base principles. Each prediction is based only on a single starting value: now the last monthly SSN value. In this new monthly scheme, predictions are calculated for 18 months following this last available SSN value. As this SSN series stops six months before the current month (last observed monthly SN) due to the 13-month symmetrical smoothing window, this thus delivers 12-month ahead predictions relative to the present, corresponding to the single one-year ahead value of the initial ML implementation. In this case, the correction factor is also derived by a least square regression over equivalent pairs of dates over all past observed cycles, but for different times intervals (number of months) instead of a single 1-year interval. Together with this upgrade, \citet{StewardOstrow1970} also included formulae giving the errors on the predictions and on the correction coefficient itself. This is the implementation described in detail in the rest of this article.

However, as new solar cycles elapsed, the ML predictions proved to fall short of expectations. When applying the method to predict the maxima of cycles up to Cycle 22 in the course of the ascending phase,  \citet{Hildner1990} found clear limitations. In particular, while the maximum SN value can be reasonably well predicted for most solar cycles, the ML method cannot reliably predict the exact time of maximum. This failure indicates that the ML method cannot account for the variable rise time and duration of solar cycles. 

Starting from a similar idea, \citet{Niehuss1996} tried to eliminate the differences in cycle lengths, which lead to a misalignment of the maxima and ending minima of the cycles. They chose to reduce all cycles to a single common length, equal to the mean length of all past cycles, using a Lagrangian interpolation scheme first introduced by \citet{HollandVaughan1984} for an application to the F10.7 radio flux time series. Moreover, in order to avoid a discontinuity when passing a cycle minimum, they add a parallel scheme that goes from maximum to maximum instead of from minimum to minimum. Although they obtain a slight improvement of predictions around the time of maximum and the first years of the declining phase of the cycle, their method still involves a base mean cycle of fixed duration, which makes predictions of the end of a cycle unrealistic. 

Almost simultaneously, \citet{Fessant1996} also reduced all past cycles to a common length using the interpolation method of \citet{HollandVaughan1984}. In addition, instead of applying the method to the whole cycle, they separated the ascending phases and descending phases of each cycle, and equalized their lengths before applying separately the ML principles to each part. This ``split at maximum'' approach brought a slight improvement, in particular in the declining part of the cycle. By comparing the ML results with a new method based on a neural network, they concluded on the superiority of the new method, mainly for cycles with a peculiar evolution. Although the neural network approach did not allow the authors to explain this better performance, they speculated that it is mainly due to a better temporal flexibility compared to the ML method.

We finally point out that those few past publications include only very limited evaluations of the ML output, often training the method on part of the ongoing cycle at the time of publication, or considering only one predictive application (next cycle maximum). Here, before conducting a thorough assessment of the reliability of ML predictions, we first describe in detail the mathematical formulation of the method and the prediction error.

\section{The Standard ML Method} \label{Sec_MLmethod}

\subsection{The Mean Cycle} \label{Sec_MeanCycle}
In the original method described by McNish and Lincoln (1949), the base element of the prediction is a mean cycle derived by using all cycles before the current (ongoing) one. In order to build this mean cycle, all cycles are aligned on the month of the minimum, and the time increments (here, months) are counted from this initial point. For each time increment, all values for that same relative time in each solar cycle are then averaged.

Let $S_{m}^{n}$ denote the value of the SSN at any month $m$ of a Cycle $n$. The sum starts at Cycle 8 up to $c-1$, where $c$ is the number of the last available cycle, thus the current ongoing cycle for which the predictions are made. The number of included cycles is thus $N_{c}=c-8$. We assume that the sequence $S_{m}^{n},\ (n=8,9, \ldots, c-1)$ is uncorrelated. The mean value, $\bar{S}_{m}$, of the SSN over those $N_c$ cycles, for the $m^{th}$ month in the cycle, is estimated as 
%\begin{linenomath}
\begin{equation}    \label{eq_A1}
\bar{S}_{m}= \frac{1}{N_{c}} \sum_{n=8}^{c-1} S_{m}^{n}  
\end{equation}
%\end{linenomath}
%
The variance of $S_{m}^{n}$ relative to this mean $\bar{S}_{m}$ for month $m$ is defined as
%\begin{linenomath}
\begin{equation}\label{eq_A9}
	\sigma_{m}^{2}= \frac{\sum_{n=8}^{c-1} \left(S_{m}^{n} - \bar{S}_{m}\right)^{2}} {N_{c}-1}   
	= \frac{\sum_{n=8}^{c-1}\left(S_{m}^{n}\right)^{2} - N_{c} \,\bar{S}_{m}^{2}} {N_{c}-1}	              
\end{equation}
%\end{linenomath}
where $S_{m}^{n}$ is the SSN value at any month $m$ of one Cycle $n$, and $\bar{S}_{m}$ is the corresponding mean SSN value at month $m$ over all past solar cycles, in the range $(n=8,9, \ldots, c-1)$.

Rather than representing observational errors, this variance thus essentially translates the full range of actual amplitudes of all solar cycles in our past record, relative to the mean cycle. It thus gives a measure of how badly the mean cycle can account for the wide diversity of actual cycles. As such, the mean cycle can be considered as the zero-order guess, in the absence of any other information. 

\begin{figure}
	\centerline{
		\includegraphics*[width= 0.85\textwidth, bb= 4mm 0mm 186mm 131mm,clip=] {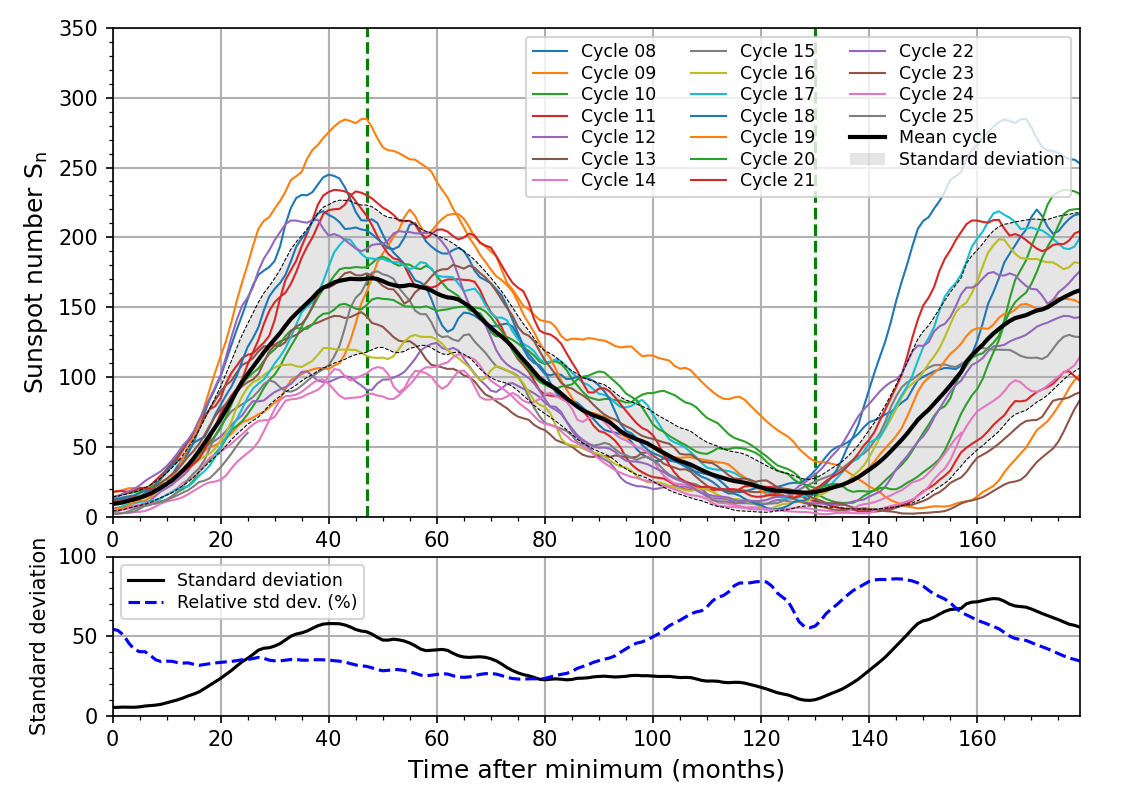}
	} 
	\caption{Construction of the mean cycle $\bar{S}_{m}$. In the upper plot showing the SSN as a function of time $m$ in months, thin colored lines correspond to all past cycles since Cycle 8, aligned on their starting minimum ($m=0$), while the thick black line is the mean $\bar{S}_{m}$ of all those cycles. All curves extend beyond the ending minimum of the mean cycle, as predictions made at the end of a solar cycle may extend beyond this mean minimum, and thus into the early rise of the next cycle. The standard deviation of all cycles around the mean cycle $\sigma_{m}$ is shown as the gray-shading in the upper plot and as the black solid curve in the lower plot, with also its relative value $\sigma_{m}/\bar{S}_{m}$, in percent of the mean-cycle value $\bar{S}_{m}$ (blue dashed line).}
		\label{Fig_MeanCycle}
\end{figure}

In Figure\,\ref{Fig_MeanCycle}, we plot the mean cycle (thick black line) together with all individual cycles included in this global mean (thin lines). The mean cycle is characterized by a broad and flat maximum peaking at SN\,=\,170 on month $m=47$, and by an ending minimum at month $m=130$ (10.8 years) with SN\,=\,17, which is higher than the starting minimum at SN\,=\,9. In the lower plot of Figure\,\ref{Fig_MeanCycle}, we can observe that the standard deviation $\sigma_{m}$ (black solid line) largely follows the modulation of the mean cycle. Like the latter, it is also larger at the ending minimum ($m=130$) than at the starting minimum ($m=0$). Moreover, the maximum of $\sigma_{m}$ after 130 months is higher than the first maximum around 40 months. 

\subsection{The ML Correction Term} \label{Sec_CorrecTerm}
As the mean cycle is a poor representation of very diverse cycles, the ML method aimed at improving this basic mean-cycle prediction by adding a correction term. The latter is based on two elements:
\begin{itemize}
	\item The actual deviation of the last observed SSN, at month $s$ (starting month for the predictions) relative to the corresponding value of the mean cycle
	\item A multiplicative gain factor based on a least-square linear regression between the values at the month of prediction $p$ and at month $s$ of the last observed SSN, over all past observed cycles
\end{itemize}
As mentioned earlier in Section\,\ref{Sec_Adaptations}, we stress again here that the reference month $s$ of the last SSN precedes by 6 months the actual moment of the prediction (last observed SN), due to the 13-month SSN smoothing window. Therefore, in operational production, month $s$ is not the actual present month. Still, as anyway $s$ corresponds to the last SSN used in the calculation and as all results are independent of this practical delay, for simplicity, we shall treat "s" as the current month in all following explanations, skipping this fixed 6-month operational shift.

Let us introduce the following residuals, denoting the difference between the SSN value and the value of the mean cycle at the reference month $m=s$ and the predicted month $m=p$ for Cycle $n$\,:
%\begin{linenomath}
\begin{equation}\label{eq_A11}
	\Delta_{s}^{n}=S_{s}^{n}-\bar{S}_{s}, \qquad (n = 8,9,... , c-1)
\end{equation}
%\end{linenomath}
%\begin{linenomath}
\begin{equation}\label{eq_A12}
	\Delta_{p}^{n}=S_{p}^{n}-\bar{S}_{p}, \qquad (n = 8,9,... , c-1)
\end{equation}
%\end{linenomath}

Then, we can form the following linear regression equation between $\Delta_{s}^{n}$ and $\Delta_{p}^{n}$
%\begin{linenomath}
\begin{equation}\label{eq_A13}
	\Delta_{p}^{n}=a_{sp}+k_{sp} \Delta_{s}^{n}+\varepsilon_{sp}^{n}
\end{equation}
%\end{linenomath}
%
Here, $\varepsilon_{sp}^{n}$ is the uncorrelated noise representing the model error.

Let us introduce the vector of differences at month $p$
%\begin{linenomath}
\begin{equation}\label{eq_A14}
	\vec{Y} =  \begin{vmatrix}
		\Delta_{p}^{8}  & \Delta _{p}^{9}  & \cdots   & \Delta _{p}^{c-1}\\
	\end{vmatrix}
	^{T}
\end{equation}
%\end{linenomath}
the vector of estimated parameters
%\begin{linenomath}
\begin{equation}\label{eq_A15}
	\vec{X_{sp}}= \begin{vmatrix}
		a_{sp}\\
		k_{sp}
	\end{vmatrix}
\end{equation}
%\end{linenomath}
the source-data matrix
%\begin{linenomath}
\begin{equation}\label{eq_A16}
	{\bf H} = \begin{vmatrix}
		1                & \ldots   & 1\\
		\Delta _{s}^{8}  & \ldots   &  \Delta _{s}^{c-1}\\
	\end{vmatrix}^{T}
\end{equation}
%\end{linenomath}
and the noise vector 
%\begin{linenomath}
\begin{equation}\label{eq_A17}
	\vec{\varepsilon} =  \begin{vmatrix}
		\varepsilon _{sp}^{8}  & \varepsilon_{sp}^{9}  & \cdots   & \varepsilon _{sp}^{c-1}\\
	\end{vmatrix}
	^{T}
\end{equation}
%\end{linenomath}

Then, the regression Equation\,(\ref{eq_A13}) can be rewritten as 
%\begin{linenomath}
\begin{equation}\label{eq_A18}
	\vec{Y} = {\bf H} \vec{X_{sp}} + \vec{\varepsilon} 
\end{equation}
%\end{linenomath}
%
The variance of the noise $\varepsilon_{sp}^{n}$, noted $(\sigma_{sp}^{\varepsilon})^{2}$, represents the scatter around the regression line.

We determine the estimate of vector $\vec{X_{sp}}$  with unknown parameters $a_{sp}$ and $k_{sp}$ on the basis of the least-square method \citep{SeberLee2003} 
%\begin{linenomath}
\begin{equation} \label{eq_A19}
	\vec{\hat{X}_{sp}} = \begin{vmatrix}
		\hat{a}_{sp}\\
		\hat{k}_{sp}\\
	\end{vmatrix}
	=\left({\bf H}^{T} {\bf H}\right)^{-1} {\bf H}^{T} \vec{Y}
\end{equation}
%\end{linenomath}
%
Then, Equation\,(\ref{eq_A19}) can be rewritten as 
%\begin{linenomath}
\begin{equation}\label{eq_A24}
	\vec{\hat{X}_{sp}} =  \begin{vmatrix}
		\hat{a}_{sp}\\
		\hat{k}_{sp}\\
	\end{vmatrix}
	= {\bf D} \vec{Y}
\end{equation}
%\end{linenomath}
%
Here, the matrix ${\bf D}$ is represented by 
%\begin{linenomath}
\begin{equation}\label{eq_A23}
	{\bf D} = \left({\bf H}^{T} {\bf H}\right)^{-1} {\bf H}^{T}=\begin{vmatrix}
		\frac{1}{N_{c}}  & \ldots   & \frac{1}{N_{c}} \\[3pt]
		\frac{\Delta_{s}^{8}}{\sum_{n=8}^{c-1}\left(\Delta_{s}^{n}\right)^{2}}  & \ldots   & \frac{\Delta_{s}^{c-1}}{\sum_{n=8}^{c-1}\left(\Delta_{s}^{n}\right)^{2}}\\
	\end{vmatrix}
\end{equation}
%\end{linenomath}

Taking into account Equations\,(\ref{eq_A12}) and (\ref{eq_A14}), the first element of vector $\vec{\hat{X}_{sp}}$ is estimated as
%\begin{linenomath}
\begin{equation}\label{eq_A23_ad}
	\hat{a}_{sp}=\frac{1}{N_{c}} \sum_{n=8}^{c-1}\left(S_{p}^{n}-\bar{S}_{p}\right)
\end{equation}
%\end{linenomath}
%
Given the definition of the mean cycle (Equation\,(\ref{eq_A1}), here with $m=p$), $\hat{a}_{sp}=0$. Therefore, in the original formulation adopted by \cite{McNishLincoln1949}, by construction, the intercept of the regression is always through the origin, and only one parameter is determined by the regression: the slope $\hat{k}_{sp}$, defined as
%\begin{linenomath}
\begin{equation}\label{eq_A22}
	\hat{k}_{sp}=\frac{\sum_{n=8}^{c-1}\Delta_{s}^{n}\Delta_{p}^{n}}{\sum_{n=8}^{c-1}\left(\Delta_{s}^{n}\right)^{2}}
\end{equation}
%\end{linenomath}
or, by re-developing explicitly 
%\begin{linenomath}
\begin{equation}
	\hat{k}_{sp} = \frac{\sum_{n=8}^{c-1}  \left(S_s^n -\bar{S}_s\right) \left(S_p^n -\bar{S}_p \right)}
	{\sum_{n=8}^{c-1} \left(S_s^n -\bar{S}_s \right)^2}
	\label{EQ_kCoeff}
\end{equation}
%\end{linenomath}
%
Thus, the vector $\vec{\hat{X}_{sp}}$ is estimated as 
%\begin{linenomath}
\begin{equation}\label{eq_A20}
	\vec{\hat{X}_{sp}} =  \begin{vmatrix}
		\hat{a}_{sp}\\
		\hat{k}_{sp}\\
	\end{vmatrix}
	= \begin{vmatrix}
		0\\
		\frac{\sum_{n=8}^{c-1}\Delta_{s}^{n}\Delta_{p}^{n}}{\sum_{n=8}^{c-1}\left(\Delta_{s}^{n}\right)^{2}}\\
	\end{vmatrix}
\end{equation}
%\end{linenomath}

Combining this result with Equations\,(\ref{eq_A12}) and (\ref{eq_A13}), we can now derive the prediction for month $p$ in Cycle $c$, based on reference month $s$:
%\begin{linenomath}
\begin{equation} \label{EQ_Predi}
	\hat{S}_{sp}^c = \bar{S}_p + \hat{k}_{sp}\,\left(S_s^c - \bar{S}_s \right) 
\end{equation}
%\end{linenomath}
where 
\begin{itemize}
	\item $\hat{S}_{sp}^c$ is the improved prediction for month $p$ in current Cycle $c$
	\item $\bar{S}_p$ is the mean-cycle value at month $p$, the target month of the prediction (from Equation\,(\ref{eq_A1}) with $m=p$)
	\item $S_s^c$ is the last observed SSN in the current cycle (acting as reference starting month $s$ for predictions)
	\item $\bar{S}_s$ is the mean-cycle value at the reference month $s$ (Equation\,(\ref{eq_A1}) with $m=s$)
	\item $\hat{k}_{sp}$ is the correction coefficient for the pair of months s -- p (Equation\,(\ref{EQ_kCoeff})). 
\end{itemize}

Equations\,(\ref{EQ_Predi}) and (\ref{EQ_kCoeff}) that we derived here match the calculations coded in the NGDC/NOAA heritage programs, confirming their exact derivation. 

\begin{figure}
	\centerline{
		\includegraphics*[width= 0.85\textwidth,bb= 5mm 0mm 186mm 124mm,clip=]{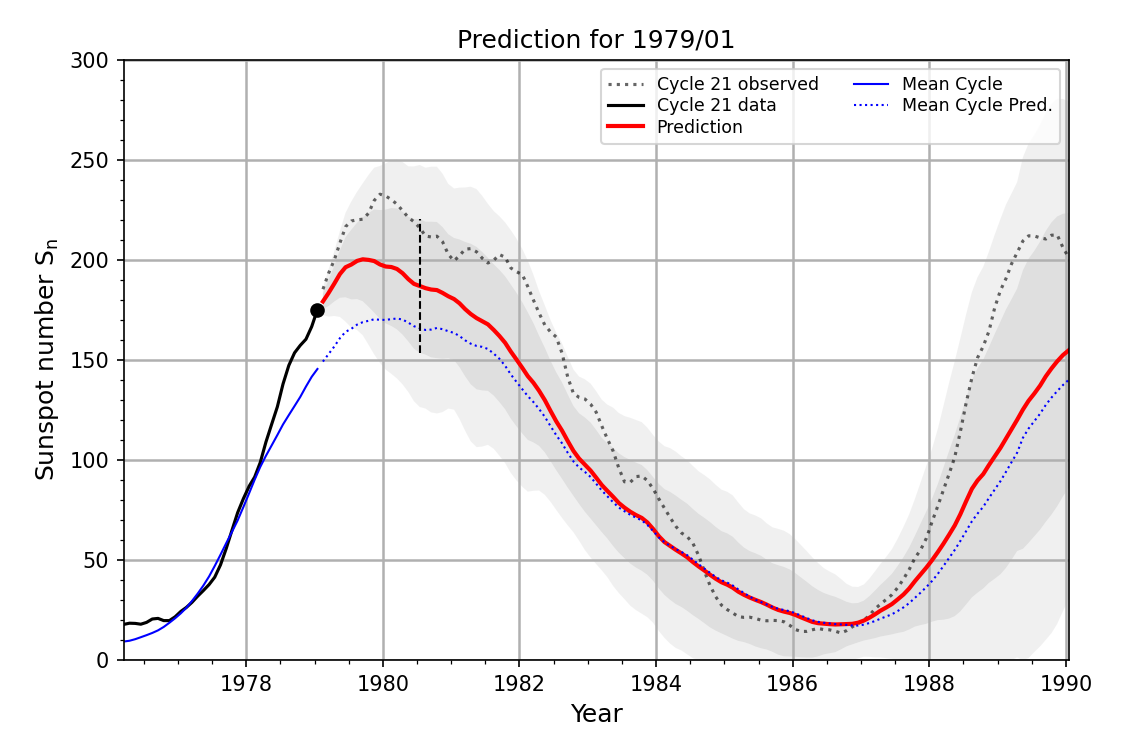}}
	\vspace{-0.53\textwidth}   % Overlays label
	\centerline{\large \bf 	\hspace{0.16 \textwidth}  \color{black}{(a)} \hfill}
	\vspace{0.48\textwidth}    % Shift back to the panel bottom 

	\centerline{
		\includegraphics*[width= 0.85\textwidth,bb= 5mm 0mm 186mm 124mm,clip=]{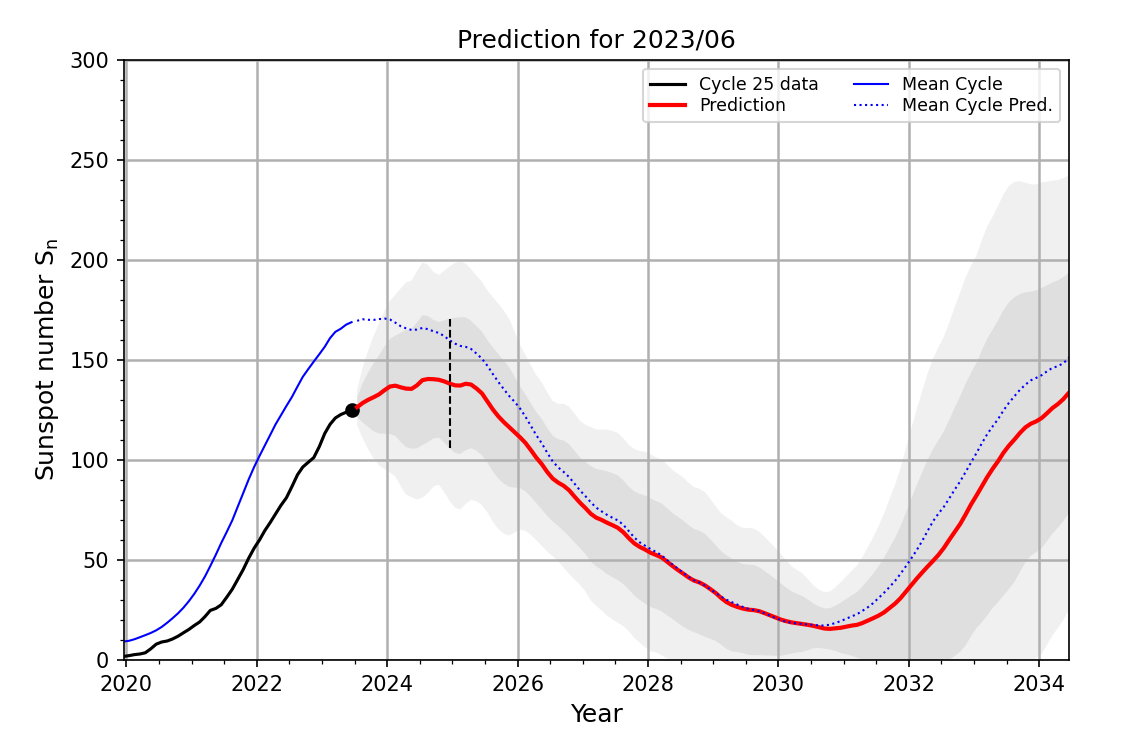}}
	\vspace{-0.53\textwidth}   % Overlays label
	\centerline{\large \bf 	\hspace{0.16 \textwidth}  \color{black}{(b)} \hfill}
	\vspace{0.48\textwidth}    % Shift back to the panel bottom 

	\caption{Example of two predictions, for a strong cycle (top (a), January 1979) and for the current rather low cycle (bottom (b), June 2023). Black line: observed SSN up to the last available month (black dot). Blue solid and dotted line: mean cycle. Red line: predictions (gray shading: the inner range is the standard deviation, and the outer range is the 90\% percentile). In the top prediction (a), the actual SSN after the prediction date is shown as a dotted black line. The vertical dashed line marks the fixed 18-month ahead limit of operational predictions as published by NOAA and WDC SILSO. }
	\label{fig2}
\end{figure}

Two examples of predictions are illustrated in Figure\,\ref{fig2}. Those plots show how the correction term in Equation\,(\ref{EQ_Predi}) ensures a seamless extension of the latest observed monthly SSN (black dot) by the adjusted predictions (red curve). As the correction coefficient $\hat{k}_{sp}$ decreases with $p$, the predictions progressively converge towards the mean cycle $\bar{S}_p$ (blue dotted line), which largely remains the ML base prediction in all cases for large $p$ ahead times. 

The bottom plot (b) in Figure\,\ref{fig2} is actually the latest ML operational prediction for Cycle 25 (at the time of manuscript submission), released in January 2024 with last SSN in June 2023. It places the upcoming Cycle 25 maximum in August 2024 at $ \rm{SN} = 140 \pm 32$, and the end of the cycle in October 2030, thus giving a cycle length of 130 months (10.8 years). As we will show later in our analysis, the exact match of this length with the duration of the mean cycle does not come by chance, but illustrates the dominant role of the mean cycle, with its fixed duration, in the ML method. 

\subsection{Some Insights on the ML Prediction Formula}
The above prediction formulae actually correspond to the original Equation\,(1) and the array of equations of page 2 in \citet{McNishLincoln1949}, for the case of a single past reference SSN value.
They thus implement the simplest version of the 1949 method. However, while the original method worked with one annual mean value, this version works on a monthly scale, with a variable time separation between the predicted month $p$ and the last SSN, which acts as fixed starting ``tie point'' $s$ for the whole sequence of monthly predictions.

The asset of the ML method thus entirely resides in the additive correction in Equation\,(\ref{EQ_Predi}), which only includes a single difference between the observed value $S_s^c$ and the mean cycle $\bar{S}_s$, for the last month available in the input SSN data. The improvement to the zero-order mean-cycle prediction thus relies entirely on this single piece of information about the actual progress of the current cycle.

Moreover, we point out that Equation\,(\ref{EQ_kCoeff}) is closely equivalent to the cross-correlation coefficient $r_{sp}$ between the monthly values $S_p^n$ at month $p$ and $S_s^n$ at month $s$ for any Cycle $n$:
%\begin{linenomath}
\begin{eqnarray} 	\label{EQ_CoeffCorr} 
	r_{sp} &=& \frac{\sum_{n=8}^{c-1} \left(S_s^n - \bar{S}_s \right) \left(S_p^n - \bar{S}_p \right)}	
	{\sqrt{\sum_{n=8}^{c-1} \left(S_s^n - \bar{S}_s \right)^2 \sum_{n=8}^{c-1} \left(S_p^n - \bar{S}_p \right)^2}} \nonumber \\ 
	&=& \frac{\sum_{n=8}^{c-1} \Delta_s^n \Delta_p^n}	
	{\sqrt{\sum_{n=8}^{c-1} \left(\Delta_s^n \right)^2  \sum_{n=8}^{c-1} (\Delta_p^n)^2}}
\end{eqnarray}
%\end{linenomath}

In $\hat{k}_{sp}$, the denominator is thus normalized only relative to the variance at month $s$ (starting month), instead of both months $s$ and $p$ in $r_{sp}$. Therefore, we can expect that the $k_{sp}$ factor will generally behave like $r_{sp}$. Indeed, as shown in a single prediction (Figure\,\ref{fig2}), $k_{sp}$ is close to unity for the first predicted months just following month $s$, and then declines to lower values for longer lead times. 
Taking the ratio of both quantities defined in Equations\,(\ref{eq_A22}) and (\ref{EQ_CoeffCorr}), we obtain:
%\begin{linenomath}
\begin{equation} 	\label{EQ_GainFac} 
	G_{sp}= \frac{\hat{k}_{sp}}{r_{sp}} = \frac{\sqrt{\sum_{n=8}^{c-1} (\Delta_s^n)^2  \sum_{n=8}^{c-1} (\Delta_p^n)^2}} {\sum_{n=8}^{c-1} ({\Delta_s^n})^{2}}	= \sqrt{\frac{\sum_{n=8}^{c-1} ({\Delta_p^n})^{2}}{\sum_{n=8}^{c-1} ({\Delta_s^n})^{2}}}
\end{equation}
%\end{linenomath}
Using Equation\,(\ref{eq_A9}) for $\sigma_{m}$, this is thus essentially the ratio between the standard deviations relative to the mean cycle at the time of the last observed SSN, $\sigma_{s}$, and at the time of the prediction, $\sigma_{p}$, respectively. 
The ML correction coefficient can thus be rewritten, as:
%\begin{linenomath}
\begin{equation} 	\label{EQ_CorrGain} 
	\hat{k}_{sp} = G_{sp}\, r_{sp}
\end{equation}
%\end{linenomath}
Therefore, the correction coefficient can be decomposed as the cross-correlation between the SSNs at times $s$ and $p$ multiplied by a gain factor $G_{sp}$. As the standard deviation $\sigma_{m}$ largely follows the variation of the mean SSN (Figure\,\ref{Fig_MeanCycle}), we thus expect the gain factor to track the growing or declining trend of the cycle between months $s$ and $p$. Therefore, the $G_{sp}$ factor can strongly differ from unity when there is a steep trend in the cycle, and thus, $k_{sp}$ may then significantly deviate from $r_{sp}$. For instance, in the early rise of the cycle, $k_{sp}$ may take values larger than 1, as the variance is much larger at the predicted month $p$ than at the starting month $s$, thus strongly amplifying the base observed difference $S_s^c - \bar{S}_s$ in Equation\,(\ref{EQ_Predi}). As explained in Section\,\ref{Sec_CorrecProperties}, we indeed observe this effect in our statistical analysis. 

\section{Prediction Errors: Derivation and Interpretation}\label{ML_error} \label{Sec_PredError}

\subsection{The Original Error Formulae}
Next to the predictions, the NOAA method, as programmed in the heritage FORTRAN programs, also computes the errors on predicted values by the following formula:
%\begin{linenomath}
\begin{equation}
	\sigma_{sp}^S = 1.812\, \sqrt{\frac{\left(\sigma_p^2 - \hat{k}_{sp}^2\, \sigma_s^2 \right)\, \left(N_c-1 \right)}
		{N_c - 2}} . 
	\sqrt{1 + \frac{1}{N_c}+ \frac{\left(S_s^c - \bar{S}_s \right)^2}{\sigma_s^2\,\left(N_c-1 \right)}}               \label{EQ_PredErr}
\end{equation}
%\end{linenomath}
where $S_s^c$ is the last SSN in the data, and $\sigma_s^2$ and $\sigma_p^2$ are the variances from Equation\,(\ref{eq_A9}), for months $m=s$ and $m=p$.

Likewise, the estimated error on the correction coefficient $k_{sp}$  is given by: 
%\begin{linenomath}
\begin{equation}
	\sigma_{sp}^k = 1.812\, \sqrt{\frac{\left(\sigma_p^2 - \hat{k}_{sp}^2\,\sigma_s^2 \right)\,\left(N_c-1 \right)}
		{N_c - 2}} / \sqrt{\sigma_s^2\,(N_c-1)}               
		\label{EQ_PredErrRel}
\end{equation}
%\end{linenomath}

Those estimated errors are an asset of this method, as most other early prediction methods lack any rigorous error estimate. In order to verify the exactness of the above programmed formulae, we reconstruct hereafter the full mathematical derivation of the above expressions.

By examining the above error formulae, we can first do the following observations.  Firstly, as the estimation of $\hat{k}_{sp}$, the slope of regression line, is made on sample data representing the cycle-to-cycle variability of solar activity, it will be affected by some uncertainty. Secondly, we can assume that true prediction errors would not match the calculated prediction errors, as the position of regression line describes only the average relation, in general. Actual values are scattered around it. Moreover, separate measurements deviated from this line in the past. Therefore, it is natural that the same deviations will happen in the future.

The errors related with the sources mentioned above may be reflected by the confidence interval of the forecast when certain assumptions about the properties of the series and the mean square forecast error are made.

\subsection{Estimation Errors of Regression Coefficients $\hat{a}_{sp}$ and $\hat{k}_{sp}$}
The covariance matrix of the estimation error of $\vec{\hat{X}_{sp}}$ is given by
%\begin{linenomath}
\begin{eqnarray} \label{eq_A25}
	cov\left(\vec{\hat{X}_{sp}}\right) &=& {\bf D} \cdot var\left(\vec{Y}\right) \cdot {\bf D}^{T} \nonumber \\ 
	&=& \left(\sigma_{sp}^{\varepsilon} \right)^{2} {\bf D} {\bf D}^{T} = \left(\sigma_{sp}^{\varepsilon} \right)^{2}  
	\begin{vmatrix}
		\frac{1}{N_{c}}  &  0\\
		0                &  \frac{1}{\sum_{n=8}^{c-1}\left(\Delta_{s}^{n}\right)^{2}}\\
	\end{vmatrix}
\end{eqnarray}
%\end{linenomath}
%
Based on the variance of $\Delta_{s}^{n}$ given by Equation\,(\ref{eq_A9}) for month $m=s$,  Equation\,(\ref{eq_A25}) can be rewritten as
%\begin{linenomath}
\begin{equation}\label{eq_A28}
	cov\left(\vec{\hat{X}_{sp}}\right)= \left(\sigma_{sp}^{\varepsilon } \right)^{2}
	\begin{vmatrix}
		\frac{1}{N_{c}}  &  0\\
		0                &  \frac{1}{\sigma_{s}^{2}\,\left(N_{c}-1\right) }\\
	\end{vmatrix}
\end{equation}
%\end{linenomath}
%
Therefore, the variances of estimates $\hat{a}_{sp}$ and $\hat{k}_{sp}$ are presented as 
%\begin{linenomath}
\begin{equation}\label{eq_A29}
	var\left(\hat{a}_{sp}\right) = \left(\sigma_{sp}^{\varepsilon} \right)^{2} \frac{1}{N_{c}}
\end{equation}
%%\end{linenomath}
%%\begin{linenomath}
\begin{equation}\label{eq_A30}
	var\left(\hat{k}_{sp}\right) =  \left(\sigma_{sp}^{\varepsilon} \right)^{2}\frac{1}{\sigma_{s}^{2}\left(N_{c}-1\right)}
\end{equation}
%\end{linenomath}

\subsection{Prediction Error} 
By re-writing Equation\,(\ref{eq_A13}), we can express the SSN value at month $p$ in Cycle $c$ as   
%\begin{linenomath}
\begin{equation}\label{eq_A31}
	{S}_{sp}^c=\bar{S}_p+a_{sp}+k_{sp}\left(S_s^c-\bar{S}_s\right) + \varepsilon_{sp}^{c} 
\end{equation}
%\end{linenomath}
%
The prediction error $\delta_{sp}$ at month $p$ is then determined by subtracting the exact value for month $p$ (Equation\,(\ref{eq_A31})) from the predicted value $\hat{S}_{sp}^c$ (Equation\,(\ref{EQ_Predi}))
%\begin{linenomath}
\begin{eqnarray} \label{eq_A33}
	\delta_{sp} &=& S_{sp}^c-\hat{S}_{sp}^c \nonumber \\
	&=& \left(a_{sp}-\hat{a}_{sp}\right)+k_{sp}\left(S_s^c-\bar{S}_{s}\right)+
		\varepsilon_{sp}^c - \hat{k}_{sp}\left(S_s^c-\bar{S}_s \right) \nonumber \\
	&=& \varepsilon_{sp}^c+ \left(a_{sp}-\hat{a}_{sp}\right) + \left(k_{sp}-\hat{k}_{sp}\right)\left(S_s^c-\bar{S}_s\right)  
\end{eqnarray}
%\end{linenomath}
%
The variance $(\hat{\sigma}_{sp}^{\delta})^{2}$ of prediction error $\delta_{sp}$ is given by
%\begin{linenomath}
\begin{equation}\label{eq_A34}
	\left(\hat{\sigma}_{sp}^{\delta} \right)^2 = var\left(\delta_{sp}\right) = \left (\hat{\sigma}_{sp}^{\varepsilon} \right)^2+var\left(\hat{a}_{sp}\right)+
	var\left(\hat{k}_{sp}\right)\left(S_s^c-\bar{S}_s\right)^{2} 
\end{equation}
%\end{linenomath}
or, by using Equations\,(\ref{eq_A29}) and (\ref{eq_A30}): 
%\begin{linenomath}
\begin{equation}\label{eq_A35}
	\left(\hat{\sigma}_{sp}^{\delta}\right)^2= \left(\hat{\sigma}_{sp}^{\varepsilon} \right)^2 \left(1+\frac{1}{N_{c}}+\frac{\left(S_s^c-\bar{S}_s\right)^{2}}
	{\sigma_s^2\left(N_{c}-1\right)}\right) 
\end{equation}
%\end{linenomath}

At this point, it is important to note that although the intercept of the regression is null in the ML formulation, as explained in Section\,\ref{Sec_CorrecTerm} (Equation\,(\ref{eq_A23_ad})), there is still a contribution from the $\hat{a}_{sp}$ term to the total prediction error, which appears as the $var\left(\hat{a}_{sp}\right)$ term in Equation\,(\ref{eq_A34}) given by Equation\,(\ref{eq_A29}).

\subsection{Estimation of Standard Deviation $\hat{\sigma}_{sp}^{\varepsilon}$}
As follows from Equation\,(\ref{eq_A13})\,:
%\begin{linenomath}
\begin{equation}\label{eq_A36}
	\varepsilon_{sp}^{c}= \Delta_{p}^{c}-k_{sp}\Delta_{s}^{c}-a_{sp}
\end{equation}
%\end{linenomath}
The variance estimation of noise  $\varepsilon_{sp}^{c}$ is presented as
%\begin{linenomath}
\begin{equation}\label{eq_A37}
	\left(\hat{\sigma}_{sp}^{\varepsilon} \right)^{2}=\frac{\sum_{n=8}^{c-1}\left(\Delta_{p}^{n}-k_{sp}\Delta_{s}^{n}\right)^{2}}{N_{c}-2}
\end{equation}
%\end{linenomath}
Here, we use $N_{c}-2$ in the denominator, as we estimate two regression coefficients  $a_{sp}$ and $k_{sp}$. As follows from Equation\,(\ref{eq_A37})\,:
%\begin{linenomath}
\begin{eqnarray} \label{eq_A38}
	\left(\hat{\sigma }_{sp}^{\varepsilon} \right)^{2} &=& \frac{\sum_{n=8}^{c-1}\left(\left(\Delta_{p}^{n}\right)^{2}-2k_{sp}\Delta_{p}^{n}	\Delta_{s}^{n}+\left(k_{sp}\Delta_{s}^{n}\right)^{2}\right)}{N_{c}-2} \nonumber \\ 
	&=& \frac{\sum_{n=8}^{c-1}\left(\Delta_{p}^{n}\right)^{2}-2k_{sp}\sum_{n=8}^{c-1}\Delta_{p}^{n}\Delta_{s}^{n}+k_{sp}^{2} \sum_{n=8}^{c-1}\left(\Delta_{s}^{n}\right)^{2}}{N_{c}-2}
\end{eqnarray}
%\end{linenomath}

Let us consider all the terms in the nominator of Equation\,(\ref{eq_A38}). According to Equation\,(\ref{eq_A9}), the first term can be presented as 
%\begin{linenomath}
\begin{equation}\label{eq_A39}
	\sum_{n=8}^{c-1}\left(\Delta_{p}^{n}\right)^{2}= \sigma_{p}^{2}\left(N_{c}-1\right)
\end{equation}
%\end{linenomath}
Let us multiply and divide the second term of Equation\,(\ref{eq_A38}) with $\sum_{n=8}^{c-1}\left(\Delta_{s}^{n}\right)^{2}$.  Then
%\begin{linenomath}
\begin{equation}\label{eq_A40}
	2\,k_{sp}\sum_{n=8}^{c-1}\Delta_{p}^{n}\Delta_{s}^{n}= 2\,k_{sp}\frac{\sum_{n=8}^{c-1}\Delta_{p}^{n}\Delta_{s}^{n}}
	{\sum_{n=8}^{c-1}\left(\Delta_{s}^{n}\right)^{2}}\sum_{n=8}^{c-1}\left(\Delta_{s}^{n}\right)^{2} 
\end{equation}
%\end{linenomath}
Taking into account Equation\,(\ref{eq_A22}) defining $k_{sp}$ and Equation\,(\ref{eq_A39}) for time $s$ instead of $p$, we can rewrite Equation\,(\ref{eq_A40}) in the following way
%\begin{linenomath}
\begin{equation}\label{eq_A41}
	2\,k_{sp}\sum_{n=8}^{c-1}\Delta_{p}^{n}\Delta_{s}^{n}= 2\,k_{sp}^{2} \sigma_{s}^{2}\left(N_{c}-1\right)
\end{equation}
%\end{linenomath}
%
Likewise, writing Equation\,(\ref{eq_A9}) for month $m=s$ as 
%\begin{linenomath}
\begin{equation}\label{eq_A43}
	\sigma_{s}^{2}=\frac{\sum_{n=8}^{c-1}\left(S_{s}^{n}\right)^{2}-N_{c}\left(\bar{S}_{s}\right)^{2}}{N_{c}-1}=
	\frac{\sum_{n=8}^{c-1}\left(\Delta_{s}^{n}\right)^{2}}{N_{c}-1} \ ,
\end{equation}
%\end{linenomath}
we can rewrite the third term of Equation\,(\ref{eq_A38}) as
%\begin{linenomath}
\begin{equation}\label{eq_A44}
	\left(k_{sp}\right)^{2}\sum_{n=8}^{c-1}\left(\Delta_{s}^{n}\right)^{2}=
	\left(k_{sp}\right)^{2}\sigma_{s}^{2}\left(N_{c}-1\right)
\end{equation}
%\end{linenomath}
Thus, Equation\,(\ref{eq_A38}) can be rewritten as 
%\begin{linenomath}
\begin{equation}\label{eq_A45}
	(\hat{\sigma}_{sp}^{\varepsilon})^{2} =\frac{\left(\sigma_{p}^{2}-2k_{sp}^{2}\sigma_{s}^{2}+k_{sp}^{2}\sigma_{s}^{2}\right)
		\left(N_{c}-1\right)}{N_{c}-2} = \frac{\left(\sigma_{p}^{2}-k_{sp}^{2}\sigma_{s}^{2}\right)\left(N_{c}-1 \right)}{N_{c}-2}
\end{equation}
%\end{linenomath}
The estimated standard deviation $\hat{\sigma}_{sp}^{\varepsilon}$ of the noise in the data thus equals
%\begin{linenomath}
\begin{equation}\label{eq_A46}
	\hat{\sigma}_{sp}^{\varepsilon}=\sqrt{\frac{\left(\sigma_{p}^{2}-k_{sp}^{2}\sigma_{s}^{2}\right)\left(N_{c}-1\right)}{N_{c}-2}}
\end{equation}
%\end{linenomath}

\subsection{Interpretation: Standard Errors and Confidence Intervals}

Taking into account Equation\,(\ref{eq_A35}), the standard deviation of the prediction is finally given by
%\begin{linenomath}
\begin{equation}\label{eq_A47}
	\hat{\sigma}_{sp}^{S}=\sqrt{\frac{\left(\sigma_{p}^{2}-k_{sp}^{2}\sigma_{s}^{2}\right)\left(N_{c}-1\right)}{N_{c}-2}}
	\sqrt{1+\frac{1}{N_{c}}+\frac{\left(S_{s}^{c}-\bar{S}_{s}\right)^{2}}{\sigma_{s}^{2}\left(N_{c}-1\right)}}
\end{equation}
%\end{linenomath}
%
Similarly, using Equation\,(\ref{eq_A30}), the standard deviation on the $k_{sp}$ coefficient is given by
% Equation had to be corrected (stdv instead of var, missing sqrt for 2nd term)
%\begin{linenomath}
\begin{equation}\label{eq_A48}
	\hat{\sigma}_{sp}^{k}=\sqrt{\frac{\left(\sigma_{p}^{2}-k_{sp}^{2}\sigma_{s}^{2}\right)\left(N_{c}-1\right)}{N_{c}-2}} 
	\sqrt{\frac{1}{\sigma_{s}^{2}\left(N_{c}-1\right)}}
\end{equation}
%\end{linenomath}

We have thus obtained here the exact mathematical derivation of the error values computed by the NOAA heritage program (Equations\,(\ref{EQ_PredErr}) and (\ref{EQ_PredErrRel})).  An equivalent derivation was also published recently in \citet{PetrovaEtal2021}, in a parallel application to radio flux predictions, together with their Kalman filter update.

The above expressions fully match the original ones, as included in the NOAA source program (respectively Equations\,(\ref{eq_A47}) and (\ref{EQ_PredErr}), and Equations\,(\ref{eq_A48}) and (\ref{EQ_PredErrRel})), except for the constant factor 1.812. This value actually corresponds to $t^{10}_{.95}$, the two-tails critical value of Student’s t-distribution for $\alpha=0.05$ (confidence level $p=1- 2\times\alpha=0.9$), and the number of degrees of freedom $k=N_c-1=10$. The latter is the number of past cycles used in the statistics, $N_c$, minus the number of estimated coefficients (here only one: $k_{sp}$). So, the original NOAA program does not provide the mean squared errors themselves, but the 90\% confidence intervals.

Now, it also means that this $t^{k}_{.95}$ factor must actually vary with time and be re-computed every time $N_c$ is incremented by 1 at each new cycle.  However, the constant hard-coded factor used in the heritage program corresponds to the fixed value $N_c=10$ and thus for predictions of Cycle 18 ($c=8+10$), i.e. the epoch when McNish and Lincoln derived their formulae. It was apparently never updated later on, leading to an overestimate of the uncertainty of the output predictions. Today, with $N_c=17$, for the same level if significance $\alpha=0.1$, the correct factor should be $t^{16}_{.95}=1.746$. Therefore, the impact of this past oversight was fortunately limited, with an overestimate by only 3.8\,\%. Following this finding, the old constant value was replaced by a calculated $t^{k}_{.95}$ value in the operational program.

\section{Key Properties of the Base ML Components} \label{Sec_KeyProp}

By injecting the actual SSN series from Cycles 8 to 25 in the above definitions of the mean cycle, correction coefficient and their respective errors, i.e. the base components of actual ML predictions, we can already infer some key properties that will help interpreting the statistics of bulk predictions presented in the next Section.

\subsection{Mean Cycle and its Dispersion: Key Properties} \label{Sec_MeanCycleProp}
Considering first the mean cycle shown in Figure\,\ref{Fig_MeanCycle}, we observe that the dispersion of cycles is lowest during the first ten months after the starting minimum, and the SN values are also lowest, starting just below 10 on the first months of the cycle. The standard deviation then grows and reaches a maximum at month 41, just before the peak of the mean cycle is reached in a flat maximum between months 44 and 50 (3.75 years), near SN= 170. The dispersion of all cycles then slowly decreases up to the second ending minimum, at month 130 (mean cycle length of 10.75 years for Cycles 8 to 24). However, the dispersion $\sigma_m$ is then twice as large as in the starting minimum. The mean value itself is also higher, at $\bar{S}_m= 17$. 

From the upper plot in Figure\,\ref{Fig_MeanCycle}, we can observe that this higher dispersion towards the end of the mean cycle is largely due to the differences in cycle lengths, which cause a temporal spread of the ending minima (from 120 to 150 months, thus over 2.5 years). Indeed, knowing that this second minimum corresponds to exactly the same set of cycle minima as in month 0, this higher standard deviation can only result from this temporal misalignment. Due to the latter, the second minimum actually includes descending or ascending sections of different cycles, instead of their actual ending minimum, thus inevitably and artificially raising this second mean minimum and its standard deviation. 

This second contribution to $\sigma_{m}$ due to temporal dispersion increases continuously with the relative time $m$, as we move away from the common tie point of the starting minimum. This temporal smearing effect is thus also already present to a lower extent at the time of the cycle maximum, around month 47. As a consequence, it also produces a rather smoothed and rounded maximum, less sharp than actual maxima, and the peak value is reduced relatively to the ascending and declining  phases of the mean cycle. After the ending mean minimum, the rise of the next cycle is also characterized by even larger errors than in the corresponding phase at the beginning of the mean cycle, thus making predictions beyond the end of a cycle particularly unreliable.

Considering now the relative error $\sigma_{m}/\bar{S}_{m}$ (blue dashed line in Figure\,\ref{Fig_MeanCycle}, lower plot), we can see that it is pretty stable over a large part of the solar cycle, at about 30\%, but late in the cycle, it grows to much higher values, in particular around the ending minimum. This results from the combination of enhanced errors late in the cycle, as explained above, with the low SSN values of the minimum phase. A notable feature is the dip at $m=130$, which marks the moment when most cycles pass their flat minimum (null or low first derivative), thus briefly quenching the effect of temporal dispersion.   

Given the above characteristics, in subsequent interpretations of the prediction performance, we must remember that the approximation of the mean cycle does not only lead to a poor match for the actual range of cycle amplitudes, but that the mean cycle is also distorted and does not vary in a very realistic way, compared to true cycles.

\subsection{Correction Coefficient: Key Properties} \label{Sec_CorrecProperties}

Now, next to this base mean cycle, the correction factor $\hat{k}_{sp}$, defined in Equation\,(\ref{EQ_kCoeff}), plays the primary role in the ML method. Like the mean cycle, it is invariable when using a fixed set of past cycles as a base (here Cycles 8 to 25), and thus $k_{sp}$ varies in the same way for each cycle prediction. 

\begin{figure}
	\centerline{
		\includegraphics*[width= 0.85\textwidth, bb= 5mm 0mm 170mm 125mm,clip=] {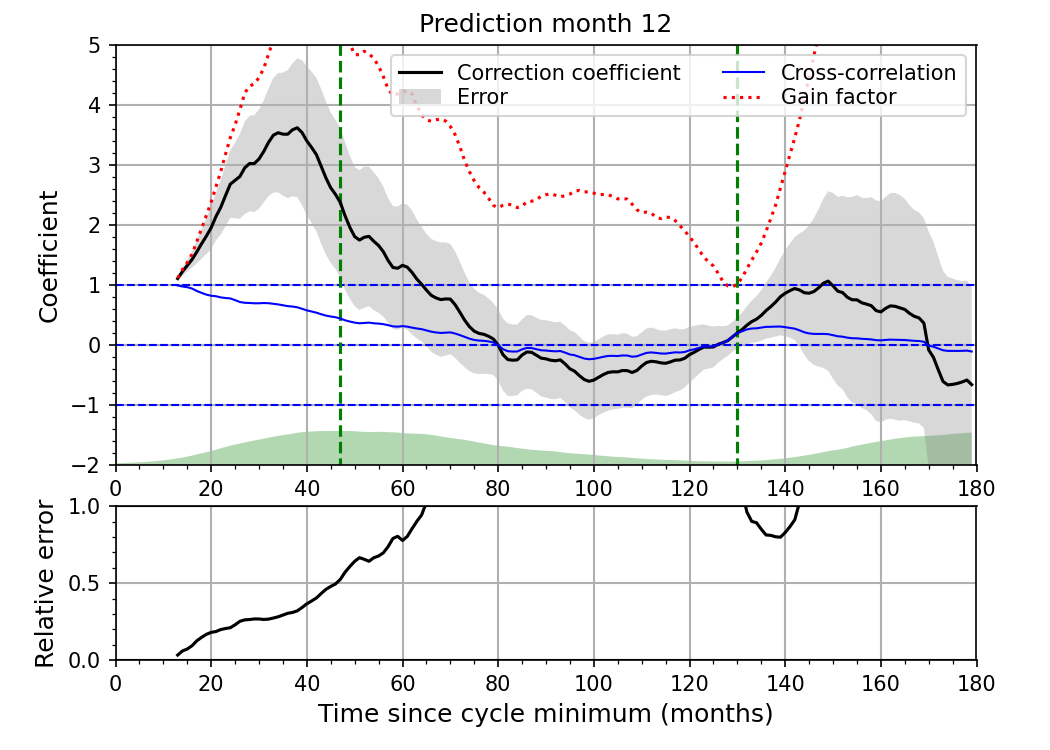}}
	\caption{Plot of the correction coefficient $\hat{k}_{sp}$ (thick black line), of the cross-correlation $r_{sp}$ (blue line), and of the gain factor $G_{sp}$ (red dashed line) for a prediction starting at month 12 after the initial cycle minimum. The error on the correction coefficient $\hat{\sigma}_{sp}^k$ is indicated by the gray shading. The green shaded curve along the horizontal axis shows the mean solar cycle (arbitrary scale) as temporal indicator, with vertical green dashed lines marking the maximum and minimum of the mean cycle. The horizontal blue dashed lines at -1, 0 and +1 highlight the maximum range for $r_{sp}$. The relative error $\hat{\sigma}_{sp}^{k}/\hat{k}_{sp}$ is plotted in the lower panel.}
	\label{Fig_PlotCoeffGain12}
\end{figure}

\begin{figure}
	\centerline{
		\includegraphics*[width= 0.85\textwidth, bb= 6mm 0mm 170mm 125mm,clip=] {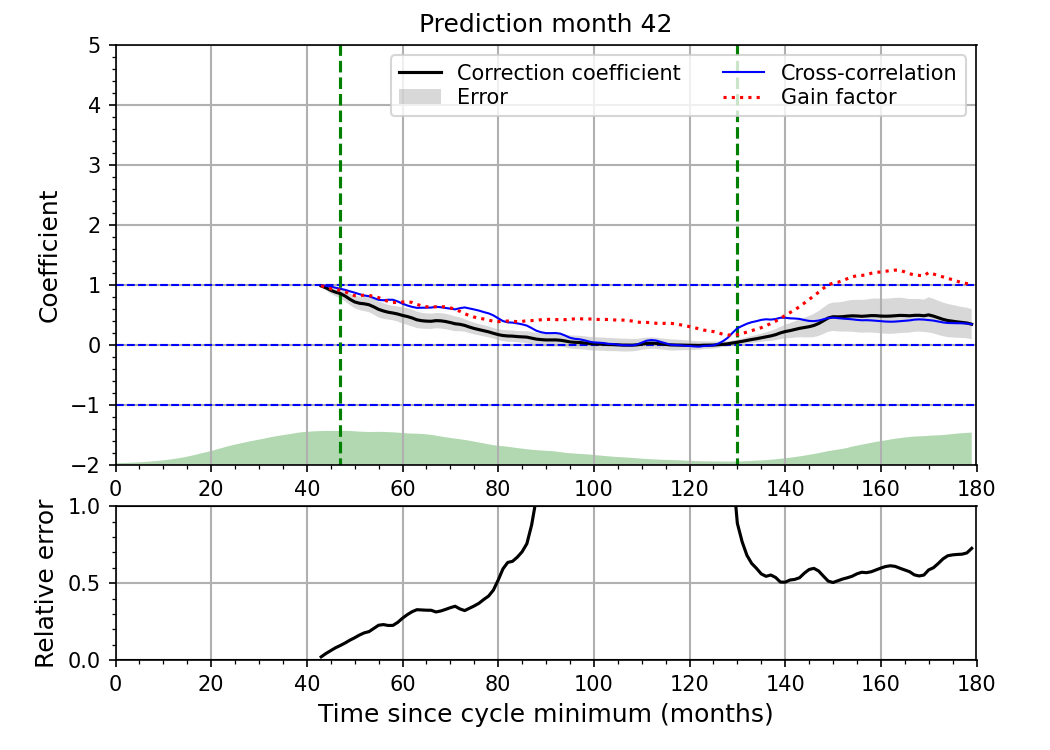}}
	\caption{Plot of the correction coefficient $\hat{k}_{sp}$ (thick black line), of the cross-correlation $r_{sp}$ (blue line) and of the gain factor $G_{sp}$ (red dashed line) for a prediction starting at month 42 after the initial cycle minimum. The contents are the same as in Figure\,\ref{Fig_PlotCoeffGain12}.}
		\label{Fig_PlotCoeffGain42}
\end{figure}

\begin{figure}
	\centerline{
		\includegraphics*[width= 0.85\textwidth, bb=6mm 0mm 170mm 125mm,clip=] {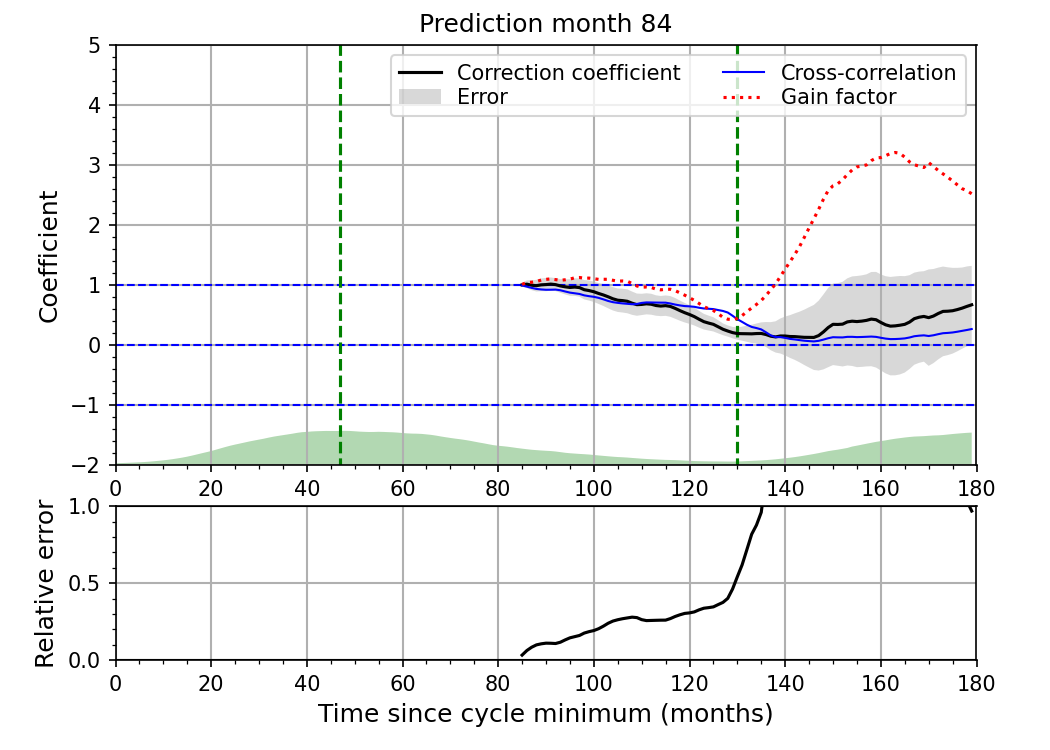}}
	\caption{Plot of the correction coefficient $\hat{k}_{sp}$ (thick black line), of the cross-correlation $r_{sp}$ (blue line) and of the gain factor $G_{sp}$ (red dashed line) for a prediction starting at month 84 after the initial cycle minimum. The contents are the same as in Figure\,\ref{Fig_PlotCoeffGain12}.}
	\label{Fig_PlotCoeffGain84}
\end{figure}

As illustration, in Figures\,\ref{Fig_PlotCoeffGain12}, \ref{Fig_PlotCoeffGain42} and \ref{Fig_PlotCoeffGain84}, we plotted the $\hat{k}_{sp}$ coefficient as a function of forward time $p$ for three starting months $s$ after the start of the cycle: respectively, month 12 early in the rising phase, month 42 near the maximum, and month 84 later in the declining phase.  In all three figures, we also decomposed $\hat{k}_{sp}$ into its primary components (Equation\,(\ref{EQ_CorrGain})): the cross-correlation $r_{sp}$ and the gain factor $G_{sp}$. We find that the properties of $\hat{k}_{sp}$, $r_{sp}$ and $G_{sp}$ change drastically over a solar cycle. For predictions made early in the cycle (starting month $s=12$, Figure\,\ref{Fig_PlotCoeffGain12}), $\hat{k}_{sp}$ is shaped primarily by $G_{sp}$ values much larger than unity, with an interval of negative $r_{sp}$ late in the cycle. However, the errors grow very quickly with $p$ and are large overall ($> 100\%$), making the corrections non-significant, thus not reliable, over most of this late part of the cycle. When reaching the maximum (month $s=42$, Figure\,\ref{Fig_PlotCoeffGain42}), $\hat{k}_{sp}$ remains close to $r_{sp}$, with a gain $G_{sp}$ mostly below 1. $r_{sp}$ and thus $\hat{k}_{sp}$ also stays close to 0 over the last three years of the cycle, thus making the predictions largely equal to the mean cycle, whatever the last observed SSN at time $s$.  For predictions made during the whole declining phase of the cycle (month $s=84$ shown in Figure\,\ref{Fig_PlotCoeffGain84}), $\hat{k}_{sp}$ remains close to $r_{sp}$, with a small error, indicating rather good predictions over this final part of the solar cycle. However, the gain and error abruptly rise again to large values beyond the ending minimum, indicating a sharp loss of predictive capability once the end of the cycle is reached.

\begin{figure}
	\centerline{
		\includegraphics*[width= 0.95\textwidth,bb=11cm 0cm 135cm 70cm,clip=] {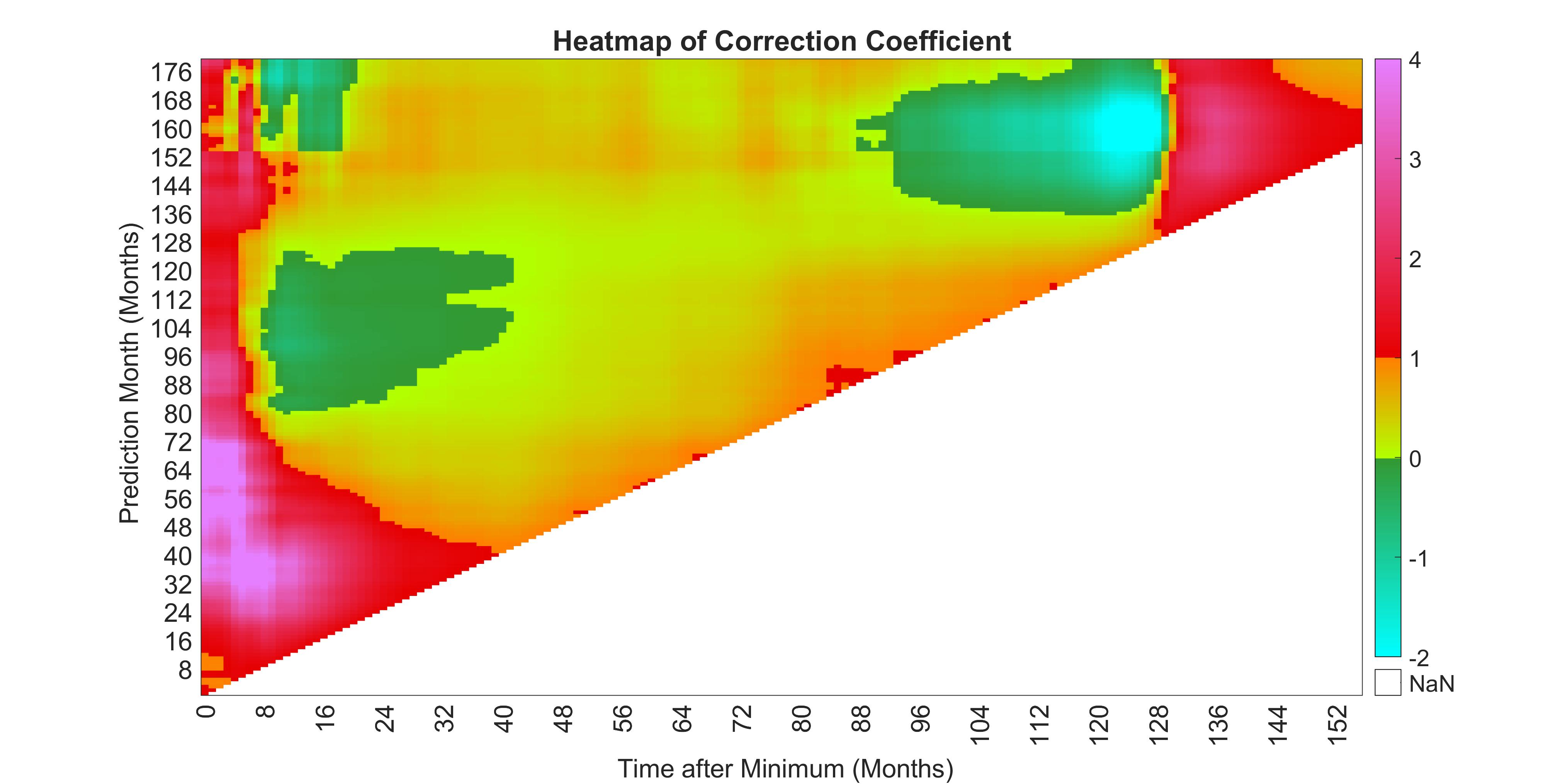} }
	\caption{2D map of the correction coefficient $\hat{k}_{sp}$ as a function of the starting time $s$ of the prediction (horizontal axis) and of the prediction time $p$ (vertical axis), both counted in months from the starting minimum of the mean cycle. In the early part of the cycle, $\hat{k}_{sp}$ takes values much larger than 1 (red zones at left and far right). Just before the maximum and the ending minimum of the mean cycle, it takes negative values (green zones).}
	\label{Fig_MapCorrCoeff}
\end{figure}

\begin{figure}
	\centerline{
		\includegraphics*[width= 0.95\textwidth,bb=11cm 0cm 135cm 70cm,clip=] {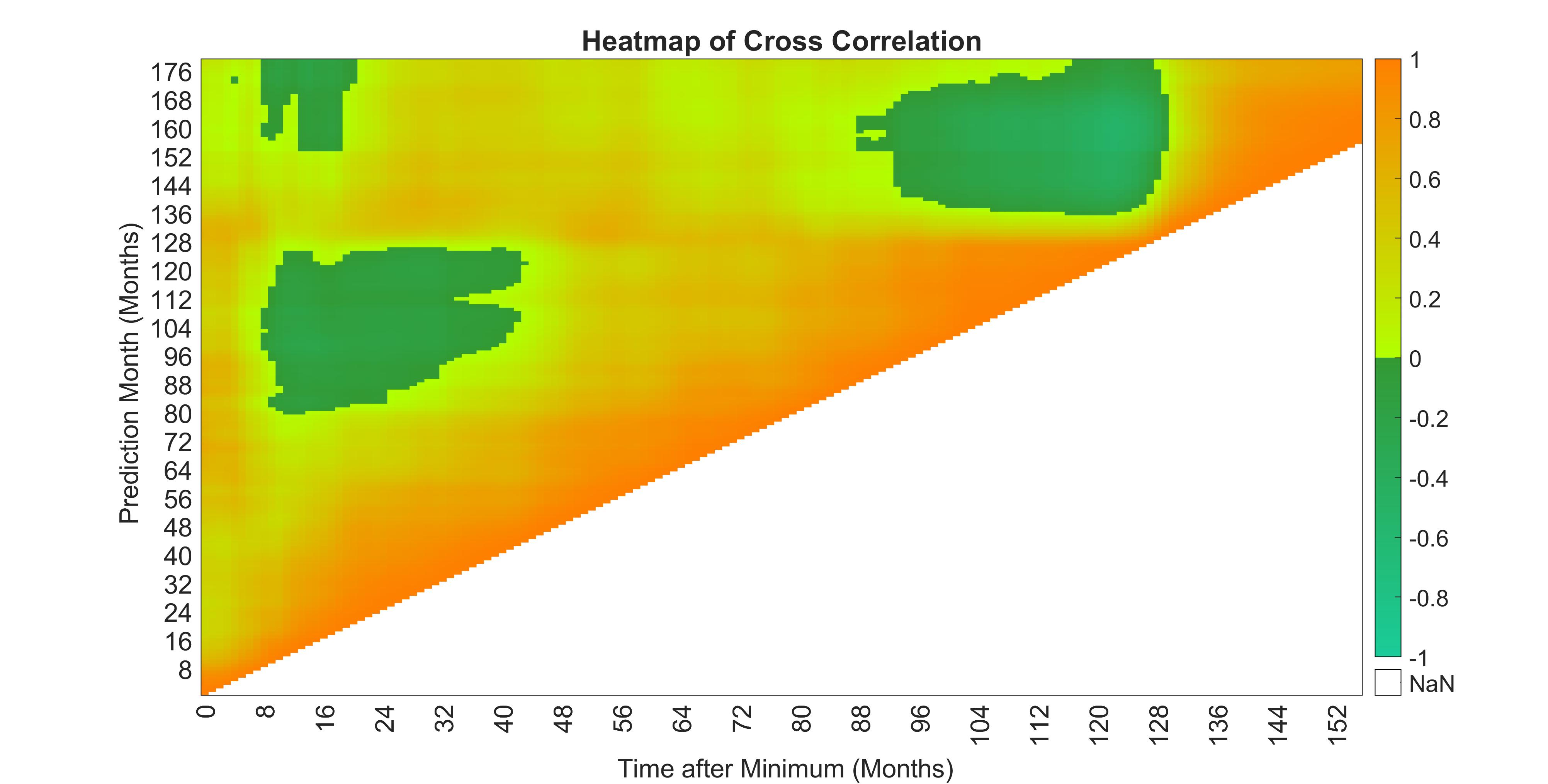} }
	\caption{2D map of the cross-correlation $r_{sp}$. We note the similarity of this map with the map of $\hat{k}_{sp}$ in Figure\,\ref{Fig_MapCorrCoeff}, except that no values exceed unity. Negative values (green zones) appear in the same time ranges, and correspond to an anti-correlation of the SSN between the time of the last observation $s$ and the prediction time $p$.}
	\label{Fig_MapCrossCorr}
\end{figure}

In order to get a more global view of those strongly varying properties along a solar cycle, we also built a two-dimensional color map of $\hat{k}_{sp}$, as a function of the starting month $s$ of the prediction and of the predicted month $p$ (Figure\,\ref{Fig_MapCorrCoeff}). In order to keep the same month $p$ aligned horizontally, we shifted each prediction up by one month for each $s$, thus referring $p$ to the start of the cycle instead of the base month $s$ of each prediction.  As a comparison, we mapped in the same way, the cross-correlation coefficient $r_{sp}$ in Figure\,\ref{Fig_MapCrossCorr}.  

As discussed in the previous Section, we can see that $\hat{k}_{sp}$ and $r_{sp}$ largely follow the same patterns.  A first prominent feature is a decreasing ramp from a value of almost 1 for the first predicted month following the last observation (lower diagonal edge) and decreasing to low values $< 0.3$, for forward times of about 2 or 3 years. This thus means that beyond that limit, long-term ML predictions largely come back to the base mean cycle.
 
However, the correction never fully vanishes and other features appear, and this time, they are fixed relative to the predicted month (vertical axis), instead of the start time (diagonal band). In particular,  negative values indicating an anti-correlation are found for predictions starting over the $s$ range 10 to 48, and $p$ ranging from 80 to 128. Therefore, it involves predictions made before the mean-cycle maximum for months in the descending phase. An even stronger anti-correlation is found at the end of the cycle for $s$ from about 90 to 128 and $p$ above 130, thus for predictions made in the late part of the declining phase, for months in the next ascending phase. Both features thus involve sections of the cycle with opposite trends. 

Such an anti-correlation is actually the signature expected from a temporal shift of the extremum. For instance, if a cycle rises faster, values at a given month $s$ will be higher during the ascending phase, but as the maximum happens earlier, a given month $p$ in the declining phase will fall a bit later in the decline, and thus be lower. We thus see that the $\hat{k}_{sp}$ factor partly accounts for this. The anti-correlation is stronger at the end of the cycle than around the maximum, again indicating the temporal shifts are growing with time $p$, as discussed above for the mean cycle. Still, we must remember that this behavior is fixed relative the mean cycle, and does not adjust to the specific evolution of each actual cycle.  

Now, there is a major feature in $\hat{k}_{sp}$ that is absent in the cross-correlation $r_{sp}$. For the first 9 months of the cycle, $\hat{k}_{sp}$ increases instead of decreasing for larger lead times $p$. It takes values much larger than 1 (up to 4.5) and remains high for all $p$ values. It thus suggests that the ML method applies strong corrections relative to the mean value for the whole temporal range of predictions. However, considering the prediction formula (Equation\,(\ref{EQ_Predi})), we must take into account that just after the initial minimum, the observed difference $(S_s^c - \bar{S}_s)$ is always small. Therefore, a high  $\hat{k}_{sp}$ amplification factor is needed to produce a correction that is large enough in the subsequent ascending and maximum phases of the cycles. This explains the extreme values of $\hat{k}_{sp}$.

\begin{figure}
	\centerline{
	\includegraphics*[width= 0.95\textwidth,bb=11cm 0cm 135cm 70cm,clip=] {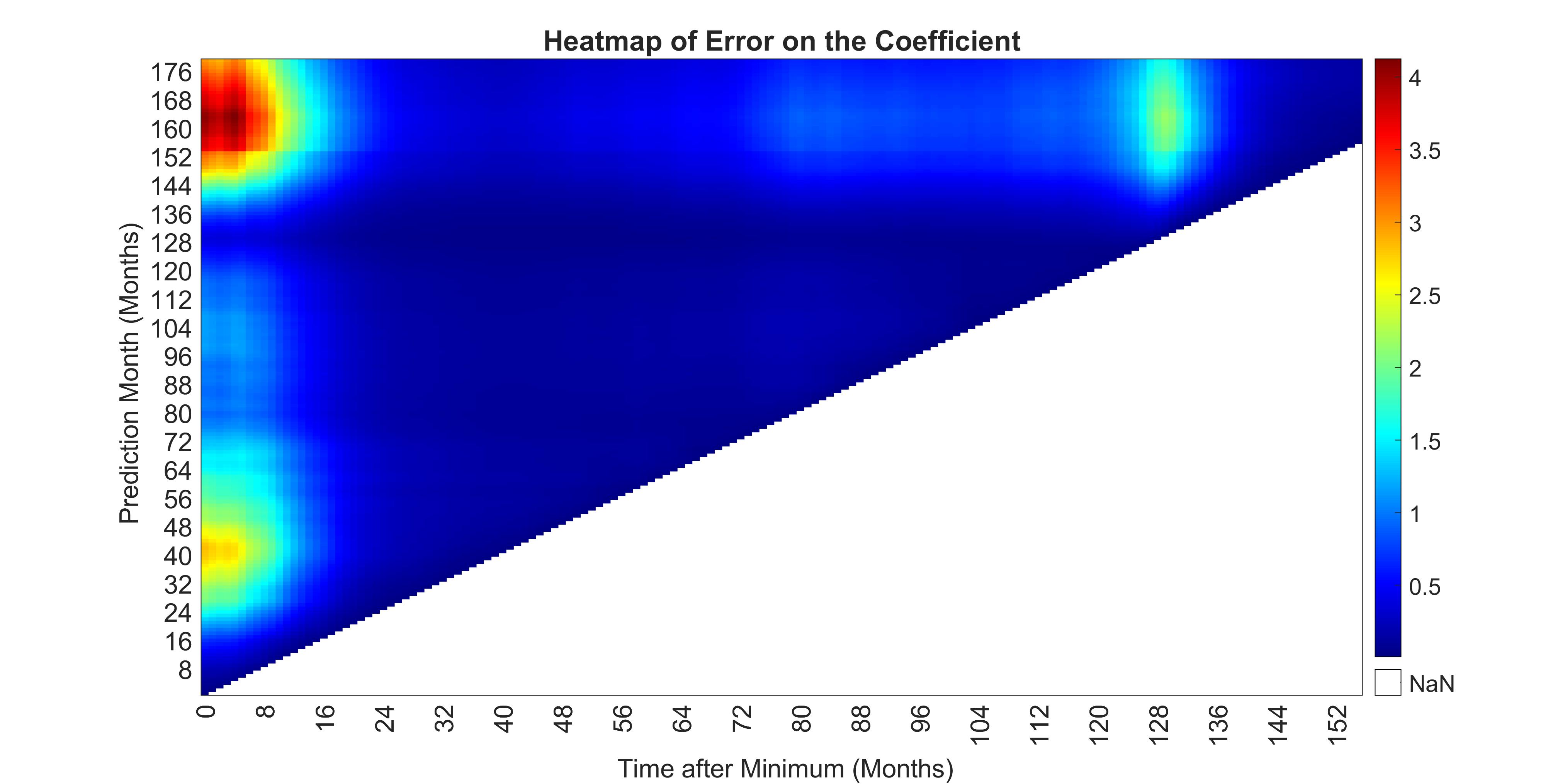}}
	\caption{2D map of the standard error $\hat{\sigma}_{sp}^k$ on the correction coefficient $\hat{k}_{sp}$, over the same coordinates as in Figures\,\ref{Fig_MapCorrCoeff} and \ref{Fig_MapCrossCorr}. This map shows that this error is particularly large during the first few months following the minimum of the cycle for predictions of the rising phase of the cycle.}
	\label{Fig_MapErrCoef}
\end{figure}

However, if we also consider the standard error $\hat{\sigma}_{sp}^{k}$ on $\hat{k}_{sp}$ (Equation\,(\ref{eq_A48})), mapped in the same way in Figure\,\ref{Fig_MapErrCoef}, we observe that the error also reaches very high values in this early part of the cycle, with peaks around months 40 and 155, i.e. the times of maxima, in which the error reaches almost 100\% of the $\hat{k}_{sp}$ value. Therefore, overall, those maps indicate that the correction term and thus the ML predictions are largely unreliable just after a new cycle is starting. The error $\hat{\sigma}_{sp}^{k}$ falls below 1 only for $s$ beyond 18 months, thus 1.5 years after the minimum  has passed, while artificially high correction factors $\hat{k}_{sp}$ for all $p$ values prevail during the first 9 months, which is thus the most unreliable part. 

Finally, one could expect a similar effect when $s$ reaches the ending minimum around month 130. However, as the minimum SN value and its standard deviation are both higher, as pointed out in Section\,\ref{Sec_MeanCycleProp}, no extreme $G_{sp}$ factor is needed to amplify those already enhanced differences. The temporal smoothing of this late part of the mean cycle also leads to a shallower rising trend in the late part following the ending minimum at $s=130$, compared to the early rise for $s$ before 40, thus further decreasing the $\hat{k}_{sp}$ factor.

All the above properties are expected to play a key role on the reliability of the resulting predictions, but those roles combine and overlap in a complex way. In order to fully assess the reliability of the ML output, we will now continue with a full simulation of predictions over multiple cycles. 

\section{Statistics of the ML Performance} \label{Sec_Statistics}

\subsection{Bulk Production of Monthly Hind-casts} \label{SubSec_Production}
As past evaluations of the reliability of the ML method have been fragmentary and often limited to a short time interval, we do not know much about the performance of ML predictions for a wide variety of solar cycles of different amplitudes and duration.

Therefore, we decided to massively produce monthly ML predictions over the past 16 cycles, for which we can then compare predictions with the actual evolution of solar activity.  For this purpose, we first converted the algorithms of the original NGDC FORTRAN programs to the Python language, which allowed modernizing the structure of the program, implementing different options in the calculation, and retrieving the input data directly from our global database instead of text files. Since 2018, those programs were run on a monthly basis at WDC SILSO in parallel with the reference FORTRAN programs as part of the operational production of cycle predictions, in order to check that the output was rigorously identical for both codes. Once validated, the core Python program was just expanded to allow running it in batch mode for any number of starting months. By using exactly the same program for this simulation of past predictions, we thus ensured that the latter are fully compliant with the current operational predictions, as well as those published previously by the NGDC/NOAA.

In our calculations, the program switches to a new cycle once the minimum of the next cycle is passed, based on the SSN series. Starting for the date of each minimum, all subsequent predictions are computed for the months counted from this new minimum. So, in other words, instead of continuing to use the late part (long $p$ times) of the mean cycle, the method jumps again to the very beginning of the mean cycle ($p=0$). This closely simulates what happens for actual operational predictions. There is only one slight compromise with reality. Indeed, for operational monthly predictions based on the latest observed data, the minimum can only be confirmed a few months after this extremum has been passed. Therefore, the switch from one cycle to the next one usually occurs with a delay of a few months after the minimum. Except for this slight displacement of the jump to a new cycle (by 2 -- 3 months), the output predictions are identical and our global statistics are not affected. 

Those abrupt transitions from one base cycle to the next at the minimum are an intrinsic feature of the ML method and they lead to two consequences in the chronology of computed ML predictions:
\begin{itemize}
	\item A discontinuity occurs at each cycle minimum, which can cause a sharp jump in the predicted SN values. The amplitude of this jump will depend on the different local mismatches of the predictions with the real SSN late and early in the cycle. It will depend on the amplitude of the ongoing cycle and the actual length of the preceding cycle, but will never exceed the prediction error.
	\item The number of predictions attached to each solar cycle is variable, and spans the actual duration of that cycle. So, when deriving our statistics for the late part of the cycles, beyond the duration of the shortest past cycle, the number of available cycles present in the data for a given prediction time $p$ decreases by unit steps for large $p$ times.
\end{itemize}
This peculiarity thus leaves visible patterns in the output predictions and our subsequent statistical results (small discrete jumps), but without strongly influencing the overall results. Those artificial discontinuities just reveal an intrinsic drawback of the ML method. When using the ML method, we just need to be aware of their presence.

Overall, the base statistical data set that was generated in this way included 2271 series of predictions, for all starting months between November 1833 and January 2023, thus spanning all cycles from Cycle 9 to the beginning of Cycle 25, currently in progress. Each starting month is treated as the last available SSN value, and we chose to compute predictions over the 13 years (156 months) that follow this starting month, which allows studying the predictions for forward times covering entirely the longest solar cycles.

We computed two versions of those bulk predictions:
\begin{enumerate}
	\item A ``strict'' simulation, where only cycles preceding the prediction date were used to build the mean cycle and to compute the correction factor.
	\item A ``homogeneous'' simulation, where all cycles from 8 to 24 are used for all predictions as a common fixed reference set, thus also including the cycles that follow the prediction date.
\end{enumerate}

By being fully faithful to reality, where only the past observed data can be used, the ``strict'' simulation is mainly useful to check the exact correspondence with actual predictions published in the past. However, for our performance simulation, it leads to varying statistical properties as a function of time, as the base sample of input data decreases as we go back in time. This leads to particularly unreliable results for the first few cycles, after Cycle 8. As anyway, no real ML prediction was published before 1949, when McNish and Lincoln invented their method, all cycles before 18 are irrelevant in this ``strict'' version of the simulation, which thus limits the statistical study to only 7 cycles (19 to 25). Finally, by using less cycles, those results are not representative of the present performance of the method. 

By contrast, in the ``homogeneous'' option where all solar cycles are included, we ensure that all predictions have the same statistical base, which is needed in our global statistical evaluation. Since we then include all elapsed cycles presently available (17 cycles, from 8 to 24), the performance of all predictions is equivalent to the operational predictions issued nowadays and in the coming years, which is what we currently need. In fact, for the last few cycles, the ``strict'' and ``homogeneous'' predictions also largely converge, as they are then based on almost the same set of past cycles. Moreover, fully reliable predictions can be obtained even for the early cycles, back to Cycle 8 in our simulation. 

Therefore, we will only discuss the results of the second simulation in our subsequent analyses below. We made many comparisons between the two sets of results, but we do not develop them in this article as they have shown that all the conclusions reached in this ``homogeneous'' analysis are also valid for the ``strict'' simulation. In fact, some properties just appear less clearly in that latter case, because they are partly altered by the more restricted and variable underlying data sample.   

\begin{figure}
	\centerline{
		\includegraphics*[width= 0.95\textwidth,bb= 11cm 2cm 101cm 45cm,clip=] {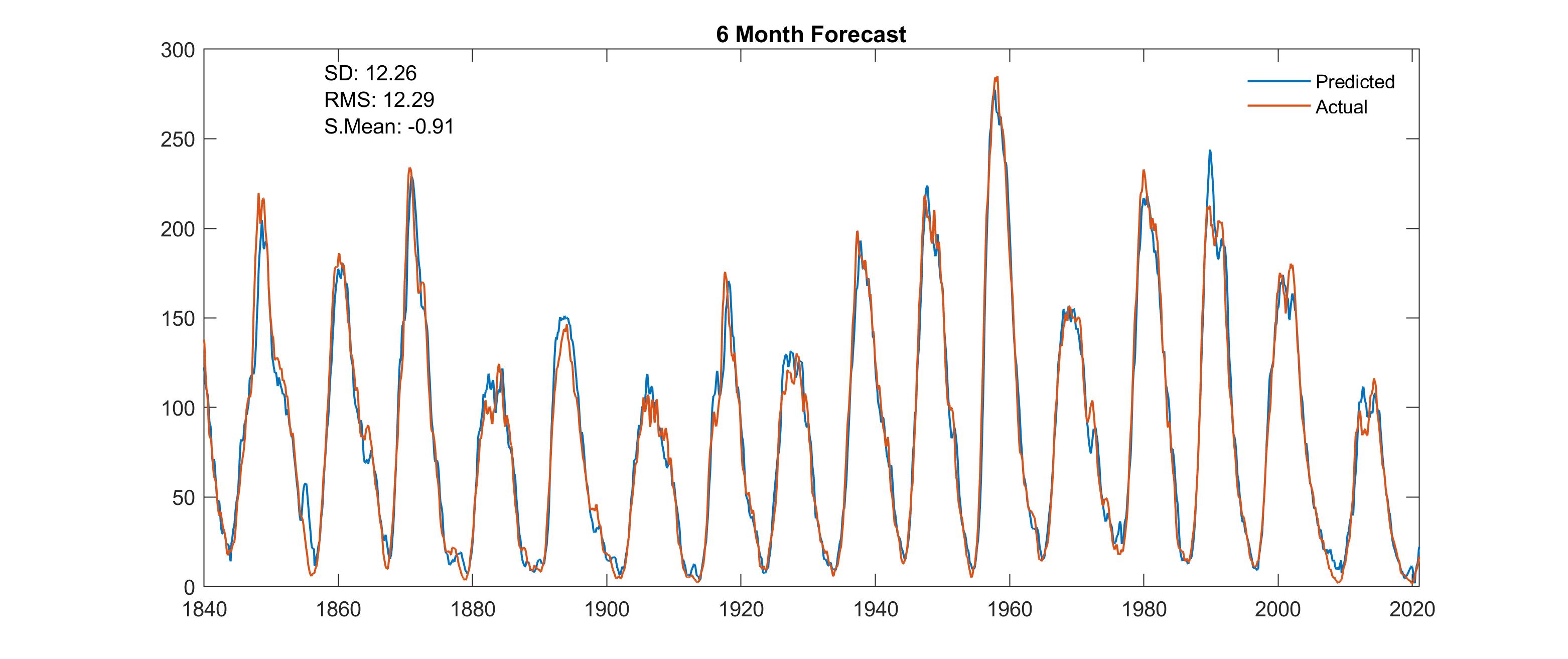} }
	\vspace{-0.42\textwidth}   % Overlays label
	\centerline{\large \bf 	\hspace{0.07 \textwidth}  \color{black}{(a)} \hfill}
	\vspace{0.40\textwidth}    % Shift back to the panel bottom 
	
	\centerline{
		\includegraphics*[width= 0.95\textwidth,bb= 11cm 2cm 101cm 45cm,clip=] {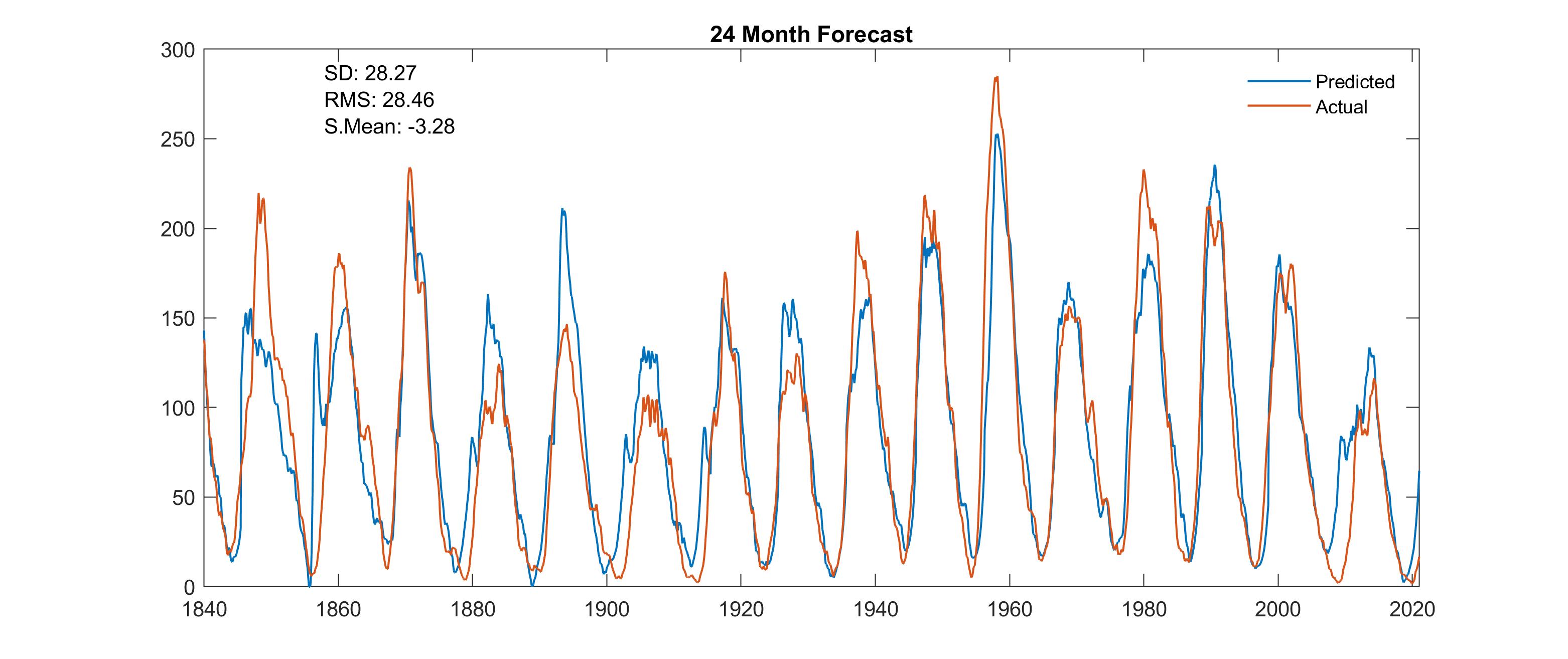} }
	\vspace{-0.42\textwidth}   % Overlays label
	\centerline{\large \bf 	\hspace{0.07 \textwidth}  \color{black}{(b)} \hfill}
	\vspace{0.40\textwidth}    % Shift back to the panel bottom 
	
	\centerline{
		\includegraphics*[width= 0.95\textwidth,bb= 11cm 2cm 101cm 45cm,clip=] {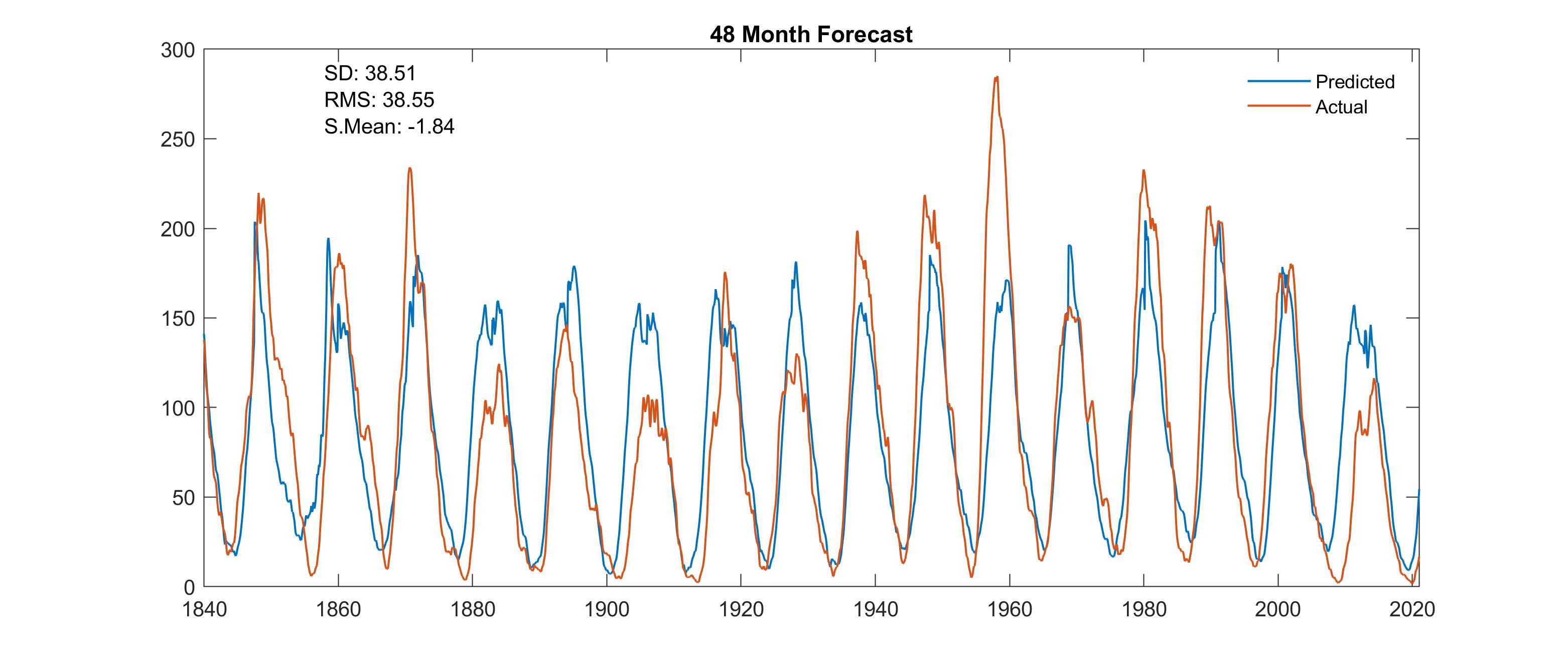} }
	\vspace{-0.42\textwidth}   % Overlays label
	\centerline{\large \bf 	\hspace{0.07 \textwidth}  \color{black}{(c)} \hfill}
	\vspace{0.40\textwidth}    % Shift back to the panel bottom 
	
	\caption{Plot of the observed SSN (red line) and predicted SSN (blue line) for a forward prediction time of 6 months (a), 24 months (b) and 48 months (c).}
	\label{Fig_O-P_06-24-48}
\end{figure}

\subsection{Observed $-$ Predicted (O$-$P) Differences: Global Properties}
In Figure\,\ref{Fig_O-P_06-24-48}, we first plot side-by-side the observed SSN over the 17 past solar cycles, and the corresponding predictions for three different forward times $p$: 6 months (short-term), 24 months (mid-term) and 48 months, which goes well beyond the limit of 18-months ahead generally adopted in the past published ML predictions.
For $p=$ 6 months (Figure\ref{Fig_O-P_06-24-48} (a)), the predictions remain very close to the actual SSN. Only the peak of Cycle 22 in 1991 is significantly overestimated. For $p =$ 24 months (Figure\ref{Fig_O-P_06-24-48} (b)), predictions start to differ significantly from the actual SSN, sometimes by more than 20\%. We observe that the predictions tend to fall below the SSN for strong cycles, while they overestimate the SSN for weak cycles. Some of the cycle minima are also overestimated. For $p =$ 48 months (Figure\ref{Fig_O-P_06-24-48} (c)), all those tendencies have grown further, and we observe that predictions become rather uniform, with only a minor modulation of the predicted amplitude from cycle-to-cycle, whatever the actual amplitude of the true cycle.

\begin{figure}
	\centerline{
		\includegraphics*[width= 0.95\textwidth,bb= 11cm 3cm 134cm 58cm,clip=] {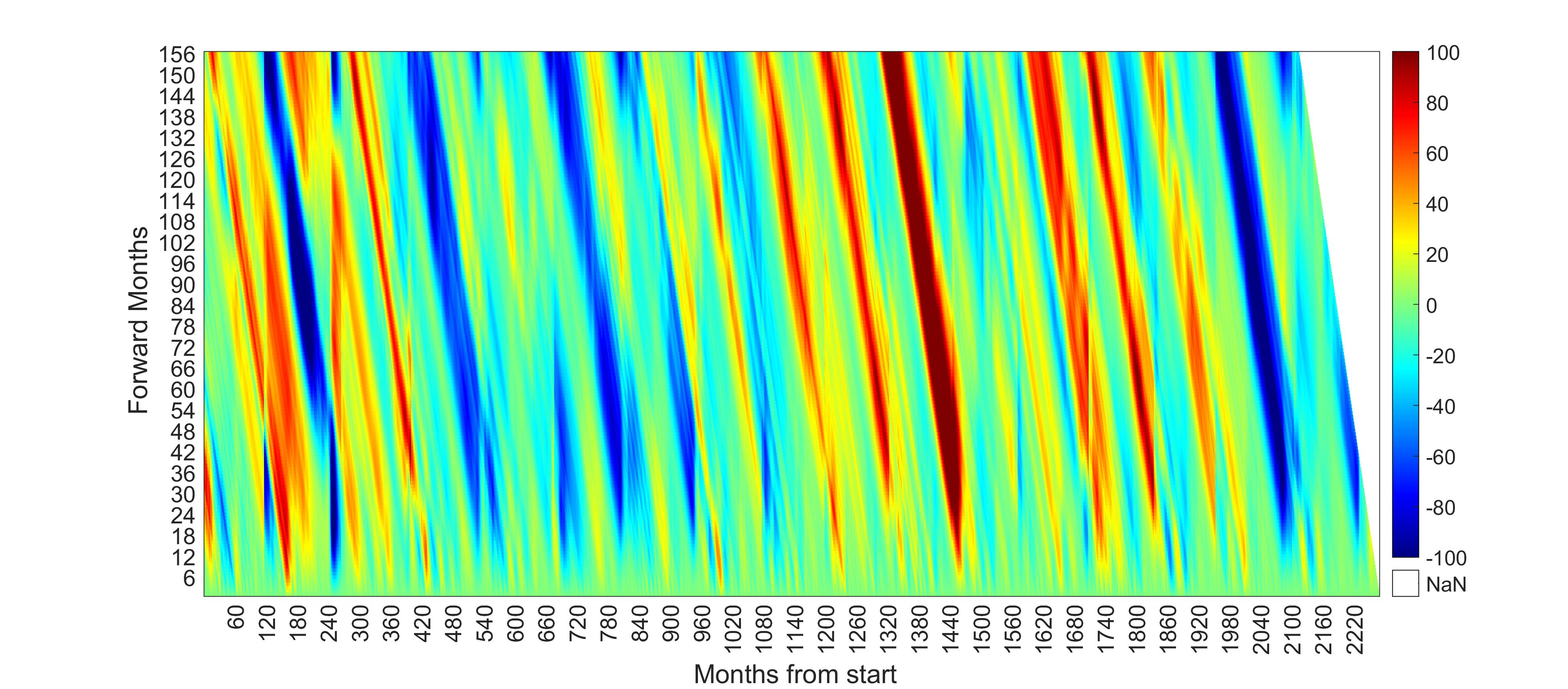} }
	\caption{Map of the difference between observed and predicted SSN, $\hat{S_p^s}-S_p^s$, as a function of the starting time of the prediction $s$ (horizontal axis) and of the prediction forward time $p$ (vertical axis). The color scale is shown at right and spans the range -100 to +100.}
	\label{Fig_MapDiffAllCycles}
\end{figure}

Another way to display this general behavior is to plot a map of the difference between the observed and predicted SSN as a function of start time $s$ and forward prediction time $p$, as in Figure\,\ref{Fig_MapDiffAllCycles}. Here, we can see that this difference is dominated by diagonal ridges of positive or negative differences, spaced by about 11 years. The diagonal pattern is actually inclined at 45°, which suggests that the primary differences observed $-$ predicted are fixed in absolute time (fixed prediction date). The ridges fade out at the bottom, at low $p$ forward times, for which they decrease towards zero, indicating an improving agreement between predictions and actual observations. On the other hand, for $p > 24$~months, the ridges already reach their full amplitude, which then remains largely constant for longer $p$ times, indicating that the prediction errors do not grow further and become largely independent from prediction time $p$.

\begin{figure}
	\centerline{
		\includegraphics*[width= 0.75\textwidth,bb= 3cm 0cm 58cm 45cm,clip=] {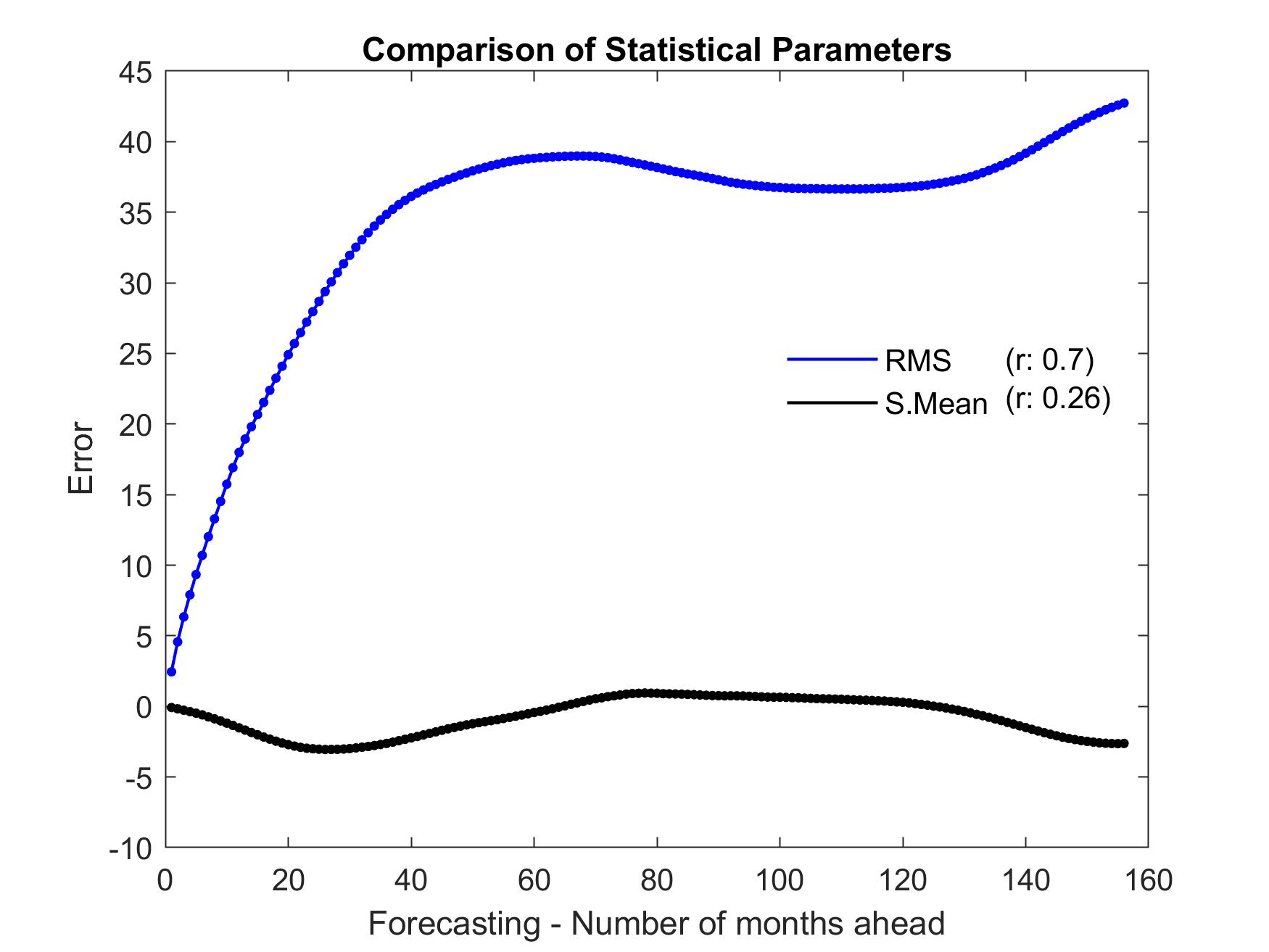} }
	\caption{Variation of the root-mean-square (RMS) difference $\delta_p^{rms}$ (blue curve, labeled ``RMS'') as a function of prediction lead time $p$ in months. As the mean difference $\delta_p$ (black, labeled ``S.Mean'') is close to 0 for all $p$ times, the standard deviation $\sigma_p^{\delta}$ is almost equal to $\delta_p^{rms}$ and is not plotted here, as it would be fully superimposed on the ``RMS'' curve.}
	\label{Fig_RMSallCycles}
\end{figure}

This overall dependency to $p$ can be summarized by computing the root-mean-square (RMS) difference between the observed and predicted SSN for ahead month $p$ , defined as:
%\begin{linenomath}
\begin{equation}
\delta_p^{rms}=\sqrt{\frac{ \sum_{n=8}^{c-1}\sum_{s=1}^{N_s} \left(\hat{S_{sp}^n}-S_{sp}^n \right)^2}{N}}      \label{EQ_RMSdiff}
\end{equation}
%\end{linenomath}
where $\hat{S_{sp}^n}$ and $S_{sp}^n$ are respectively the predicted and observed SSN for the same month, and the sum is over all starting times $s$ of all predictions in each cycle, and over all input Cycles 8 to 24 (number of predictions $N=2271$), with a given fixed prediction forward time $p$, thus along an horizontal line in the map of Figure\,\ref{Fig_MapDiffAllCycles}. 

A measure of systematic deviations of predictions relative to observed SSN is provided by the mean difference:
%\begin{linenomath}
\begin{equation}
\delta_p = \frac{\sum_{n=8}^{c-1}\sum_{s=1}^{N_s} \left(\hat{S_{sp}^n}-S_{sp}^n \right)}{N}         \label{EQ_SignedDiff}
\end{equation}
%\end{linenomath}
and the random dispersion around this mean difference is given by the standard deviation:
%\begin{linenomath}
\begin{equation}
\sigma_p^\delta=\sqrt{\frac{\sum_{n=8}^{c-1}\sum_{s=1}^{N_s}\left(\left(\hat{S_{sp}^n}-S_{sp}^n \right) - \delta_p \right)^2}{N-1}}    \label{EQ_SD}
\end{equation}
%\end{linenomath}

Figure\,\ref{Fig_RMSallCycles} shows that the RMS difference $\delta_p^{rms}$ increases steadily over the first 40 months before stabilizing at a value of about 38, which is the overall dispersion of all cycles in the base sample over the whole duration of the cycle (thus mixing all phases of the cycle). The mean difference $\delta_p$ is almost null for all times $p$, indicating that no significant systematic deviation of predictions relative to the observed SSN exists, over all 17 past solar cycles and all starting times along those cycles. Consequently, the standard deviation  $\sigma_p^\delta$ coincides with $\delta_p^{rms}$ (blue curve in Figure\,\ref{Fig_RMSallCycles}), which tells us that the overall uncertainty measured by  $\delta_p^{rms}$ is entirely due to random cycle-to-cycle variability, rather than to systematic biases in the ML prediction method.

\subsection{O$-$P Differences : Single Cycles}
In order to better quantify the above properties, we are now going to consider in more detail single solar cycles. As illustration, we present three different sample cycles with low, intermediate and high maximum amplitude, respectively Cycles 14, 23 and 21.

\begin{figure}
	\centerline{
		\includegraphics*[width= 0.75\textwidth,bb= 2cm 1cm 88cm 65cm,clip=] {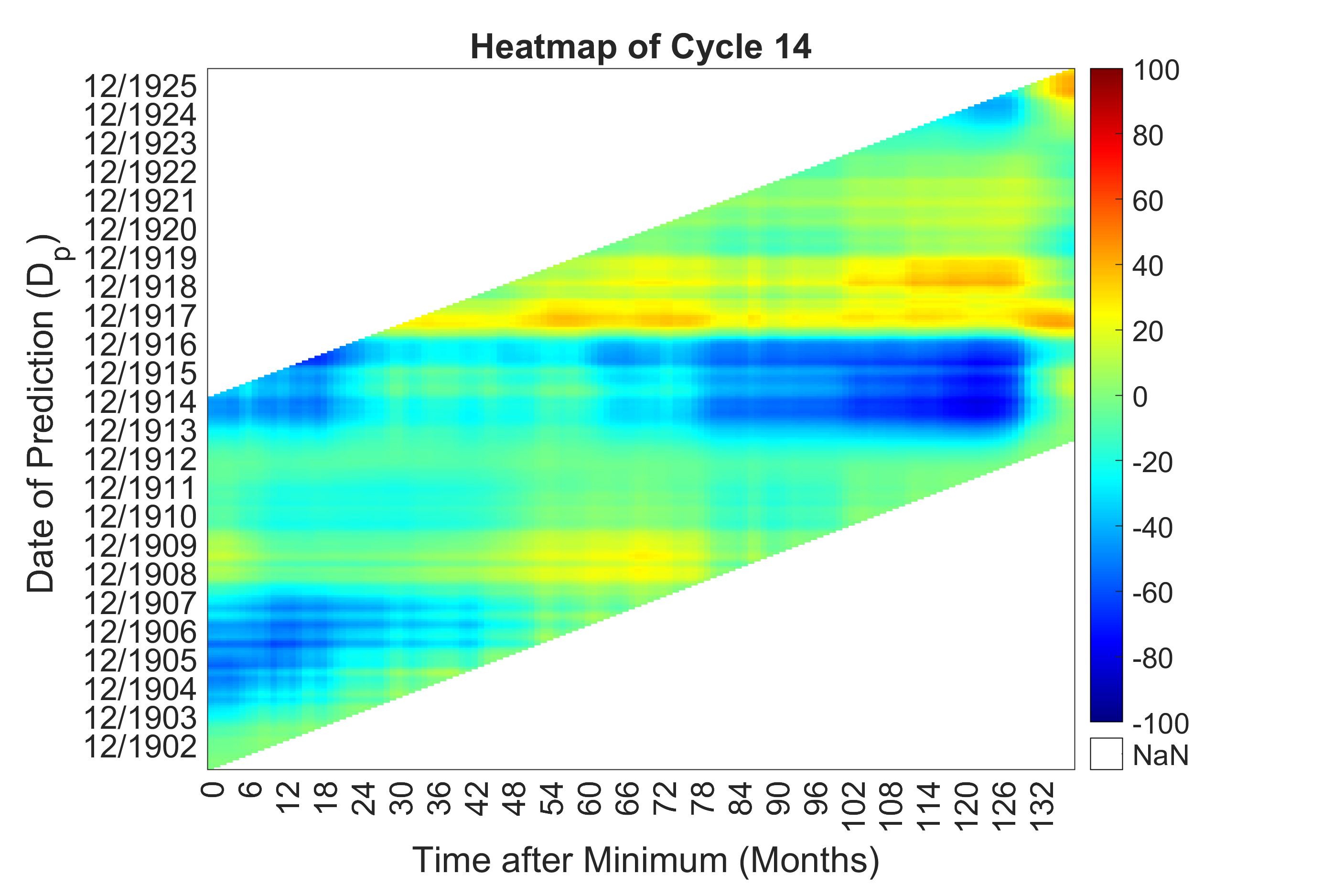} }
	\vspace{-0.51\textwidth}   % Overlays label
	\centerline{\large \bf 	\hspace{0.26 \textwidth}  \color{black}{(a)} \hfill}
	\vspace{0.48\textwidth} 

	\centerline{
		\includegraphics*[width= 0.75\textwidth,bb= 2cm 1cm 88cm 65cm,clip=] {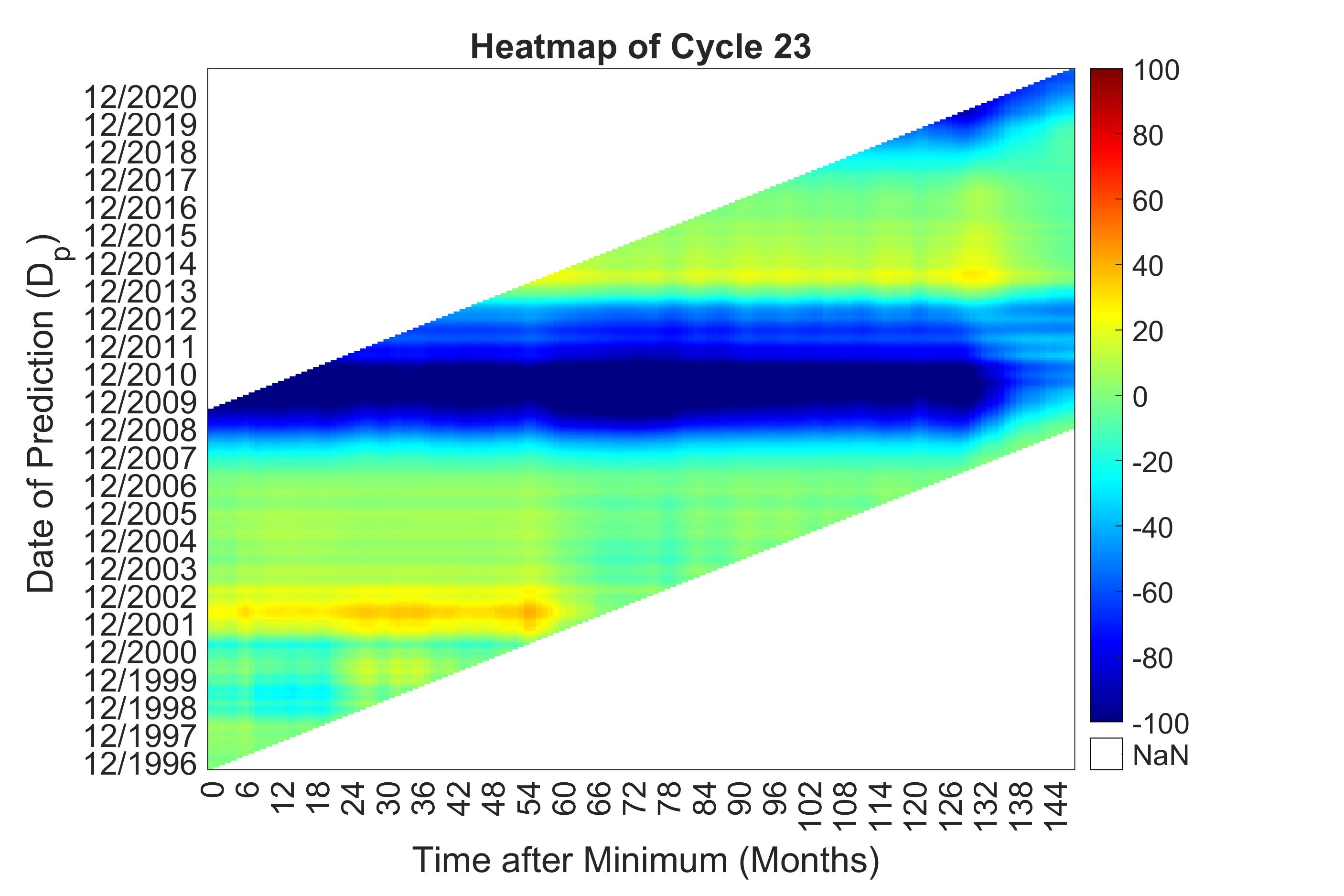} }
	\vspace{-0.51\textwidth}   % Overlays label
	\centerline{\large \bf 	\hspace{0.26 \textwidth}  \color{black}{(b)} \hfill}
	\vspace{0.48\textwidth} 

	\centerline{
		\includegraphics*[width= 0.75\textwidth,bb= 2cm 1cm 88cm 65cm,clip=] {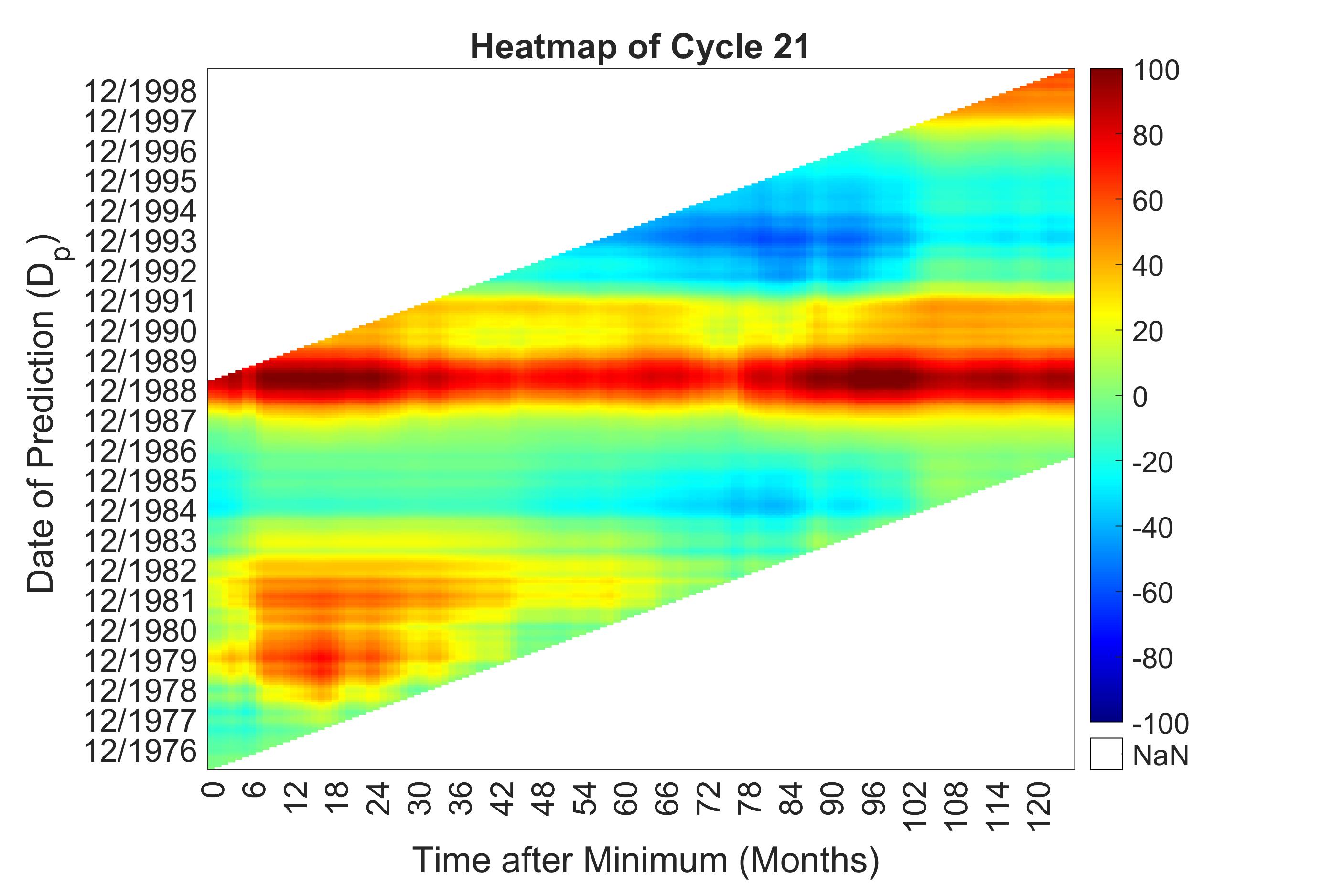} }
	\vspace{-0.51\textwidth}   % Overlays label
	\centerline{\large \bf 	\hspace{0.26 \textwidth}  \color{black}{(c)} \hfill}
	\vspace{0.48\textwidth} 

	\caption{Maps of the difference between observed and predicted SSN for Cycle 14 (low maximum, (a)), Cycle 23 (intermediate maximum, (b)) and Cycle 21 (high maximum, (c)). Here, the vertical axis spans 18 years, and thus extends well past the end of the predicted cycle, as we wish to produce predictions up to the end of the cycle, which thus extend into the next cycle. In practice, only the lower half of the map (about 11 years) is actually used in operational predictions.}
	\label{Fig_MapDiff_cycle14-23-21}
\end{figure}

For each cycle, we plotted a map of the difference between observed and predicted SN  (Figure\,\ref{Fig_MapDiff_cycle14-23-21}), in the same way as in Figure\,\ref{Fig_MapDiffAllCycles}, but now extracting vertical slices spanning only  a single solar cycle in the horizontal temporal dimension. However, here, the vertical axis now gives the date for which the prediction is made, instead of simply the forward time $p$ relative to the last observed month, like in Figure\,\ref{Fig_MapDiffAllCycles}. Therefore, the predictions now fill a diagonal band, with its lower edge corresponding the first month following the starting month of the prediction (horizontal axis). The upper edge of the band correspond to the 155-month (13 year) range over which we calculated predictions. Therefore, only the lower half of the plot corresponds to predictions of the cycle in progress. The upper part, which extends into the next cycle, would be used only for predictions made shortly before the actual end of the cycle, as explained in Section\,\ref{SubSec_Production}.

We find that the differences are lowest along the lower edge and grow upwards, but here, the differences stabilize at values that are constant relative to the date for which the prediction is made, independently of the forward time of the prediction, i.e. of the starting date of the prediction. This produces horizontal bands of rather constant values in the maps. Moreover, the differences are mostly negative for Cycle 14 (Figure\,\ref*{Fig_MapDiff_cycle14-23-21} (a)), which is lower than the mean cycle, mostly positive for Cycle 21, a high cycle (Figure\,\ref*{Fig_MapDiff_cycle14-23-21} (c)), and fluctuating around 0 for Cycle 23 (Figure\,\ref*{Fig_MapDiff_cycle14-23-21} (b)), which matches fairly well the mean cycle. This thus shows that beyond the first months following the starting month, the prediction errors are predominantly the difference relative to the fixed mean cycle, and they are largely independent of the starting date of the predictions. 

\begin{figure}
	\centerline{
		\includegraphics*[width= 0.90\textwidth,bb= 4mm 2mm 173mm 115mm,clip=] {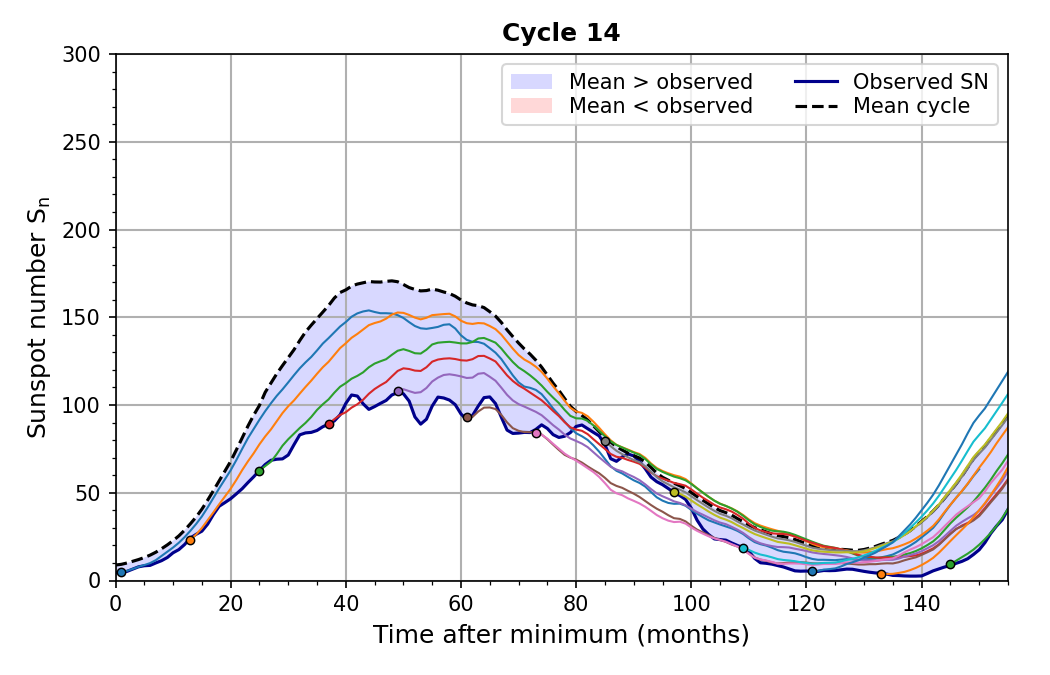} }
	\caption{Difference between the SSN in individual solar cycles (black solid curve) and the mean cycle (dashed line) for a weak solar cycle (Cycle 14). The red (blue) shading indicates that the true cycle is higher (lower) than the mean cycle. The colored thin curves are actual monthly predictions taken every 12 months (starting month marked by a colored dot). }
	\label{Fig_CompMeanCyc14}
\end{figure}

\begin{figure}
	\centerline{
		\includegraphics*[width= 0.90\textwidth,bb= 4mm 2mm 173mm 115mm,clip=] {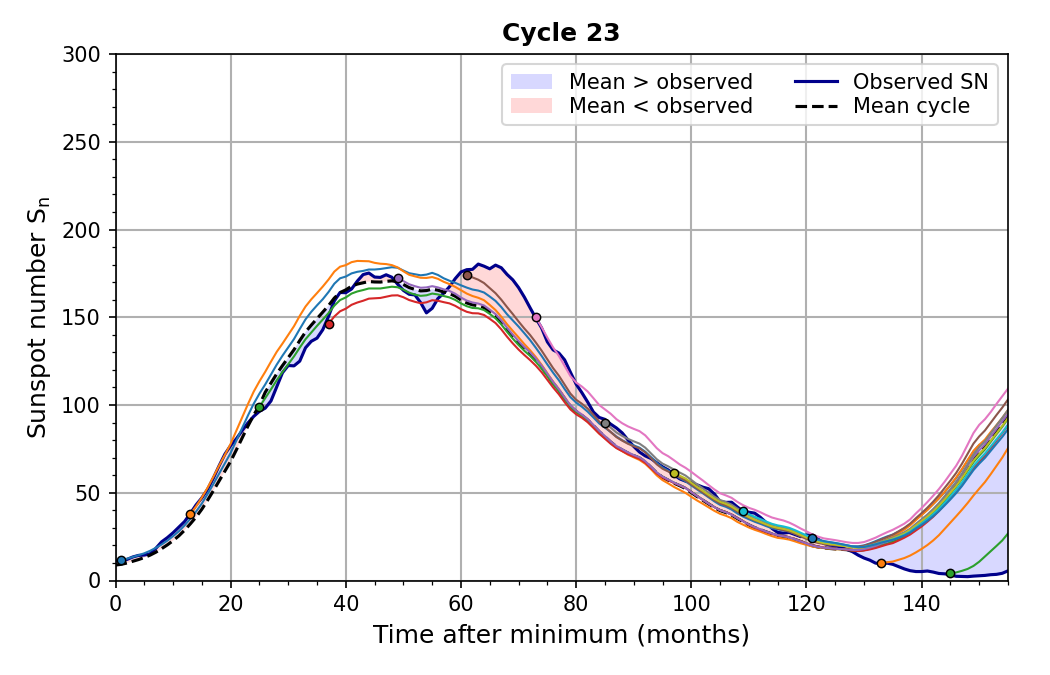} }
	\caption{Difference between the SSN in individual solar cycles (black solid curve) and the mean cycle (dashed line) for a cycle of medium amplitude (Cycle 23). The curves are presented like in Figure\,\ref{Fig_CompMeanCyc14}.}
	\label{Fig_CompMeanCyc23}
\end{figure}

\begin{figure}
	\centerline{
		\includegraphics*[width= 0.90\textwidth,bb= 4mm 2mm 173mm 115mm,clip=] {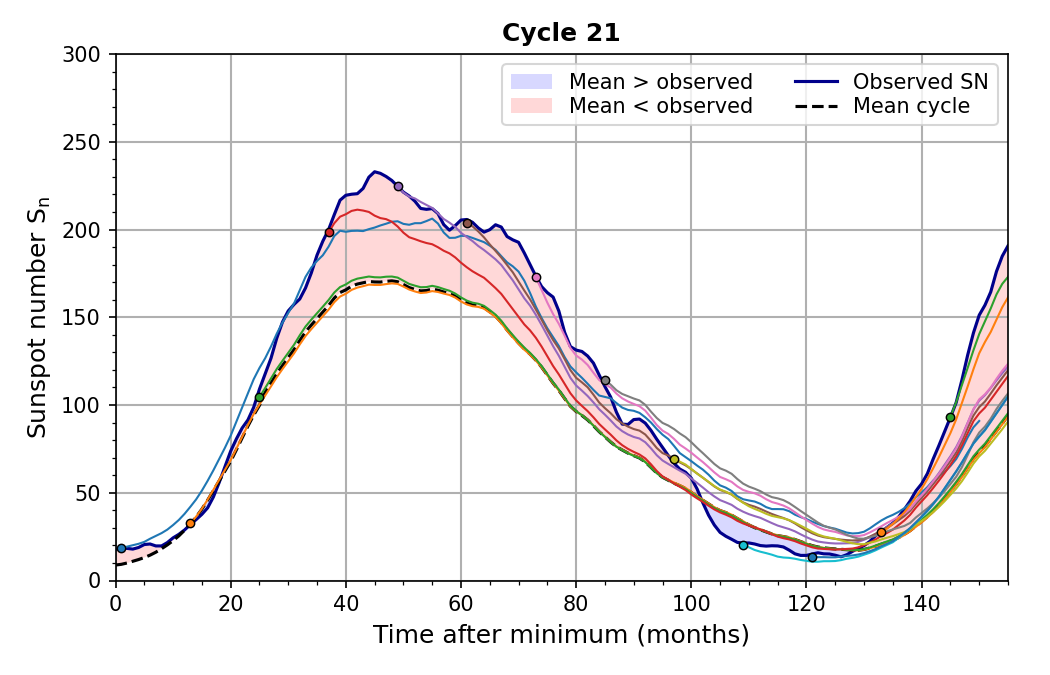} }
	\caption{Difference between the SSN in individual solar cycles (black solid curve) and the mean cycle (dashed line) for a cycle of high amplitude (Cycle 21). The curves are presented like in Figure\,\ref{Fig_CompMeanCyc14}.}
	\label{Fig_CompMeanCyc21}
\end{figure}

Comparing the maps in Figure\,\ref{Fig_MapDiff_cycle14-23-21} with the simple difference between the three cycles and the mean cycle (Figures\,\ref{Fig_CompMeanCyc14}, \ref{Fig_CompMeanCyc23} and \ref{Fig_CompMeanCyc21}), we can see that the latter (shaded area between the curves) indeed matches the fixed variations of the differences along the vertical dimension in the maps of differences (lower half, spanning 155 months). In particular, the differences are mostly negative for the lowest cycle (\#14, Figure\,\ref{Fig_CompMeanCyc14}) and mostly positive for the highest cycle (\#21, Figure\,\ref{Fig_CompMeanCyc21}). Accordingly, individual predictions (thin curves) are systematically overestimating the actual activity level for Cycle 14, and underestimating it for Cycle 21. For Cycle 23 (Figure\,\ref{Fig_CompMeanCyc23}), which has almost exactly the same amplitude as the mean cycle, the differences are lower and fluctuate around zero. In that case, individual predictions match fairly closely the true cycle over its full duration, though they still largely fail beyond the ending minimum. In addition, we note that the largest mismatch near the maximum is actually due to the fact that the maximum of Cycle 23 comes later than the maximum of the mean cycle. Likewise, the large negative difference beyond month 130 is almost entirely due to a strong mismatch between the length of Cycle 23, which was much longer than the fixed duration of the mean cycle. 

Overall, those results illustrate how the excess or deficit of the actual SSN relative to the mean cycle leads to a corresponding upwards or downwards difference in the individual predictions, and also to a corresponding artificial trend over the first two years of the predictions. Such a trend seems to be intrinsic to the ML principles. Namely, the predictions track the observed cycle at $p=0$ and progressively drift away by converging towards the invariable mean cycle as $p$ increases. This upward or downward bias thus starts already at $p=1$, suggesting that it is a primary contribution to the prediction error. For large $p$ times beyond 12 to 24 months into the future, the prediction errors fully reduce to the discrepancy between the actual solar cycle and the fixed mean cycle, thus producing a fixed pattern matching the history of the actual solar cycles. Those long-range errors are characterized by the following properties:
\begin{itemize}
	\item they are independent of the forward time of the prediction, beyond the first few months following the starting time of the prediction.
	\item they do not grow indefinitely for increasing forward times, but fluctuate within the range of differences with the mean cycle.
	\item as the mean cycle provides the baseline prediction, the ML method suffers from a downward bias when the actual cycle has a high amplitude, and an upward bias for weak solar cycles. 
\end{itemize}

Moreover, we observe that the O$-$P differences are due both to the amplitude of the actual cycle but also to the different times of the cycle maximum and ending minimum, i.e. the rise time and the cycle length. Around those two extrema, this temporal mismatch may become the dominant cause of the prediction error.

\subsection{RMS Prediction Errors over Cycle Phases (Single Cycles)}
Based on the above insight, we can derive a representative value of the RMS prediction error valid for all cycles, taking into account the two main properties found in the above results:
\begin{itemize}
	\item the prediction error is always close to zero for the first predicted months, and grows over 12 to 24 month, before stabilizing for longer forward prediction times $p$.
	\item for long forward times, the prediction errors are constant for fixed prediction dates. Therefore, as the starting date of the prediction $s$ progresses, the same error will occur for shorter forward times relative to this starting date, thus drifting on the scale of the relative forward prediction times $p$.
\end{itemize}

Therefore, we derived $\delta_{sp}^{rms}$, $\delta_{sp}$ and $\sigma_{sp}^{\delta}$ in the same way as in Equations\,(\ref{EQ_RMSdiff}), (\ref{EQ_SignedDiff}) and (\ref{EQ_SD}), but instead of including all starting times, we limited the sum to starting times $s$ within a limited time range, to avoid the merging of errors in the different fixed phases of the solar cycle. In other words, we prevent the smearing of the diagonal patterns shown in Figure\,\ref{Fig_MapDiffAllCycles} by taking sufficiently narrow vertical time slices taken at the same phase in all cycles. This is equivalent to averaging across multiple vertical strips (one per cycle) instead of the whole width of the map, which led to Figure\,\ref{Fig_RMSallCycles}.  However, given the 13-month smoothing of the input SN, it is pointless to restrict the width of our time ranges below 12 months. We thus settled for a 1-year width.

\begin{figure}
	\centerline{
		\includegraphics*[width= 0.82\textwidth, bb= 1cm 0cm 96cm 65cm,clip=] {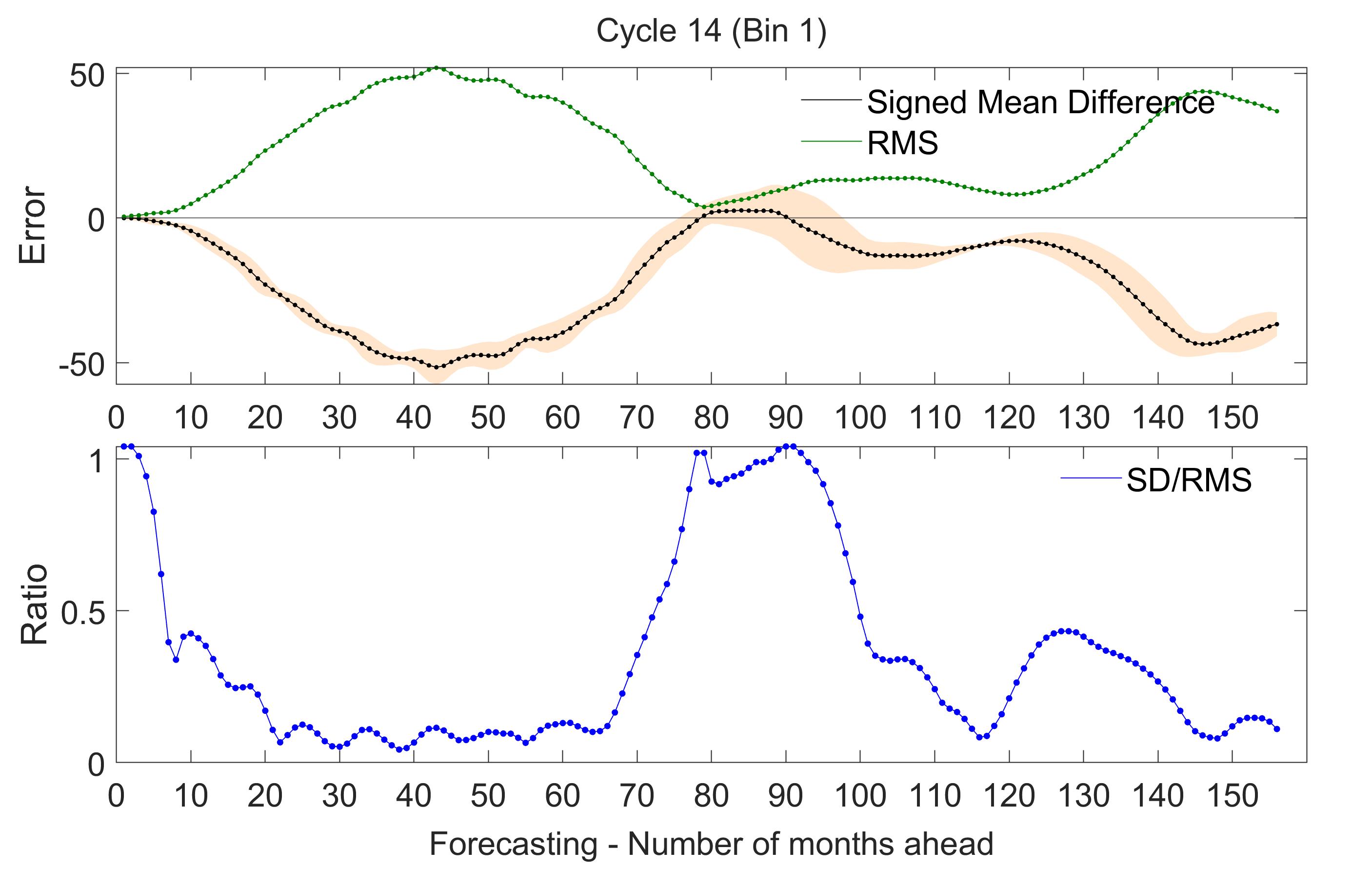} }
	\vspace{-0.55\textwidth}   % Overlays label
	\centerline{\large \bf 	\hspace{0.15\textwidth} \color{black}{(a)} \hfill}
	\vspace{0.50\textwidth} 

	\centerline{
		\includegraphics*[width= 0.82\textwidth, bb= 1cm 0cm 96cm 65cm,clip=] {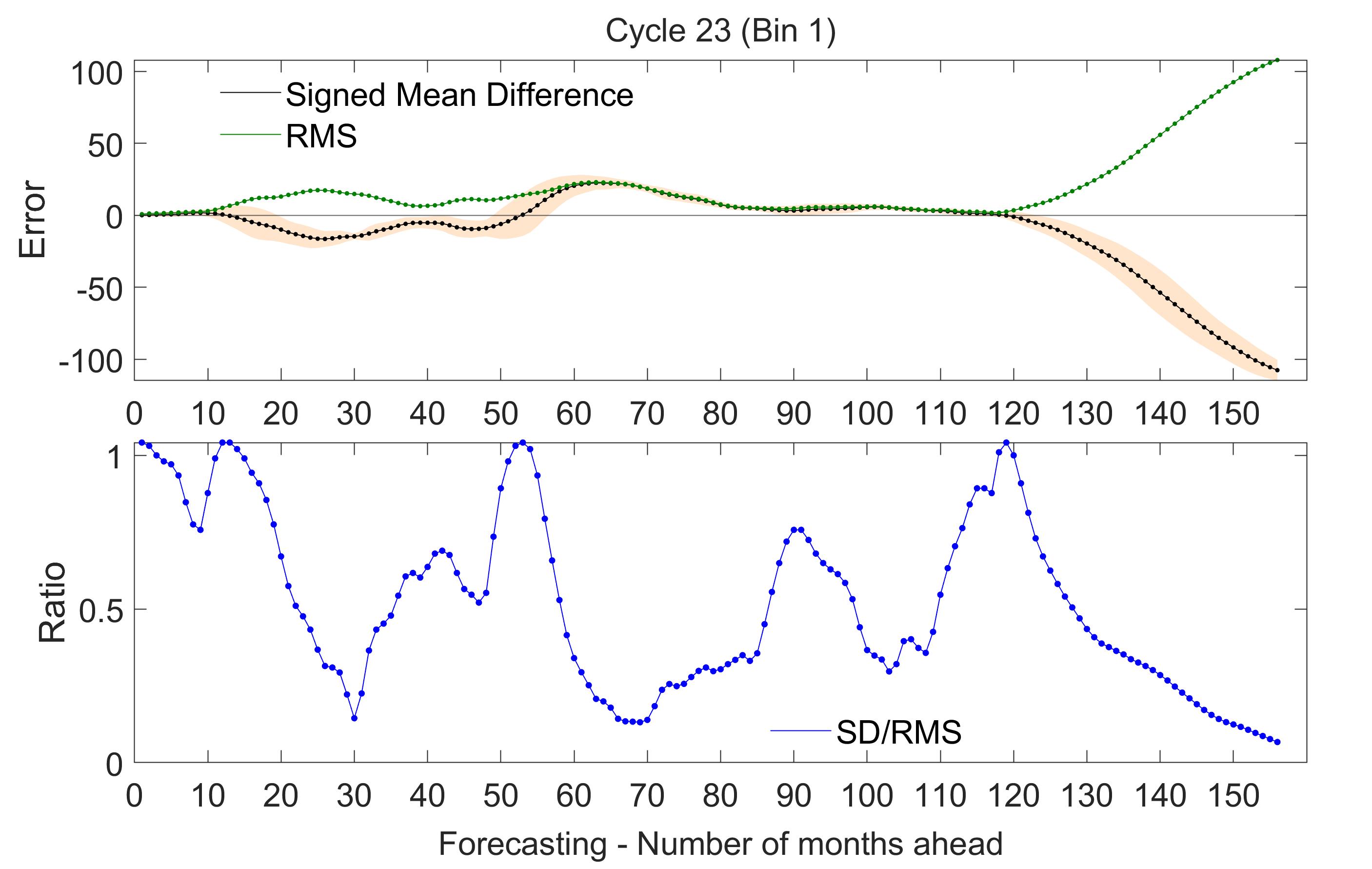} }
	\vspace{-0.55\textwidth}   % Overlays label
	\centerline{\large \bf 	\hspace{0.15\textwidth} \color{black}{(b)} \hfill}
	\vspace{0.50\textwidth} 

	\centerline{
		\includegraphics*[width= 0.82\textwidth, bb= 1cm 0cm 96cm 65cm,clip=] {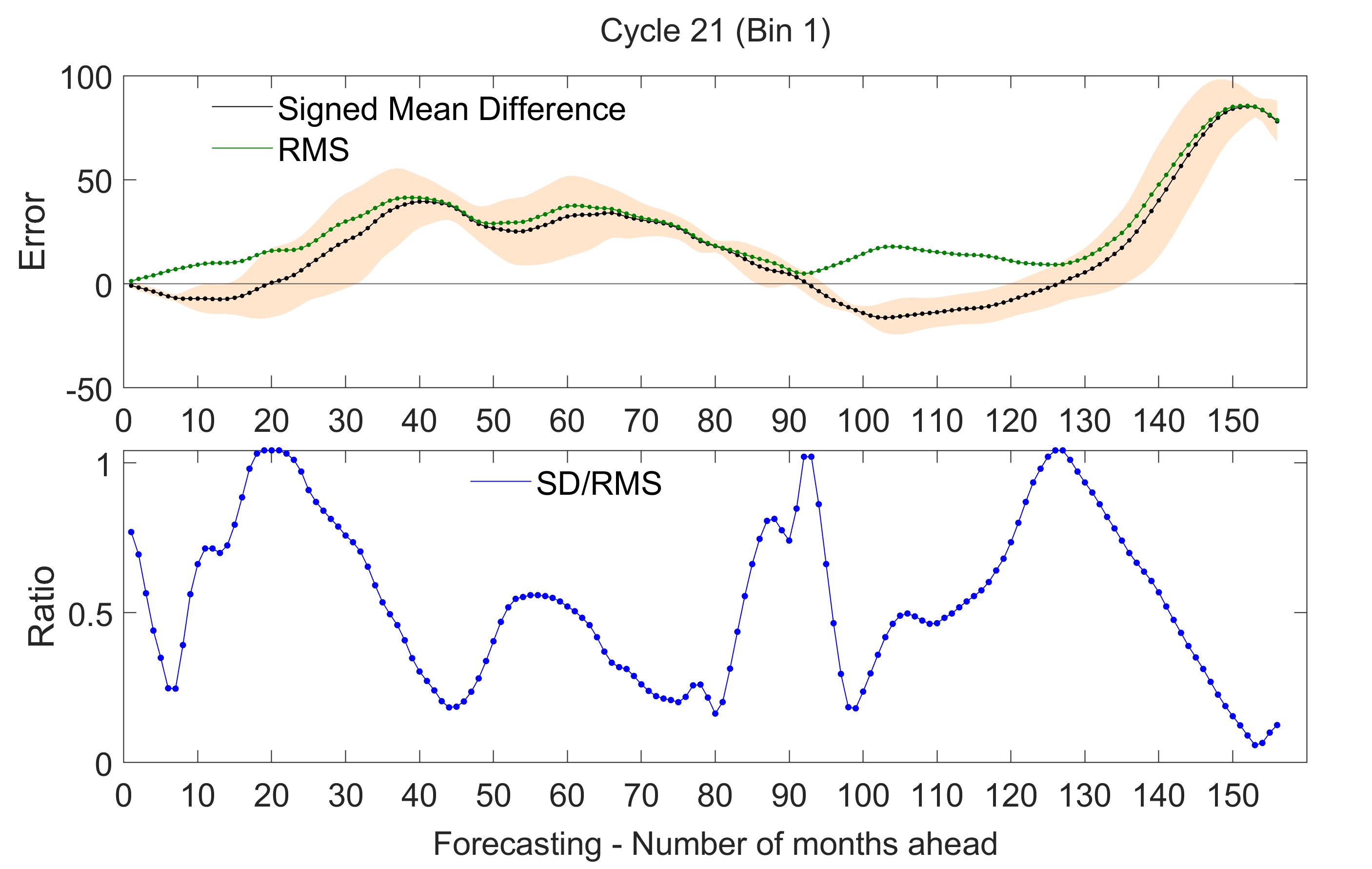} }
	\vspace{-0.55\textwidth}   % Overlays label
	\centerline{\large \bf 	\hspace{0.15\textwidth} \color{black}{(c)} \hfill}
	\vspace{0.50\textwidth} 

	\caption{Prediction error as a function of forward prediction time $p$, for Cycle 14 (low maximum, (a)), Cycle 23 (intermediate maximum, (b)) and Cycle 21 (high maximum, (c)), averaged for starting times in year 1 of the cycle (thus over  months $s= 1, ...\,, 12$). In the upper plot of each sub-figure, the RMS error $\delta_{sp}^{rms}$ is plotted in green and the signed mean $\delta_{sp}$ difference is plotted in black, with the standard deviation $\sigma_{sp}^{\delta}$ (SD) around this difference shown as the shaded band. The lower plot shows the ratio $\sigma_{sp}^{\delta}/\delta_{sp}^{rms}$.}
	\label{Fig_RMSy1Cyc}
\end{figure}

We first show $\delta_{sp}^{rms}$, $\delta_{sp}$ and $\sigma_{sp}^{\delta}$ for the three cycles illustrated in the previous Section, for predictions with a starting day in the first year of the cycle (Figure\,\ref{Fig_RMSy1Cyc}). We can see that the mean difference $\delta_{sp}$ (black curve) varies in accordance with the patterns found above for each cycle. In addition, we plot the RMS error $\delta_{sp}^{rms}$ as the green curve, and the standard deviation $\sigma_{sp}^\delta$ relative to the mean of all predictions within the 1-year time bin as the shaded interval around $\delta_{sp}$. The latter provides a measure of the random errors in the predictions and of the variation of predictions over the 1-year time interval (evolution of the cycle), while $\delta_{sp}^{rms}$ also includes systematic deviations between predictions and observations. Therefore, we also plot in the lower panel the $\sigma_{sp}^{\delta}/\delta_{sp}^{rms}$ ratio. This ratio varies between 0 and 1, and it indicates which fraction of $\delta_{sp}^{rms}$ is random. A low ratio thus corresponds to intervals in which the RMS prediction error is dominated by systematic deviations of the predictions. 

We can see that the RMS error are low and the $\sigma_{sp}^{\delta}/\delta_{sp}^{rms}$ ratio is high over the first 6 to 24 months of the predictions. The small errors are thus dominated by the uncertainty of the last observed value and the change of solar activity during the 1-year time window. After this early period, the ratio falls to a low value of 0.2 or below, indicating that the error is mainly due to a systematic deviation, as is confirmed by a large mean signed difference, which deviates from 0 by much more than $\sigma_{sp}^\delta$,  shown in the upper plot. Depending on the cycle, the $\sigma_{sp}^{\delta}/\delta_{sp}^{rms}$ ratio peaks again near 1 in the late part of the cycle, every time the actual SSN comes close to the mean cycle. In particular, this occurs near the end of the predicted cycle, as all cycles decrease to a limited range of low SSN values.

\begin{figure}
	\centerline{
		\includegraphics*[width= 0.82\textwidth, bb= 0cm 0cm 96cm 65cm,clip=] {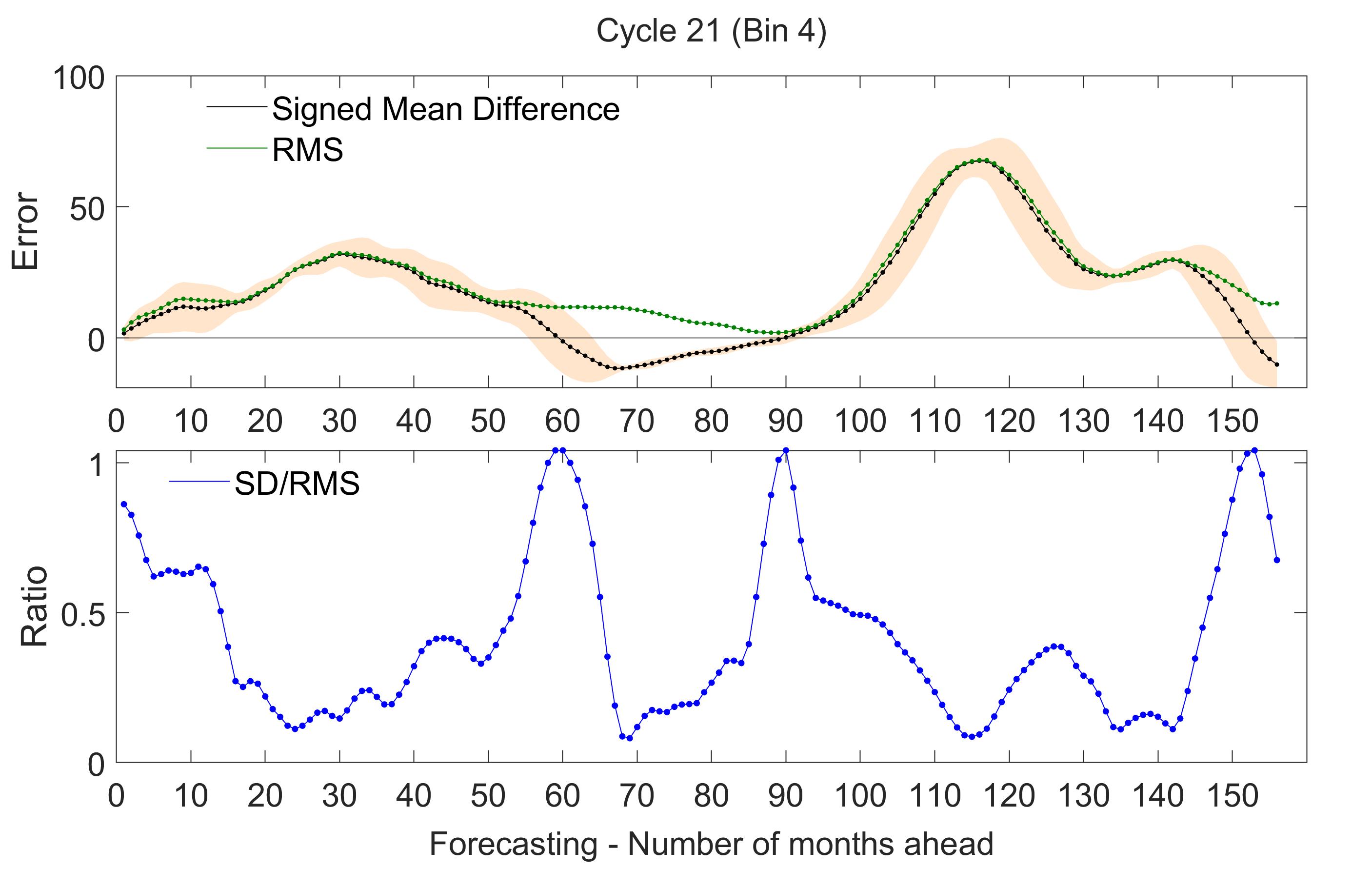} }
	\vspace{-0.55\textwidth}   % Overlays label
	\centerline{\large \bf 	\hspace{0.15\textwidth} \color{black}{(a)} \hfill}
	\vspace{0.50\textwidth} 

	\centerline{
		\includegraphics*[width= 0.82\textwidth, bb= 0cm 0cm 96cm 65cm,clip=] {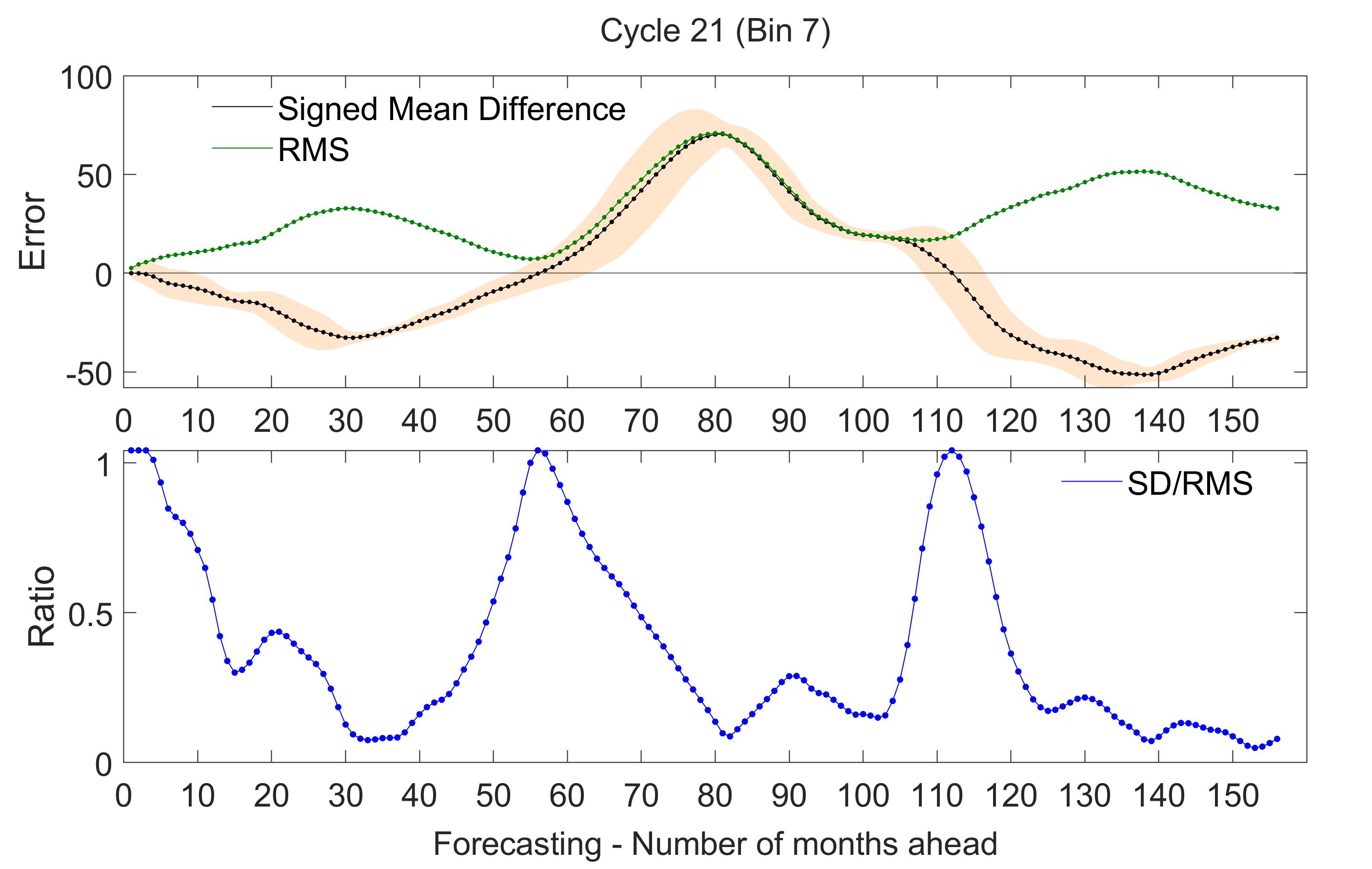} }
	\vspace{-0.55\textwidth}   % Overlays label
	\centerline{\large \bf 	\hspace{0.15\textwidth} \color{black}{(b)} \hfill}
	\vspace{0.50\textwidth} 

	\caption{Prediction error as a function of ahead months $p$, for Cycle 21, averaged for starting times in year 4 (a) and in year 7 (b) of the cycle. See bottom plot of Figure\,\ref{Fig_RMSy1Cyc} for year 1, with the same curve layout.}
	\label{Fig_RMSCyc21_y1-4-7}
\end{figure}

Now, looking at the same error measures for a single cycle (here, Cycle 21) but for predictions starting in year 1, 4 and 7 (Figure\,\ref{Fig_RMSy1Cyc} (c), and Figure\,\ref{Fig_RMSCyc21_y1-4-7} (a), (b)), we observe that in accordance to the previous Section, the variations of errors as a function of forward time, remain largely unchanged, but are each time-shifted by 36 months to the left, confirming the largely invariant and temporally-fixed long-range errors. On the other hand, the curve is always ``tapered'' on the left, for the first predicted months, where $\delta_{sp}^{rms}$ and $\sigma_{sp}^\delta$ decrease towards 0 for the starting date, and the corresponding $\sigma_{sp}^\delta/\delta_{sp}^{rms}$ ratio rises to 1.

\subsection{RMS Prediction Errors over Cycle Phases (All Cycles)}

\begin{figure}
	\centerline{\includegraphics*[width= 0.82\textwidth, bb= 1cm 0cm 96cm 65cm,clip=] {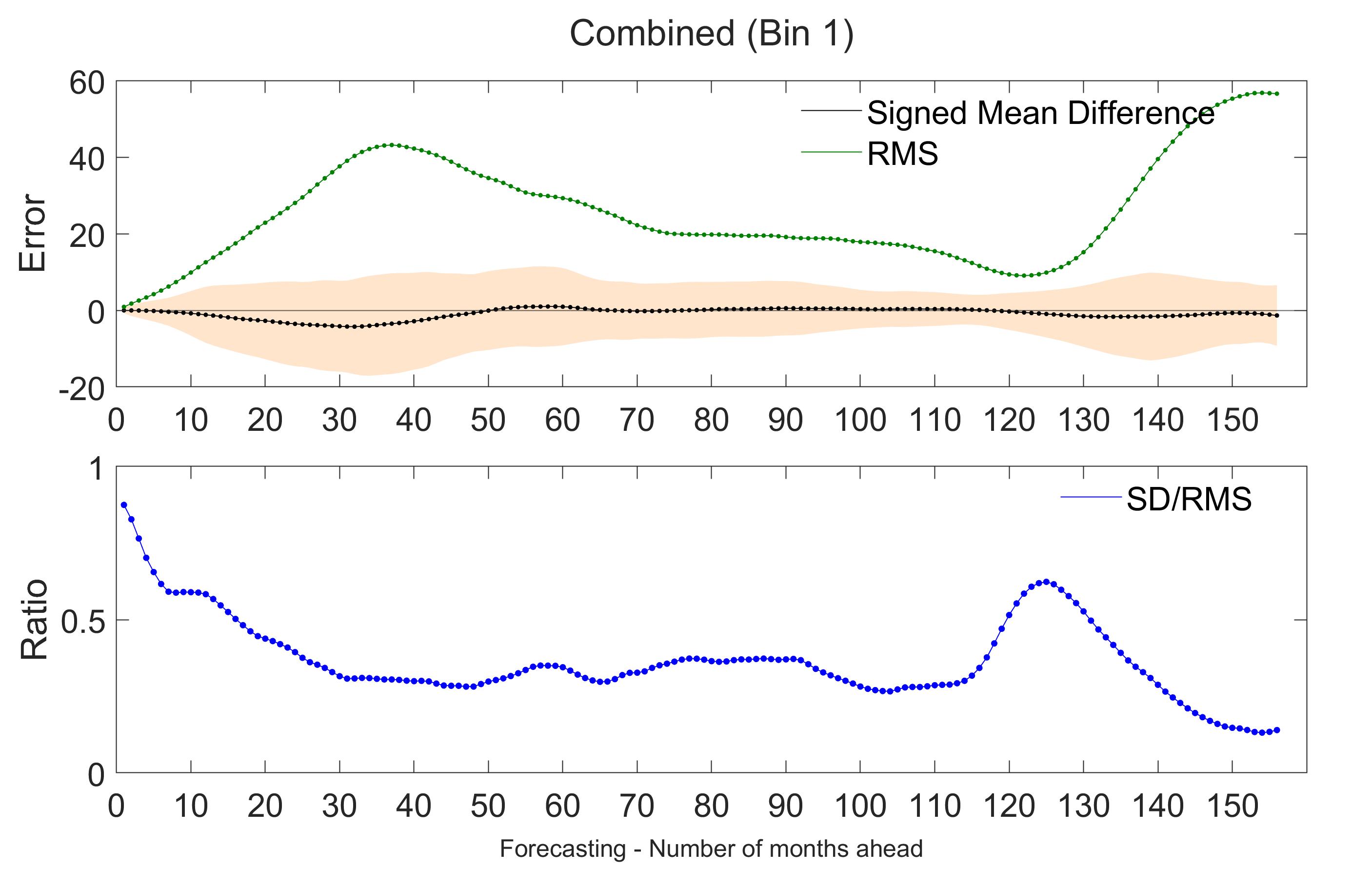} }
	\vspace{-0.55\textwidth}   % Overlays label
	\centerline{\large \bf 	\hspace{0.15\textwidth} \color{black}{(a)} \hfill}
	\vspace{0.50\textwidth} 

	\centerline{\includegraphics*[width= 0.82\textwidth, bb= 1cm 0cm 96cm 65cm,clip=] {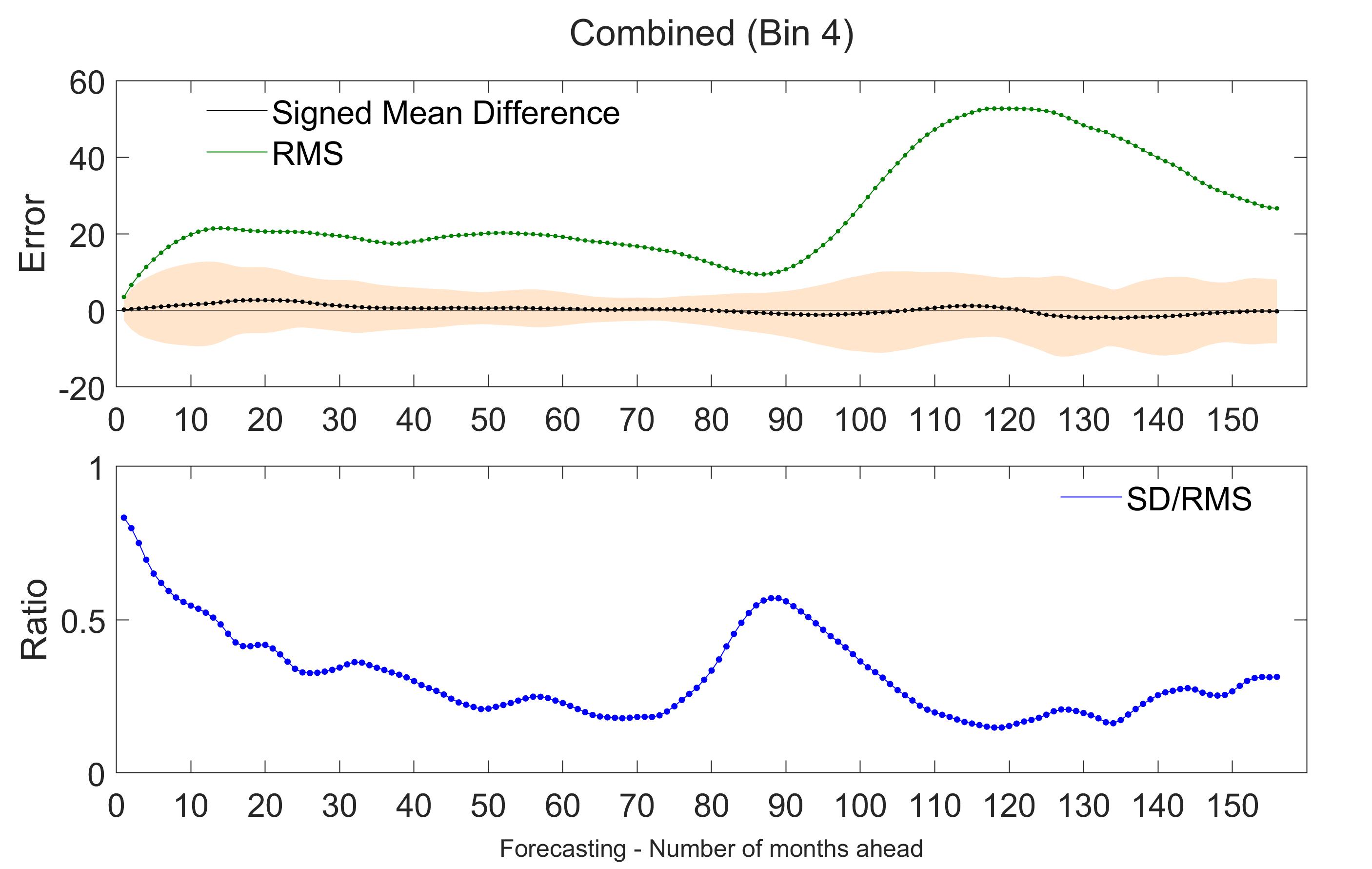} }
	\vspace{-0.55\textwidth}   % Overlays label
	\centerline{\large \bf 	\hspace{0.15\textwidth} \color{black}{(b)} \hfill}
	\vspace{0.50\textwidth} 

	\centerline{\includegraphics*[width= 0.82\textwidth, bb= 1cm 0cm 96cm 65cm,clip=] {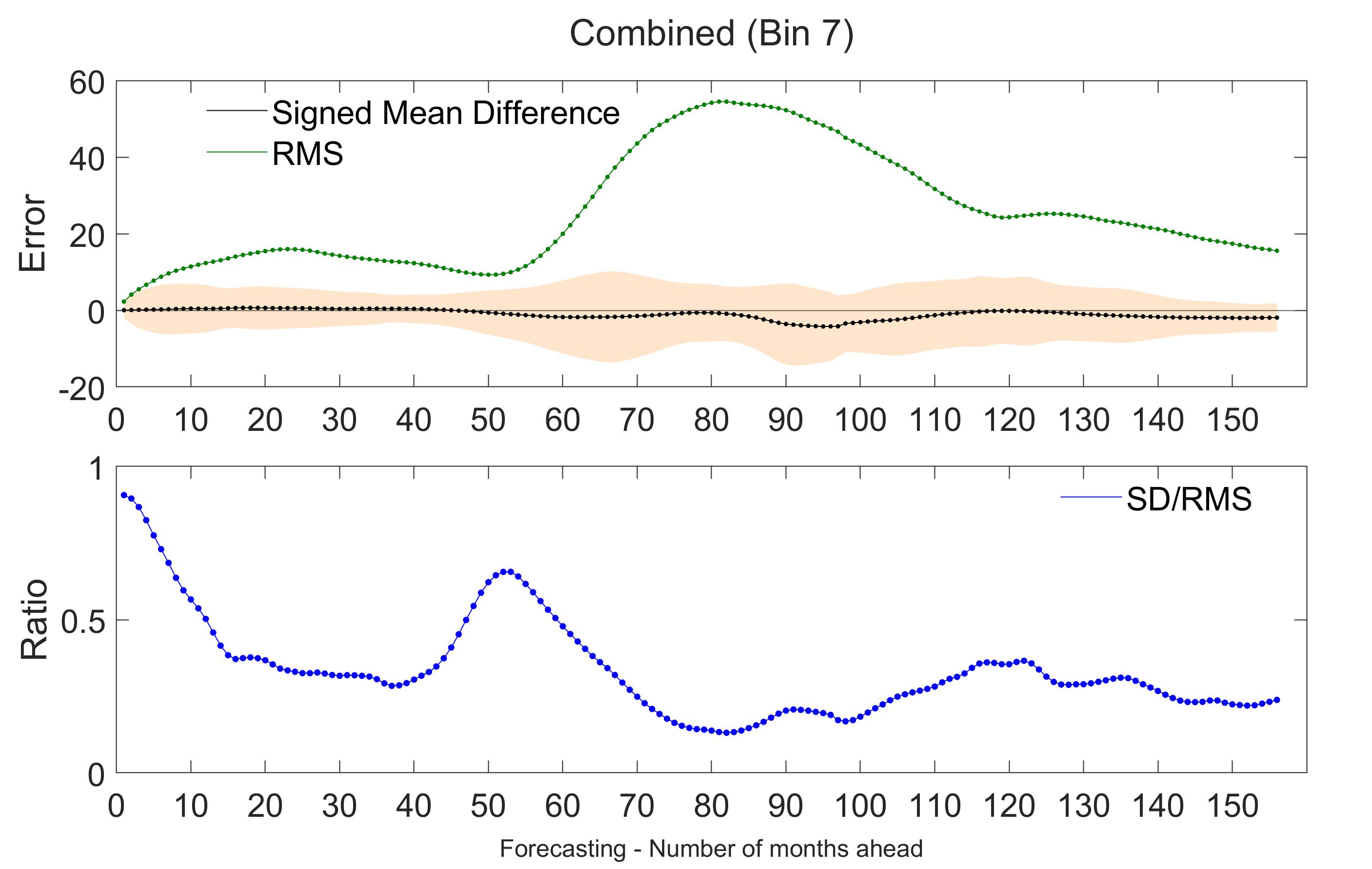} }
	\vspace{-0.55\textwidth}   % Overlays label
	\centerline{\large \bf 	\hspace{0.15\textwidth} \color{black}{(c)} \hfill}
	\vspace{0.50\textwidth} 

	\caption{Prediction error as a function of forward months, for all cycles, averaged for starting times in year 1 (a), 4 (b) and 7 (c) of the cycle. The plotted quantities are the same as in Figures\,\ref{Fig_RMSy1Cyc} and \ref{Fig_RMSCyc21_y1-4-7} for single solar cycles, with sums including here all predicted cycles (9 to 24).}
	\label{Fig_RMS_y1-4-7_Combi}
\end{figure}

In actual predictions, we cannot use the true observed cycle as a reference to evaluate the prediction errors. Therefore, the only applicable measure of uncertainty must be derived from the mean of all past predicted cycles. In Figure\,\ref{Fig_RMS_y1-4-7_Combi}, we show the same error measures, again for 1-year bins (year 1, 4 and 7), but where the RMS error $\delta_{sp}^{rms}$ is summed over all cycles from 10 to 24, as:
%%\begin{linenomath}
	\begin{equation}
		\delta_{p}^{rms}={\sqrt{\frac{\sum_{n=8}^{c-1} \sum_{s=1}^{N_{bin}} \left(\hat{S}_{sp}^n - S_{sp}^n \right)^2} {N_{c}\,N_{bin}}}} \label{EQ_RMSdiffbin}
	\end{equation}
%%\end{linenomath}
where $\hat{S}_{sp}^n$ and $S_{sp}^n$ are respectively the predicted and observed SSN for the same month, and the sum is over all starting times $s$ of all predictions in each 1 year bin (13 bins), and over all input cycles 8 to 24 for every bin (total number of predictions $N_t= N_{c}\,N_{bin} = 204$), with a given fixed prediction forward time $p$.

Here, the first important difference with the above RMS errors for individual cycles is that the mean difference $\delta_{sp}$ is low and does not deviated significantly from 0 (within one $\sigma_{sp}^{\delta}$). Therefore, the systematic deviations specific to each solar cycles are indeed efficiently eliminated. However, now, those cycle-to-cycle differences contribute to the standard deviation, which is now higher than for single cycles. As a consequence, the $\sigma_{sp}^{\delta}/\delta_{sp}^{rms}$ ratio also decreases from near-unity at month 1, but remains at a plateau between 0.2 and 0.3 for longer forward times. We note however that the ratio decreases steadily over the first 40 months, which is longer than for individual cycles. This suggests that when considering the average deviations of multiples cycles instead of a single cycle, the systematic bias associated with the ML default mean cycle is improved for predictions up to month $p$=40, thanks to the correction factor discussed in Sections\,\ref{Sec_CorrecTerm} and \ref{Sec_CorrecProperties}. 

$\delta_{sp}^{rms}$ and $\sigma_{sp}^{\delta}$ now also largely follow the the variation of the mean cycle, with a maximum around month 40 and a minimum around month 126. The ending minimum, and subsequent rise of the next cycle that follows, are higher, which corresponds to the higher dispersion of the SSN at the end of the cycle, and the corresponding artificially high mean-cycle, as diagnosed in Section\,\ref{Sec_MeanCycle}. This rise of errors towards the end of the predicted cycle can thus be traced to the growing contribution of cycle-length differences as the prediction date, counted from the start of the cycle (initial minimum), increases.This is the temporal smearing effect intrinsic to the single mean cycle that we described in Section\,\ref{Sec_MeanCycle}. 

Because of this effect, although the range of SN values decrease at the ending minimum, the standard deviation remains high. Therefore, the second minimum is marked by a secondary peak in the $\sigma_{sp}^{\delta}/\delta_{sp}^{rms}$ ratio, but the latter remains well below 1 (0.7). Finally, like for individual cycles, the whole pattern of variations shifts left by 36 months for predictions on year 1, 4 and 7, while remaining essentially constant. Only the first 20 to 70 months feature the same ramp to 0 for errors and 1 for the $\sigma_{sp}^{\delta}/\delta_{sp}^{rms}$ ratio, as mentioned above.
  
\subsection{RMS Error $\delta_{sp}^{rms}$ versus Calculated $\hat{\sigma}_{sp}^S$}

The above RMS error $\delta_{sp}^{rms}$ for all past cycles over 1-year intervals for predictions at starting times $s$ along the solar cycle should provide a good estimate of the statistical error that must be attached to each prediction at a specific time $s$.

\begin{figure}
	\centerline{
		\includegraphics*[width= 0.82\textwidth, bb= 1cm 0cm 96cm 65cm,clip=] {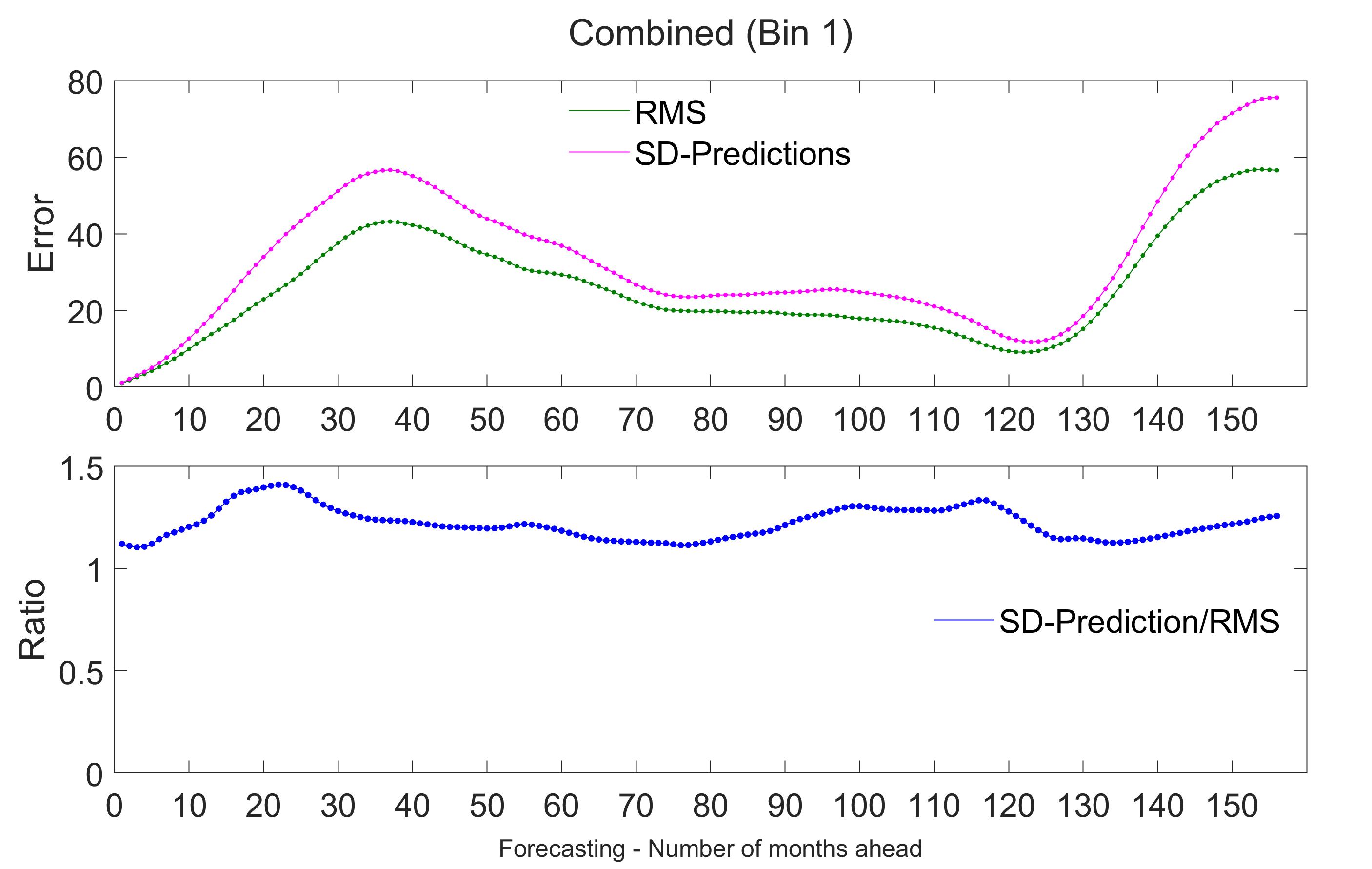} }
	\caption{Comparison of the RMS error $\delta_{sp}^{rms}$ over Cycles 10 to 24 with the standard deviation of predictions $\hat{\sigma}_{sp}^S$ given by Equation\,(\ref{eq_A47}), for year 1 of the predicted cycle. The upper plot shows together the RMS error (black) and the standard deviation $\hat{\sigma}_{sp}^S$ as a function of forward prediction time. The lower plots shows the ratio $\hat{\sigma}_{sp}^S/\delta_{sp}^{rms}$.}
	\label{Fig_RMSsigma_y1}
\end{figure}

\begin{figure}
	\centerline{
		\includegraphics*[width= 0.82\textwidth, bb= 1cm 0cm 96cm 65cm,clip=] {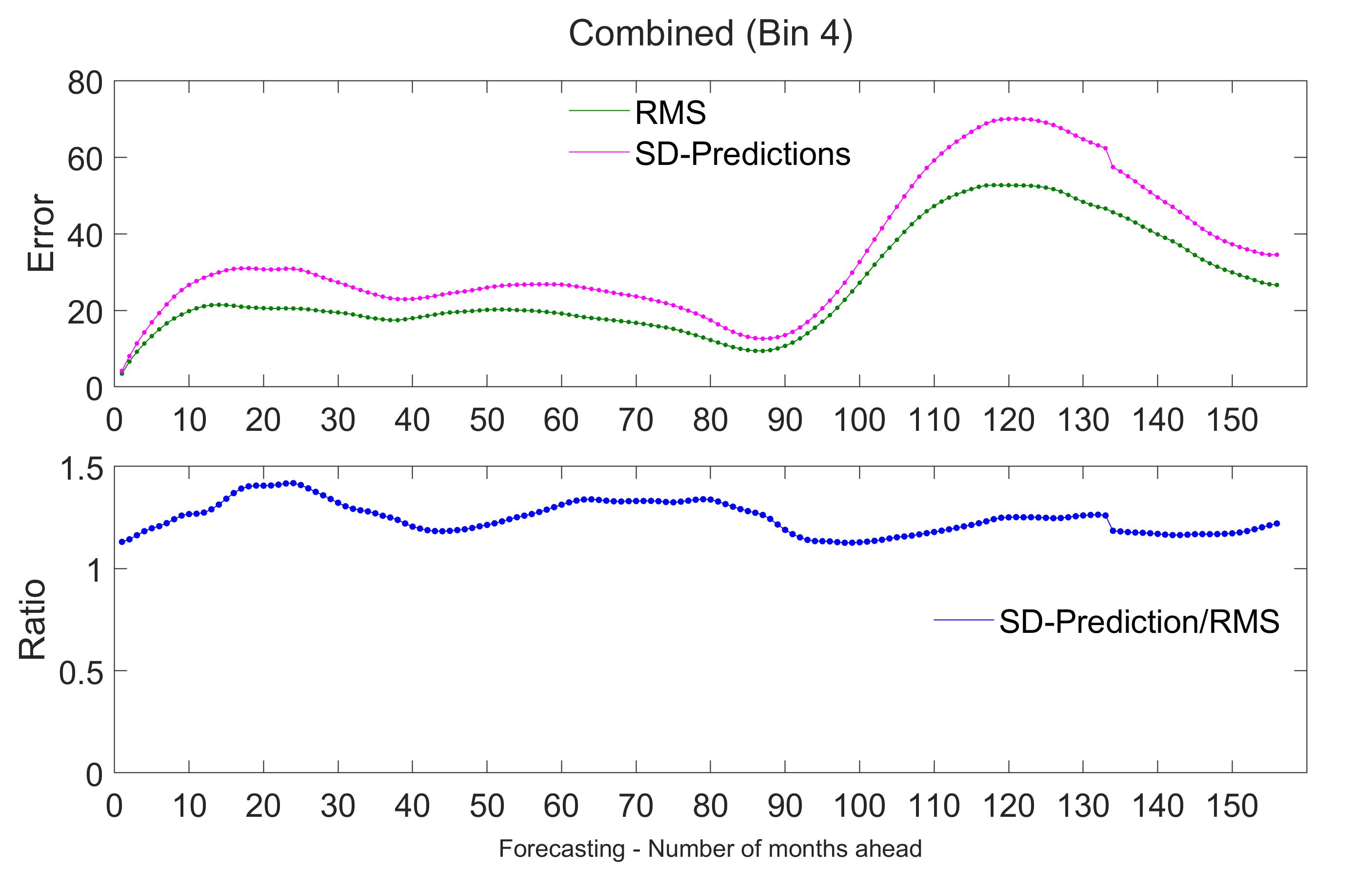} }
	\caption{Comparison of the RMS error $\delta_{sp}^{rms}$ over Cycles 10 to 24 with the standard deviation of predictions $\hat{\sigma}_{sp}^S$, for year 4 of the predicted cycle. The layout is the same as in Figure\,\ref{Fig_RMSsigma_y1}.}
	\label{Fig_RMSsigma_y4}
\end{figure}

\begin{figure}
	\centerline{
		\includegraphics*[width= 0.82\textwidth, bb= 1cm 0cm 96cm 65cm,clip=] {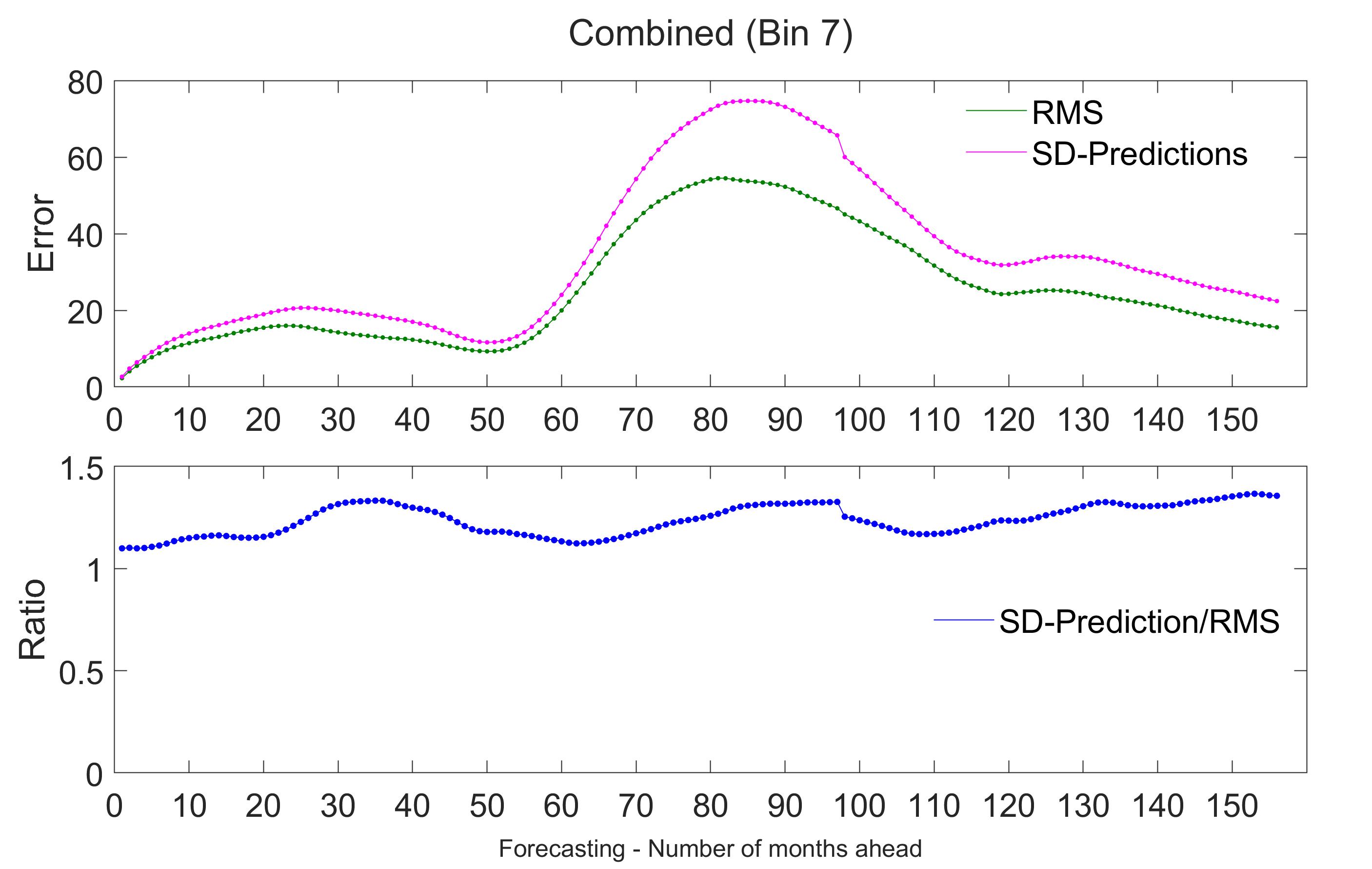} }
	\caption{Comparison of the RMS error $\delta_{sp}^{rms}$ over Cycles 10 to 24 with the standard deviation of predictions $\hat{\sigma}_{sp}^S$, for year 7 of the predicted cycle.  The layout is the same as in Figure\,\ref{Fig_RMSsigma_y1}.}
	\label{Fig_RMSsigma_y7}
\end{figure}

We can thus compare this statistics, based on Cycles 10 to 24, with the standard error derived mathematically in Section\,\ref{Sec_PredError} (Equation\,(\ref{eq_A47})). In Figures\,\ref{Fig_RMSsigma_y1}, \ref{Fig_RMSsigma_y4} and \ref{Fig_RMSsigma_y7}, we plot the value of the RMS error derived in the previous Section and the uncertainty $\hat{\sigma}_{sp}^S$ calculated for the mid-point of each 1-year interval, as well as the ratio between the two values. We show again the predictions for years 1, 4 and 7 into the predicted cycles.

We find a good match between the two error values in all cases. They vary in the same way over the solar cycle. This leads to a largely constant ratio, with a mean value of 1.3. The ratio is slightly higher during the ascending and descending phases of the cycle, and lower during the extrema, which indicates that the larger rms differences are associated with the evolution of solar activity during the 1-year intervals used in the statistics. The trend of solar activity during the temporal window just adds a contribution to the rms dispersion.

We thus obtain a good confirmation that $\hat{\sigma}_{sp}^S$ given by Equation\,(\ref{eq_A47}) provides a valid measure of the uncertainty of operational ML predictions. 

\section{Discussion: Practical Prediction Limits} \label{Sec_Discussion}
In their original article, \citet{McNishLincoln1949} limited the range of prediction to 18 months, though they do not explain how they chose this limit, which thus seems to be largely empirical. As in operational predictions, the SSN is only determined up to 6 months before the current month (last SN data), this thus simply corresponds to predictions over the coming 12 months.

Based on our above analysis, we may thus wonder what is the actual temporal limit up to which the ML predictions remain usable. The global evolution of the overall RMS error (Figure\,\ref{Fig_RMSallCycles}) suggested that the correction factor brings an actual improvement up to month 40. Beyond this limit, the RMS error ceases to increase, and stabilizes at a value corresponding to the range of values taken by the real cycles relative to the base mean cycle. Therefore, the predictions do not become entirely meaningless, but the long-range uncertainty (RMS error) is definitely large, at $\pm 35$ on average for all cycles and all prediction times $s$. 

A similar upwards ramp for small $p$ is also found in the RMS errors within 1-year bins. However, as can be seen in Figure\,\ref{Fig_RMS_y1-4-7_Combi}, the RMS error actually varies differently for predictions made at different times along the solar cycle, which suggests a dependency of the maximum prediction range on the phase of the solar cycle. 

\begin{figure}
	\centerline{
		\includegraphics*[width= 0.75\textwidth, bb= 7cm 0cm 97cm 65cm,clip=] {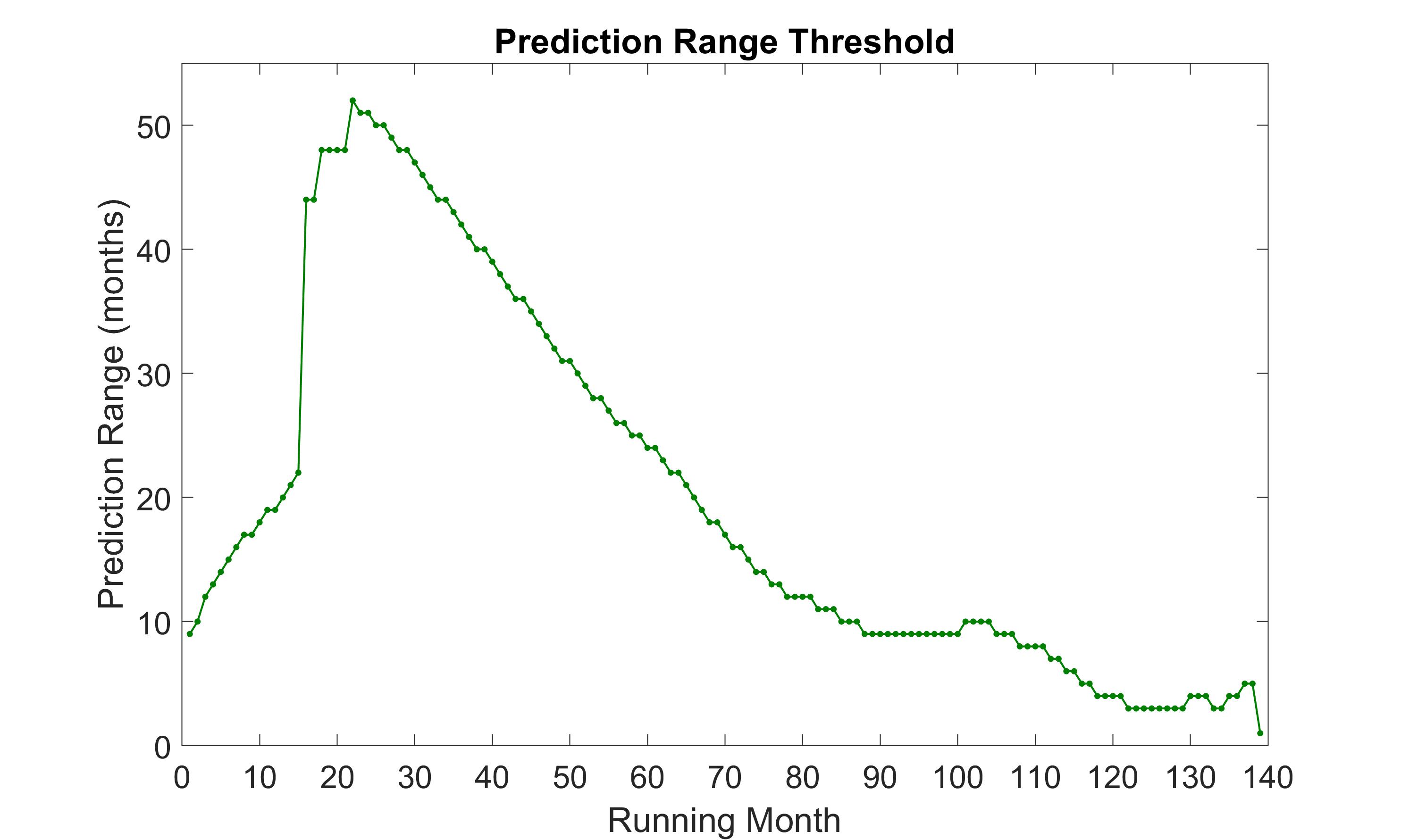} }
	\caption{Prediction range, in forward months,  for a 20\% percentage threshold on the prediction error $\hat{\sigma}_{sp}^S$.}
	\label{Fig_PredRange_Percent}
\end{figure}

For operational purposes, we can define a maximum tolerance on the predictions in percent of the predicted value, thus based on the $\hat{\sigma}_p/\hat{S}_p^c$ ratio. Adopting empirically a tolerance of 20\%, we derived the forward month $p$ for which this threshold is reached for each base month $s$ in the course of the cycle (Figure\,\ref{Fig_PredRange_Percent}). We find that the range largely exceeds 18 months during the main middle part of the cycle from 18 to 65 months, i.e. from 1.5 years in the early rising phase to 2 years past the mean maximum in the declining phase. Therefore, with this 20\% tolerance, reliable predictions can thus be derived well beyond the fixed 18-month limit, up to 50 months, i.e. 4 years forward. On the other hand, the range drops below 18 months, in the first year following the starting minimum of a cycle, and again at the ending minimum. In the latter case, the range essentially stops at the fixed moment of the ending minimum, because of the steep rise of the error after the minimum. Such short ranges mean that even short-term extrapolation of the last observed SSN are completely unreliable, and that predictions are useless well before the conventional 18-month range. This is consistent with the corresponding maxima of the error on the correction coefficient diagnosed in Section\,\ref{Sec_CorrecProperties}.

This adaptive prediction range could thus be implemented in operational predictions in order to provide the longest prediction possible at any time, to better fulfill the needs of Sun -- Earth applications (solar activity, geomagnetic impacts, radio propagation, etc.).

\section{Conclusions: Beyond the ML Method} \label{Sec_Conclusions}
In this article, we first gathered a full chronology of the development of the ML method, and its different implementations. We also developed the full mathematical derivation of the prediction uncertainty, confirming the original formula implemented in the heritage NOAA programs. This thus fully documents one of the earliest prediction methods developed for operational use, where detailed descriptions were either missing or scattered over different publications. Based on the established formulae, we could then derive key properties of the two main components on which the ML method is based, namely the mean of all past cycles and the correction term based on the last observed monthly SSN. Finally, based on hind-casts produced by the standard reference software for all months from Cycle 9 to 25, we then derived global statistics of the observed$-$predicted differences based on the actual record of the past SSN in order to determine the actual performance of the ML method, in particular as a function of the phase in the course of the solar cycle.

Our main conclusions are the following:
\begin{itemize}
	\item The original of the ML method, and in particular of the prediction errors is mathematically reconstructed and verified. The bulks statistics of hind-casts are in good agreement with the calculated error, further confirming its validity.
	\item The mean cycle is the dominant component in the ML predictions, in particular for large forward times $p>40$. As this mean does not take into account the varying duration of actual solar cycles, it suffers from a temporal smearing that increases from the beginning to the end of the cycle. It leads to an artificially flattened maximum and to a strong degradation of predictions around and beyond the ending minimum at month 130.
	\item The standard deviation of the mean cycle purely results from the differences between actual past cycles for a given time $s$, which result both from the different cycle amplitudes and different cycle lengths.  
	\item The ML correction term largely follows the cross-correlation $r_{sp}$ between the time $s$ of the last observed SSN and the forward prediction time $p$, but with a gain factor that can be much larger than unity. This occurs in the early and late phases of the solar cycle, creating an extreme sensitivity of the ML predictions to small variation of the SSN during the minima of the cycle, associated with large uncertainties.
	\item Overall, for long forward times $p$, the prediction errors are entirely dependent on the difference between the actual cycle and the mean cycle. They create a fixed pattern in absolute time counted from the starting minimum. This pattern is similar to the standard deviation of the mean cycle values. This property leads to a systematic downward bias of the predictions for cycles amplitude larger than the mean cycle, and an upward bias for cycles of lower amplitude than the mean cycle. In other words, high cycle maxima tend to be underestimated, while low maxima are overestimated.
	\item Only for short forward times $p$, the correction term helps to reduce the prediction error, which ramps up from almost 0 at $p=1$ to the standard deviation of the mean cycle for $p$ ranging from 10 to 50 months, depending on the phase of the cycle. The errors for the first predicted months are essentially random, and result mostly from the short-term randomness of solar activity, via the last observed monthly SSN, while for long-range predictions they become more systematic as mentioned above.
	\item The prediction errors vary in the course of the solar cycle by the combination of two primary effects:
	\begin{itemize}
		\item A shift of the long-term errors towards smaller forward times $p$ when the prediction time $s$ increases. Those errors reach high values around and past the ending minimum.
		\item Variations of the decline rate of the cross-correlation $r_{sp}$ for small forward $p$ times. This decline is particularly steep at the starting and ending minimum of the cycle.
	\end{itemize}
	\item Finally, we determined the forward time range over which usable predictions can be obtained, i.e. the precision falls within acceptable limits (operational tolerance set at 20\%, as described in Section\,\ref{Sec_Discussion}). We find that the maximum prediction range varies widely over the duration of the cycle, due to the complex combination of the above-mentioned effects, from less than 10 months up to 50 months (4 years). Therefore, the forward prediction time range is shorter than the conventional fixed 18-month range adopted by NOAA, at the beginning and the end of a cycle. On the other hand, over the middle of the cycle, this usable range proves to be much longer. 
	
\end{itemize}

Overall, as expected for this rather simple method, the ML approach suffers from several strong limitations that were already suspected by earlier studies \citep{HollandVaughan1984, Hildner1990, Fessant1996}. We thus find that the method essentially breaks down when transiting from one cycle to the next. Only short-term extrapolations are then possible, over time ranges that may be even shorter than 18 months. On the other hand, in the course of a cycle, roughly during 5 years surrounding the maximum ($s=10$ months to 60 months), longer-term predictions can be considered, but systematic biases appear for particularly high or low cycles. We conclude that those limitations are mostly tied to the fact that the ML predictions are founded on the use of the mean cycle, a single global average of all past cycles. The ML correction term can only partly compensate adverse properties of the underlying mean cycle:
\begin{itemize}
	\item A fixed amplitude that leads to systematic biases in predictions
	\item A fixed duration, which leads to a temporal mismatch with the actual times when the maximum and the transition to the next cycle is actually occurring. 
	\item The fixed duration also produces a smoothing of the extrema of the cycle and an elevation of the  minimum level at the end of the cycle. 
\end{itemize} 

Earlier attempts tried to reduce those limitations, with limited gains. By bringing all cycles to the same length, \cite{Niehuss1996} and \citet{Fessant1996} reduced the third effect, which helps to obtain a cycle profile more consistent with actual cycles. However, although it is a bit more realistic, this single mean cycle cannot be representative of the different rising and declining profiles of high and low cycles, and this mean cycle has still a fixed length, which does not solve the first two limitations. By also moving the initial node point from the starting minimum of the cycle to its maximum, \citet{Fessant1996} indeed slightly reduces the temporal mismatch at the end of the cycle, where it is maximum. However, as the mean cycle still has a fixed declining time, the three above limitations still largely remain.

Therefore, we conclude that improving the ML method and achieving a meaningful gain in its predictive performance requires abandoning at least part of its base principles. In order to go further while still using the past record of the SN as single base input, at least three main ingredients must be included in the prediction strategy, each addressing one of the three above limitations, namely:
\begin{itemize}
	\item Using a base mean cycle with an amplitude that itself is predicted on the base of the last observed SN, thus only including past cycles with amplitudes similar to the predicted maximum.  
	\item Taking into account the known relation between the rise time to the maximum and also of the cycle length with the amplitude of the cycle, i.e. the so-called Waldmeier effect \citep{Waldmeier1935}
	\item Using as template the actual shape of cycle in different amplitude ranges, without mixing cycles of different amplitude.
\end{itemize}   
However, such adaptations do not have much in common with the original ML scheme, and must be considered as distinct methods that just belong to the same general category of ``climatology-based'' prediction methods, as explained in our introduction. 

It turns out that such more advanced methods were implemented independently a long time ago and have been in routine operational use over the past decades, thus are also forming heritage standards against which other methods can be compared, including the ML method. The first one is the Standard Curve (SC) method by \citet{Waldmeier1968}, based on a set of averaged and interpolated mean solar cycles covering the whole range of possible amplitudes, which was created much earlier \citep{Waldmeier1935, Waldmeier1937PhDT}. The other one is the Combined Method (CM) introduced more recently by \cite{DenkmayrCugnon1997}. Beyond the adaptive derivation of a mean cycle of proper amplitude, this method also uses a precursor parameter, the aa geomagnetic index, to obtain a prediction of the next maximum when passing the end of the predicted cycle, and thus when purely sunspot-based methods fail completely, like we saw here with the ML method.

The discussion of those other heritage methods and a throughout comparison of their performance with the ML predictions go beyond the scope of this paper and should become the topic of future articles submitting the other methods to the same bulk bench-marking techniques as the one developed in this ML study. In this perspective, all tools developed and trained here on the ML method will prove invaluable for probing other methods in a fully equivalent way.

%%%%%%%%%%%%%%%%%%%%%%%%%%%%%%%%%%%%%%%%%%%%%%%%%%%%%%%%%%%%%%%%%%%
%% Acknowledgements

\begin{acks}
This work, and F.~Clette in particular, benefited from the base support to the World Data Center SILSO, which produces and distributes the international sunspot number used in this study, provided by Belgian Solar-Terrestrial Center of Excellence (STCE). T.~Podladchikova and S.~Jain acknowledge financial support from the European Union’s Horizon 2020 research and innovation program in the framework of the SOLARNET project.
\end{acks}

\vspace{\baselineskip}

%\textbf{Disclosure of Potential Conflicts of Interest}\\

%% Available additional data environments:
%% required: authorcontribution, fundinginformation, dataavailability
%% optional: materialsavailability, codeavailability
\begin{authorcontribution}
F.~Clette reconstructed, documented and re-programmed the original ML method. Using the resulting operational software, he did the bulk calculation of all past ML predictions, and contributed to their interpretation. F.~Clette also wrote a large part of the manuscript. 
S.~Jain processed the output of the bulk predictions, derived the global predicted$-$observed statistics and created many of the corresponding figures used in Section 5. T.~Podladchikova carried out the complete mathematical derivation of the error calculation, presented in Sections 3 and 4, and jointly with S.~Jain, contributed to the statistical diagnostics of the ML prediction uncertainties.

\end{authorcontribution}
\begin{fundinginformation}
The World Data Center SILSO, which produces the international sunspot number and mid-term solar-cycle predictions, is supported on an annual basis by Belgian Solar-Terrestrial Center of Excellence (STCE) funded by the Belgian Science Policy Office (BelSPo). T.~Podaldchikova and S.~Jain received financial support from the European Union’s Horizon 2020 research and innovation program under grant agreement No. 824135 (SOLARNET).
\end{fundinginformation}

\begin{dataavailability}
The ML operational predictions are freely distributed by the World Data Center SILSO, via the Forecast section of its Web site, at \url{https://www.sidc.be/SILSO/forecasts}.
Specific output data sets from this benchmarking study can be obtained by direct request to the corresponding author.
\end{dataavailability}
\begin{ethics}
\begin{conflict}
The authors declare that they have no conflicts of interest.
\end{conflict}
\end{ethics}

%%%%%%%%%%%%%%%%%%%%%%%%%%%%%%%%%%%%%%%%%%%%%%%%%%%%%%%%%%%%%%%%%%%%
%% Bibliography

% Using BibTeX
\bibliographystyle{spr-mp-sola}
\bibliography{MLmethod}  

\end{document}